\documentstyle[eqsecnum,preprint,aps]{revtex}
\tighten
\draft
\begin{document}
\widetext
\preprint{CLNS 96/1413}
\bigskip
\bigskip
\title{Asymmetric Orbifolds and Grand Unification}
\medskip
\author{Zurab Kakushadze\footnote{e-mail: 
zurab@hepth.cornell.edu \\
Address after September 1, 1996: Lyman Laboratory, Harvard University, 
Cambridge, MA 02138.} and S.-H. Henry Tye\footnote{e-mail:
tye@hepth.cornell.edu}}
\bigskip
\address{Newman Laboratory of Nuclear Studies,
Cornell University,
Ithaca, NY 14853-5001, USA}

\bigskip
\date{July 16, 1996}
\bigskip
\medskip
\maketitle
\begin{abstract}

{}We generalize the rules for the free fermionic string construction
to include other asymmetric orbifolds with world-sheet bosons.
We use these rules to construct various grand unified string models
that involve level-$3$ current algebras. We present the explicit 
construction of three classes of $3$-chiral-family grand unified models 
in the heterotic string theory. Each model has $5$ left-handed, and two 
right-handed families, and an adjoint Higgs. Two of them are $SO(10)$ and 
the third is $E_6$. With Wilson lines and/or varying moduli, we show how 
other $3$-family grand unified models, such as $SO(10)$ and $SU(5)$, 
may be constructed from them. 
\end{abstract}


\widetext
\section{Introduction}
\bigskip

{}The construction of string models has a long history. The number 
of consistent string models is clearly very large (one may consider 
various string models as different classical string vacua of a single 
theory; in this case, we are talking about the construction of classical 
vacua). The best understood string models are probably those obtained
via toroidal compactification, along with their orbifolds\cite{DHVW}. 
However, the orbifold construction is still not fully
explored. This is in part due to the lack of simple rules 
for constructing such models, 
in particular, asymmetric orbifolds\cite{NSV}.

{}The first class of asymmetric orbifold string models are the
free fermionic string models\cite{KLT}. Although this class of models 
is rather restrictive (allowing only multiple ${\bf Z}_2$ twists), 
the rules for such model building are quite simple. 
As a result, rather complicated models 
can be readily constructed. For example, the first level-$2$ grand unified
models were constructed using these rules\cite{lewellen}.
It is natural to attempt to generalize these
rules to include more general orbifolds. 
It is not surprising that the rules become more involved as one tries to 
include more general types of orbifolds.
At certain point, the rules will become too complicated to be of any 
practical use. It seems some intermediate ground is called for.
In this paper, we present explicit rules for the construction of 
a rather large class of, but not all, asymmetric orbifold models. 
There are many interesting orbifold and free fermionic string 
constructions in the literature\cite{interesting,try}. 
The rules should be useful 
for the book-keeping
of complicated model building, as is the case in the search for 
three-family grand unified string models.

{}The rules for the asymmetric orbifold construction attempt to combine
the approach of Ref\cite{NSV} with
those for the free fermionic string models\cite{KLT}. Here we shall
follow the notations of Ref\cite{KLST}. We shall 
impose the consistency requirements on the string one-loop partition 
function in the light-cone gauge: ($\em i$) one-loop modular invariance; 
($\em ii$) world-sheet supersymmetry (if present), 
which insures space-time Lorentz invariance in the covariant gauge; 
and ($\em iii$) the physically 
sensible projection; this means the contribution of a
space-time fermionic (bosonic) degree of freedom to the partition function 
counts as minus (plus) one. In all cases that can be checked, this 
condition plus the one-loop modular invariance and factorization imply 
multi-loop modular invariance. 

{}The building blocks of one-loop partition function of any specific 
string model are the (appropriate) characters of the 
world-sheet fermions and bosons. The fermion
characters are the same as the ones used in the free fermionic 
string models. The characters for bosons are combined from two types, 
those for twisted chiral bosons and those for chiral boson lattices. 
A key to obtaining relatively simple rules is the 
choice of basis for the chiral lattices. They are chosen so that, up to 
phases, all the chiral lattice characters are permuted under any modular 
transformation, as is the case for the chiral fermion characters. 
Our discussion shall focus on heterotic
strings compactified to four space-time dimensions.

{}As in the free fermionic string model constructions, the rules may
be used to build new models without direct reference to their original
partition functions and/or the characters. This turns out to be
useful because the partition function can get rather complicated,
while the rules are relatively easy to use.
In this paper, we consider abelian orbifolds.
For a general lattice, the sublattice invariant under a twist may be
difficult to identify. Sometimes, it is easier to start with a lattice
whose invariant sublattice is obvious, and then introduce background fields,
in particular Wilson lines, that commute with the twists. 
In fact, this
approach is very useful in the symmetric orbifold construction\cite{INQ}.
Here, our work can be considered as a generalization of their work to
the asymmetric orbifold case.
If the Wilson lines do not commute with the twists, we may first construct 
the model with the Wilson lines, and then perform the twists on the 
resulting model, as we shall do in this paper. Alternatively, we may
use the non-abelian orbifold construction. 
The rules given here can be used as a basis for further generalization to
the non-abelian orbifold case, which will be discussed elsewhere\cite{shiu}.

{} After summarizing the rules, we use them to construct various grand 
unified models in the $4$-dimensional heterotic string theory.
It is well known that, in field theory, adjoint Higgs
(or other appropriate higher dimensional) representation is
necessary for a grand unified
gauge group to break spontaneously to the
$SU(3) \otimes SU(2) \otimes U(1)$ gauge group of the standard model.
It is also known that, for current algebras at level 1, space-time
supersymmetry with chiral fermions does not co-exist with massless
scalar fields in the adjoint or higher dimensional representations of
the gauge group in heterotic string models. From these 
facts, one concludes that a grand unified model in the
superstring theory is possible only if the grand unified gauge
group comes from current algebras at levels higher than 1.
Grand unified models with level-$2$ current algebras are not difficult to 
construct\cite{lewellen,erler}. All models of this type have an even 
number of chiral families. 
There are many attempts searching for grand unification with $3$ chiral 
families, so far unsuccessful\cite{try}. 
Our own investigation is consistent with the observation that 
level-$2$ models do not yield $3$ chiral families.

{}The next obvious step is to go to higher levels. The central charge
of higher-level current algebras increases quite rapidly. 
Intuitively, the feature of three chiral families seems to come most 
naturally with a ${\bf Z}_3$ orbifold. String consistency naturally 
leads one to level-$3$ current algebras. As a result of this
consideration, a level-$3$ current algebra looks most promising.
Our search indicates that a ${\bf Z}_3$ orbifold alone does not yield 
$3$ families. However, a ${\bf Z}_6$ orbifold can yield 3 chiral families;
more specifically, it yields 5 left-handed and 2 right-handed families.
In the construction of $3$-family grand unification, we shall impose an
additional condition to help shrink the size of the moduli space of 
models: the presence of an asymtotically-free hidden sector
which can become strong above the electroweak scale.
In this paper, we consider three different asymmetric ${\bf Z}_6$ 
orbifolds. The largest grand unified gauge symmetries in these three 
orbifolds are $SO(10)$, $SO(10)$ and $E_6$, respectively. From 
these models, we may obtain other models by one of the three (not 
inequivalent) ways: 
(1) changing the moduli, 
(2) introducing/modifying the Wilson lines, and 
(3) giving some of the massless Higgs fields non-zero vacuum 
expectation values in the effective field theory.
In particular, it is easy to convert the $E_6$ model to another $SO(10)$
model. The massless spectra of the three $SO(10)$ models coming from the
three different asymmetric ${\bf Z}_6$ orbifolds are given in Table I. 
There, each of the $U(1)$ charges is normalized so that the lowest
allowed values are $\pm 1$, with conformal highest weight $r^2/2$,
where $r$ is the compactification
radius of the corresponding chiral world-sheet boson.
The states are grouped into sectors, the untwisted $U$ and the twisted
$T$ sectors. This grouping follows from the string construction, and
implements the selection rules for string couplings.

{}Exploring the moduli space of the first $SO(10)$ model, we find other
closely related $SO(10)$ models. Their massless spectra are given in 
Table II. The second and the third $SO(10)$ models in Table I are 
somewhat isolated points in their respective moduli space, since changing 
the moduli typically damages the strong-interacting feature
of the hidden sector. By turning 
on an appropriate Wilson line on the any of the $SO(10)$ models, 
an $SU(5)$ model appears. 
The same Wilson line acting on the $E_6$ model yields an $SU(6)$ model. 
The massless spectra of the $E_6$ model, the $SU(6)$ model, 
and their closely related $SU(5)$ model are given in Table III.
The same procedure can be used to obtain 
other variations of the above models. For example, we may break the
grand unified gauge symmetry to that of the standard model. We note that 
all the 
above models share the following features: \\
(1) $5$ left-handed and $2$ right-handed chiral families of the grand
unified gauge symmetry, \\
(2) one adjoint Higgs field, and \\
(3) an asymptotically-free $SU(2)$ hidden sector, whose fine structure 
coupling at the string scale is three times that of the grand unified 
gauge group.\\
Furthermore, many of them have a non-trivial (but not asymptotically-free)
gauge symmetry that may 
play the role of a horizontal symmetry, or as a messenger/intermediate 
sector. Although these models do not contain Higgs fields in higher 
representations ({\em i.e.}, other than adjoint) 
of the grand unified gauge group, it is possible that
higher-dimensional operators present in the string theory may play
equivalent roles. Clearly, detailed analyses are necessary to see if
this happens in any of the above models, and if supersymmetry is 
dynamically broken.

{}It is clear that the three types of ${\bf Z}_6$ orbifolds 
considered in this paper are quite similar. Although these three are
probably not exhaustive, a distinct feature of grand unification in 
string theory is the very limited number of possibilities, as compared
to standard model building. At the same time, the grand unified models 
presented here are quite complex.
This complicated structure hopefully bodes well for phenomenology.

{}The construction of the first supersymmetric 
$SO(10)$ grand unified model with $5$ left-handed and $2$ right-handed 
families was presented in Ref\cite{three}. This model has gauge symmetry
$SU(2)_1 \otimes SU(2)_3 \otimes SO(10)_3 \otimes U(1)^3$. 
Here, we give the details of that construction, using the rules developed
in this paper. Although the derivation of the rules is somewhat complicated,
the rules themselves, summarized at the end of section III, are relatively 
easy to use. So the reader may skip the derivation at first reading.
To get a better feeling of the model as well as the rules,
we present an alternative way to construct this model. 

{}In section II, we briefly review the fermion and the boson characters 
and other background material that we shall use later. In section III, 
we derive the rules for model building. To be specific, we consider 
only heterotic strings compactified to four spacetime dimensions. 
The rules for model building are summarized. For most of the applications
in this paper, a simplified version of the rules is used. 
The simplified rules are also summarized at the end of section III.
In section IV, we use these rules to construct the three-family $SO(10)$
grand unified string model given in Ref\cite{three}.
The massless spectrum of this model is given in the first column in Table I.
In section V, we discuss an alternative construction of the same model.
The quantum numbers of the particles are easier to identify in 
this construction. Section VI contains a discussion of the moduli 
space of the above construction, 
in which the above model is a special point in the moduli space. 
Another special point yields a model with gauge symmetry
$SU(2)_1 \otimes SO(10)_3 \otimes U(1)^4$. Generic points in this 
moduli space have smaller intermediate sectors. Their massless spectra are 
given in Table II. The second $SO(10)$ model is obtained by a 
slightly different ${\bf Z}_6$ orbifold. 
Although it has the same gauge symmetry as the first $SO(10)$ model, it has
a different spectrum. This model is 
discussed in section VII, while its massless spectrum is shown in
the second column in Table I. Section VIII discusses the third $SO(10)$
model, whose massless spectrum is shown in
the third column in Table I. The third $SO(10)$ model is intimately 
related to the $E_6$ model, which is discussed in section IX. 
There we also discuss how $SU(5)$, $SU(6)$ and other models 
may be obtained from the above models.
Section X contains the summary and remarks.
Some of the details are relegated to the appendices. 

\widetext
\section{Preliminaries}
\bigskip

{}String models are constructed by considering consistent one-loop
partition functions, which are built of particular linear combinations of
characters, which in turn are products of characters of 
world-sheet free fermions and (un-)twisted chiral bosons. After a 
general discussion of the framework, the characters are reviewed.
We shall discuss the particular form of the bosonic characters that
are suitable for our purpose.

\subsection{Framework}
\medskip

{}In this subsection we set up the framework for the remainder of this paper.
To be specific, we consider heterotic strings compactified to four
space-time dimensions. In the light-cone gauge, which we adopt, we have
the following world-sheet degrees of freedom:
One complex boson $\phi^0$ (corresponding to two transverse
space-time coordinates); three right-moving complex bosons 
$\phi^\ell_R$, $\ell=1,2,3$ (corresponding to six internal coordinates);
four right-moving complex fermions $\psi^r$, $r=0,1,2,3$
($\psi^0$ is the world-sheet superpartner of the right-moving component of
$\phi^0$, whereas $\psi^\ell$ are the world-sheet superpartners of 
$\phi^\ell_R$, $\ell=1,2,3$);
eleven left-moving complex bosons $\phi^\ell_L$, 
$\ell=4,5,...,14$ (corresponding to twenty-two internal coordinates). 
Before orbifolding, the 
corresponding string model has $N=4$ space-time supersymmetry and the
internal momenta span an even self-dual Lorentzian lattice $\Gamma^{6,22}$. 
   
{}It is convenient to organize the string states into sectors labeled by the
monodromies of the string degrees of freedom. Thus, consider
the sector where
\begin{eqnarray}\label{group1}
 &&\psi^r ({\overline z} e^{-2\pi i} ) =\exp (-2\pi i V^r_i)
 \psi^r ({\overline z}) ~,\nonumber\\
 &&\phi^\ell_R ({\overline z} e^{-2\pi i} )=\exp (-2\pi i T^\ell_i)
 \phi^\ell_R ({\overline z}) -U^\ell_i~,~~~\ell=1,2,3 ~,\\
 &&\phi^\ell_L ({z} e^{2\pi i} )=\exp (-2\pi i T^\ell_i ) 
 \phi^\ell_L ({z}) +U^\ell_i ~,~~~\ell=4,...,14 \nonumber
\end{eqnarray}
(Note that $\phi^0 ({z} e^{2\pi i}, {\overline z} e^{-2\pi i} )=\phi^0 ({z},
{\overline z})$ since $\phi^0$ corresponds to space-time coordinates).
These monodromies can be combined into a single vector
\begin{equation}\label{V_i}
 V_i =(V^0_i (V^1_i ~~  (T^1_i ,U^1_i ))(V^2_i ~~  (T^2_i ,U^2_i ))
 (V^3_i ~~  (T^3_i ,U^3_i ))\vert\vert (T^4_i ,U^4_i)...
 (T^{14}_i, U^{14}_i )) ~.
\end{equation}
The double vertical line separates the right- and left-movers.
Without loss of generality we can restrict the values of $V^r_i$ and 
$T^\ell_i$ as follows: $-{1\over 2} \leq
V^r_i < {1\over 2}$; $0\leq T^\ell_i <1$. So $V^r_i=0~(-{1\over 2})$
corresponds to a Neveu-Schwarz (Ramond) fermion.
(A complex boson (fermion) with boundary condition
$T^\ell_i~(V^r_i)=0$ or $1\over 2$ can be split into two real bosons
(fermions)). The shifts $U^\ell_i$ can be
combined into a real $(6,22)$ dimensional Lorentzian
vector ${\vec U}_i$ defined up to
the identification ${\vec U}_i \sim {\vec U}_i+{\vec P}$,
where ${\vec P}$ is an arbitrary vector of $\Gamma^{6,22}$.

{}The monodromies (\ref{group1}) can be viewed as fields $\Phi$ (where $\Phi$
is a collective notation for the fields $\psi^r$, $\phi^\ell_R$ and 
$\phi^\ell_L$) being periodic $\Phi ({z} e^{2\pi i}, {\overline z} 
e^{-2\pi i} )=\Phi ({z},{\overline z})$ up to the identification
$\Phi \sim g(V_i) \Phi g^{-1} (V_i)$, where $g(V_i)$ is an element of the
orbifold group $G$. In this paper we will only consider the cases where
$G$ is abelian. For two elements $g(V_i)$ and $g(V_j)$ 
to commute, we must have
\begin{equation}
 (1-\exp(-2\pi i T^\ell_i ))U^\ell_j -(1-\exp(-2\pi i T^\ell_j ))U^\ell_i
 \in \Gamma^{6,22}~.
\end{equation}
In this paper we will confine our attention to cases where 
$U^\ell_i =0$ if $W^\ell_j \not=0$, and $U^\ell_j =0$ if $W^\ell_i \not=0$.
(The more general case
can be conveniently considered in the framework of non-abelian orbifolds 
discussed in \cite{shiu}).

{}This leads us to a simpler form of 
$V_i$ where instead of having 
a double entry $(T^\ell_i ,U^\ell_i)$ for each complex boson
we will specify
either $T^\ell_i$ (whenever $T^\ell_i \not=0$, in which case $U^\ell_i =0$),
or $U^\ell_i$ (whenever $U^\ell_i \not=0$, in which case $T^\ell_i =0$).
To keep track of whether a given entry 
corresponds to a twist or a shift, it is convenient to 
introduce {\em auxiliary} vectors
\begin{equation}
 W_i=(0 (0~W^1_i) (0~W^2_i) (0~W^2_i)
 \vert\vert 
 W^4_i ~...~W^{14}_i )~.
\end{equation}  
The entries $W^\ell_i$ are defined as follows: $W^\ell_i={1\over 2}$ if  
$T^\ell_i \not=0$; $W^\ell
=0$, otherwise. For example, 
\begin{eqnarray}
 &&V_i =(V^0_i (V^1_i ~T^1_i)(V^2_i ~T^2_i)(V^3_i ~T^3_i)\vert\vert
 U^4_i ~...~U^{13}_i ~T^{14}_i )~,\\
 &&W_i=(0 (0~{1\over 2})^3 \vert\vert 0^{10}~{1\over 2})~,
\end{eqnarray} 
where $T^1_i$, $T^2_i$, $T^3_i$ and $T^{14}_i$ correspond to the twists, 
$U^4_i$,...,$U^{13}_i$ correspond to the shifts, and $V^r_i$, $r=0,1,2,3$,
specify the fermionic spin structures.

{}The notation we have introduced proves convenient in describing the 
sectors of a given string model based on the orbifold group $G$. For $G$ to be 
a finite discrete group, the element $g(V_i)$ must have a finite order $m_i
\in {\bf N}$,
{\em i.e.} $g^{m_i} (V_i)=1$. This implies that $V^r_i$ and $T^\ell_i$ must be
rational numbers, and the shift vector ${\vec U}_i$ must be a rational 
multiple of a vector in $\Gamma^{6,22}$; that is, $m_i V^r_i , m_i T^\ell_i 
\in {\bf Z}$, and $m_i {\vec U}_i \in \Gamma^{6,22}$. To describe all the 
elements of the group $G$, it is convenient to introduce the set of 
generating vectors $\{ V_i \}$ such that 
${\overline {\alpha V}}={\bf 0}$ if and only if
$\alpha_i \equiv 0$. Here ${\bf 0}$ is the null vector: 
\begin{equation}
 {\bf 0}=(0 (0~0)^3 \vert\vert 0^{11})~.
\end{equation}
Also, $\alpha V \equiv \sum_i \alpha_i V_i$
(The summation is defined as $(V_i +V_j )^\ell=V^\ell_i +
V^\ell_j$), $\alpha_i$ being integers that 
take values from $0$ to $m_i -1$. The overbar notation is defined as follows:
${\overline {\alpha V}} \equiv \alpha V -\Delta(\alpha)$, and the components
of ${\overline {\alpha V}}$ satisfy 
$-{1\over 2}\leq {\overline {\alpha V}}^{\, r} <{1\over 2}$, 
$0\leq {\overline {\alpha T}}^{, \ell}<1$; 
here $\Delta^r (\alpha),\Delta^\ell (\alpha)  
\in {\bf Z}$. So the elements of the group $G$ are in one-to-one 
correspondence with the vectors ${\overline {\alpha V}}$ and will be denoted
by $g({\overline {\alpha V}}$). It is precisely the abelian nature of $G$ that
allows this correspondence (by simply taking all possible linear
combinations of the generating vectors $V_i$).

{}Now we can identify
the sectors of the model. They are labeled by the vectors
${\overline {\alpha V}}$, and in a given sector ${\overline {\alpha V}}$ the
monodromies of the string degrees of freedom are given by 
$\Phi ({z} e^{2\pi i}, {\overline z} 
e^{-2\pi i} )=g({\overline {\alpha V}}) \Phi (z,{\overline z}) g^{-1} 
({\overline {\alpha V}})$.   

{}$G$ is a symmetry of the
Hilbert space of the original string model with $N=4$ supersymmetry
compatible with the operator algebra of the underlying 
(super)conformal field theory. If $\vert 
\chi \rangle$ is a state in the original Hilbert space, 
$g({\overline {\alpha V}})\vert \chi \rangle$ 
(where there must exist a representation of $g({\overline {\alpha V}})$ via
the vertex operators of the theory)
also belongs to the same Hilbert space. One
consequence of this requirement is that $G$ must (anti-)commute with the 
(super-)Virasoro algebra.
This implies the following {\em supercurrent} constraint
\begin{equation}\label{supercurrent}
 V^\ell_i + T^\ell_i =V^0_i \equiv s_i ~(\mbox{mod}~1)~,~~~\ell=1,2,3~.
\end{equation}
Here $s_i$ is the monodromy of the supercurrent 
${\overline S}({\overline z}e^{-2\pi i} )=\exp(2\pi is_i)
{\overline S}({\overline z})$, which must satisfy $s_i \in {1\over 2}{\bf Z}$.
Then the sectors with ${\overline {\alpha V}}^{\, 0} =0$ 
give rise to space-time bosons, while
the sectors with ${\overline {\alpha V}}^{\, 0}
=-{1\over 2}$ give rise to space-time fermions. 

{}Now we may write the one-loop partition function as
\begin{equation}\label{orbifold_partition_function}
 {\cal Z} = {1\over {\prod_{i} m_i }} \sum_{\alpha ,\beta} \mbox{Tr} 
  ( q^{H^L_{\overline {\alpha V}}}
 ~{\overline q}^{H^R_{\overline {\alpha V}}} g^{-1} ({\overline {\beta V}}))\\
       ={1\over {\prod_{i} m_i }} \sum_{\alpha ,\beta}
 C^{\overline {\alpha V}}_{\overline {\beta V}}
 {\cal Z}^{\overline {\alpha V}}_{\overline {\beta V}} ~,
\end{equation}
where $H^L_{\overline {\alpha V}}$ and $H^R_{\overline {\alpha V}}$ are the
left- and right-moving Hamiltonians, respectively. The trace is 
taken over the states in the Hilbert space corresponding to the sector
${\overline {\alpha V}}$.
The sum over $\alpha$ sums all the different sectors while the sum
over $\beta$ provides the appropriate GSO projections.
We shall first discuss the characters 
${\cal Z}^{\overline {\alpha V}}_{\overline {\beta V}}$. 
They are constructed so that modular transformations simply permute them.
Then we shall solve for the coefficients 
$C^{\overline {\alpha V}}_{\overline {\beta V}}$ to yield a consistent
${\cal Z}$.

\subsection{Fermion and Boson Characters}
\medskip

{}To derive relatively simple rules for building asymmetric orbifold models,
we will confine our attention to the orbifolds
with twists whose orders are co-prime numbers
(The  order of a twist generated by a vector ${\overline {\alpha V}}$
is defined as
the smallest positive integer $t({\overline {\alpha V}})$, such that 
$\forall\ell~t({\overline {\alpha V}}) {\overline {\alpha T}}^{\,\ell}
\in {\bf Z}$; note that $t({\overline {\alpha V}})$ is a divisor of 
$m({\overline {\alpha V}}))$.
In this case, a given model can be 
viewed as being generated by a single twist $V^*$ of order $t^*=\prod_i
t_i$, such that ${\overline {\alpha V}}=
{\overline {\alpha^* V^*}}$,
where $\alpha^* /t^* =\sum_i \alpha_i /t_i~({\mbox{mod}~1})$, 
and $\alpha^*$ takes values $0,1,...,t^* -1$. 
Although it is possible to work 
in the $V^*$ basis, it will prove convenient to do so in the $\{V_i \}$ basis. 

{}In a given sector ${\overline {\alpha V}}$, the right- and left-moving
Hamiltonians are given by the corresponding sums of the Hamiltonians for 
individual string degrees of freedom. The Hilbert space in the 
${\overline {\alpha V}}$ sector is given by
the momentum states $\vert {\vec P}_
{\overline {\alpha V}} +\alpha {\vec U}, {\bf n} \rangle$, and also 
the states obtained from these states by the actions of the fermion and boson
creation operators (oscillator excitations). $\bf n$ is a collective
notation for $n_i=0,1,...,t_i -1$. 
In the untwisted sectors, that 
is, sectors ${\overline {\alpha V}}$ with $t({\overline {\alpha V}})=1$, 
we have ${\vec P}_
{\overline {\alpha V}} \in \Gamma^{6,22}$. In the twisted sectors 
${\overline {\alpha V}}$ with $t({\overline {\alpha V}}) 
\not=1$, we have ${\vec P}_
{\overline {\alpha V}} \in 
{\tilde I}({\overline {\alpha V}})$, where ${\tilde I}({\overline {\alpha V}})$
is the lattice dual to the lattice $I({\overline {\alpha V}})$, which
in turn is the sublattice of $\Gamma^{6,22}$ invariant under the action of
the twist part of the group element $g({\overline {\alpha V}})$. 
The ground states $\vert {\vec 0}, {\bf n} \rangle$ in the 
${\overline {\alpha V}}$ sector  
appear with certain degeneracies $\xi({\overline {\alpha V}}, {\bf n})$, 
which are non-negative integers.
In the untwisted sectors $\xi({\overline {\alpha V}}, {\bf n}) =1$
if ${\bf n}={\bf 0}$, and zero otherwise. 
In the twisted sectors the situation is more involved.
First, the sum of $\xi({\overline {\alpha V}}, {\bf n})$ 
over all ${\bf n}$ is the number
of fixed points $\xi({\overline {\alpha V}})$
in the ${\overline {\alpha V}}$ sector \cite{NSV}:
\begin{equation} 
 \label{xigo}
 \sum_{\bf n} \xi({\overline {\alpha V}}, {\bf n}) 
 =\xi({\overline {\alpha V}}) =
 (M({\overline {\alpha V}}))^{-1/2} \prod_\ell[2\sin
 (\pi {\overline {\alpha T}}^{\, \ell})] ~,
\end{equation}
where the product over $\ell$ does not include the terms with 
${\overline {\alpha T}}^{\, \ell} =0$, and $M({\overline {\alpha V}})$ is the 
determinant of the metric of $I({\overline {\alpha V}})$. Second,
since different fixed points generically
have different phases (monodromies) under the
action of twists, the states corresponding to different fixed points
contribute with different phases into the partition function. These 
contributions are related by the requirements of modular
invariance and physically sensible projection. These constraints unambiguously
fix the degeneracies $\xi({\overline {\alpha V}}, {\bf n})$, and we 
will give the procedure to determine them in a moment.

{}Since the momentum 
in the twisted sectors belongs to a shifted 
${\tilde I}({\overline {\alpha V}})$ lattice, to satisfy the level matching 
condition, this lattice must have a prime
$N({\overline {\alpha V}})$, where $N({\overline {\alpha V}})$ 
is the smallest positive integer such that 
for all vectors ${\vec P}\in {\tilde I}({\overline {\alpha V}})$, 
$N({\overline {\alpha V}}) {\vec P}^2\in 2{\bf Z}$; 
moreover, for the corresponding characters to have the correct modular 
transformation properties ({\em i.e.}, so that they are permuted under both 
$S$- and $T$-transformations), either $N
({\overline {\alpha V}}) =1$ (in which
case $I({\overline {\alpha V}})$ is an even self-dual lattice), or 
$N({\overline {\alpha V}})=
t({\overline {\alpha V}})$ (in which case $I({\overline {\alpha V}})$ is 
even but not self-dual).

Now the characters ${\cal Z}^{\overline {\alpha V}}_{\overline {\beta V}}$
can be computed as products of
building blocks, or contributions of individual string degrees of freedom,
which are reviewed in Appendix A. The result 
can be written as a product of the corresponding fermion and boson 
characters:
\begin{equation}
 {\cal Z}^{\overline {\alpha V}}_{\overline {\beta V}}=
 {\overline Z}^{\overline {\alpha V}}_{\overline {\beta V}}
 {\cal Y}^{\overline {\alpha V}}_{\overline {\beta V}} ~.
\end{equation}
The fermion characters 
${\overline Z}^{\overline {\alpha V}}_{\overline {\beta V}}$ read:
\begin{equation}
 {\overline Z}^{\overline {\alpha V}}_{\overline {\beta V}}=
 \prod_{r} {\overline Z}^{{\overline {\alpha V}}^{\, r}}_
 {{\overline {\beta V}}^{\, r}}
\end{equation}
(The characters ${\overline Z}^v_u$ for a right-moving fermion are complex
conjugates of the characters $Z^v_u$ for a left-moving fermion given by
(\ref{fermionZ})).

{}The boson characters
${\cal Y}^{\overline {\alpha V}}_{\overline {\beta V}}$ read:
\begin{equation}
 {\cal Y}^{\overline {\alpha V}}_{\overline {\beta V}}=
 \xi^{\overline {\alpha V}}_{\overline {\beta V}}
 Y^{\overline {\alpha V}}_{\overline {\beta V}}
 \prod_{\ell=1}^3
 {\overline X}^
 {{\overline {\alpha T}}^{\,\ell}}_{{\overline {\beta T}}^{\, \ell}} 
 \prod_{\ell=4}^{14}
 X^{{\overline {\alpha T}}^{{\,\ell}}}_{{\overline {\beta T}}^{{\, \ell}}}
\end{equation}
(The characters ${\overline X}^v_u$ for a right-moving boson are complex
conjugates of the characters $X^v_u$ for a left-moving boson
given by (\ref{bosonX})). 
The product over $\ell$ does {\em not} include terms with 
${\overline {\alpha T}}^{\,\ell} ={\overline {\beta T}}^{\, \ell} =0$.

{}The characters $Y^{\overline {\alpha V}}_{\overline {\beta V}}$ read:
\begin{equation}
 Y^{\overline {\alpha V}}_{\overline {\beta V}}=
 {1\over{\eta^d  {\overline \eta}^{d^\prime} }}
 \sum_{{\vec P}\in {\cal I}^* ({\overline {\alpha V}}, {\overline {\beta V}})}
 q^{{1\over 2}({\vec P}^L+\alpha{\vec U}^L )^2}
 {\overline q}^{{1\over 2}({\vec P}^R+\alpha{\vec U}^R )^2}
 \exp(-2\pi i(\beta{\vec U} \cdot {\vec P}+{1\over 2} \nu({\overline 
 {\alpha V}}, {\overline {\beta V}}) {\vec P}^2))~, 
\end{equation}
where ${\vec P}^L$, 
${\vec P}^R$, and ${\vec U}^L$, ${\vec U}^R$, are the 
left- and right-moving parts of the momentum and shift vectors, respectively.
The lattice ${\cal I}^* ({\overline {\alpha V}}, {\overline {\beta V}})$
(with Lorentzian metric $((-)^{d^\prime},(+)^d)$) 
is the sublattice of ${\tilde I}({\overline  
{\alpha V}})$ invariant under the twist part of ${\overline {\beta V}}$.
Here, $\nu ({\overline {\alpha V}} ,{\overline {\beta V}})$ is
an integer taking value between $0$ and $N({\overline {\alpha V}})-1$,
defined as
\begin{equation}
 \alpha^* \nu({\overline {\alpha V}} ,{\overline {\beta V}}) =
 \beta^* ~(\mbox{mod}~N({\overline {\alpha V}}))~,
 ~~~\alpha^* \not=0~,
\end{equation}
and $\nu({\overline {\alpha V}},{\overline {\beta V}})\equiv 0$, otherwise.
We note that we have included the phase
$\nu({\overline {\alpha V}} ,{\overline {\beta V}})$
${\vec P}^2_{\overline {\alpha V}}/2$ in the above definition of
$Y^{\overline {\alpha V}}_{\overline {\beta V}}$
so that the resulting characters have simple modular 
transformation properties; in particular, they are permuted
under $S$- and $T$-transformations (See Appendix A for a more detailed
discussion).

{}The fixed point phase $f_i ({\overline {\alpha V}}, {\bf n})$ of the
$\xi({\overline {\alpha V}},{\bf n})$ fixed points is given by
\begin{equation}\label{fixedpointphases}
 f_i ({\overline {\alpha V}}, {\bf n}) =n_i /t_i ~.
\end{equation}
Then the coefficients $\xi^{\overline {\alpha V}}_{\overline {\beta V}}$
in ${\cal Y}^{\overline {\alpha V}}_{\overline {\beta V}}$ are given by:
\begin{equation} 
 \label{xito}
 \xi^{\overline {\alpha V}}_{\overline {\beta V}}=\sum_{\bf n} 
 \xi({\overline {\alpha V}},{\bf n}) \exp(-2\pi i \beta 
 f({\overline {\alpha V}},{\bf n}) )~.
\end{equation}
Following Eq.(\ref{xigo}), we see that 
$\xi^{\overline {\alpha V}}_{\overline {\beta V}}=\xi({\overline {\alpha V}})$
for $t({\overline {\beta V}})=1$.
Note that $\xi^{\overline {\alpha V}}_{\overline {\beta V}}$ are
non-zero integers. 

{}Now we can collect the phases in 
${\cal Z}^{\overline {\alpha V}}_{\overline {\beta V}}$
in a given sector ${\overline {\alpha V}}$,
\begin{equation}\label{totalchar}
 {\cal Z}^{\overline {\alpha V}}_{\overline {\beta V}} \equiv
 \mbox{Tr} ( q^{H^L_{\overline {\alpha V}}}
 ~{\overline q}^{H^R_{\overline {\alpha V}}} h^{-1} ({\overline {\beta V}})) ~,
\end{equation}
where $g^{-1}({\overline {\beta V}})=
C^{\overline {\alpha V}}_{\overline {\beta V}} h^{-1}({\overline {\beta V}})$,
and
\begin{equation}
 h^{-1} ({\overline {\beta V}}) =
 \exp (-2\pi i (\beta V \cdot {\cal N}_{\overline {\alpha V}}
 +\beta f({\overline {\alpha V}}, {\bf n}) +{1\over 2}
 \nu({\overline {\alpha V}} ,{\overline {\beta V}})
 {\vec P}^2_{\overline {\alpha V}} ))~.
\end{equation}
Here ${\cal N}_{\overline {\alpha V}}=(
N^r_{\overline {\alpha V}},J^\ell_{\overline {\alpha V}},
{\vec P}_{\overline {\alpha V}})$, {\em i.e.}, 
${\cal N}_{\overline {\alpha V}}$ is a Lorentzian vector whose components are
the right-moving complex fermion
number operator $N^\ell_{\overline {\alpha V}}$,
the generators of twists $J^\ell_{\overline {\alpha V}}$, and the shifts
${\vec P}_{\overline {\alpha V}}$ (see Appendix A). Note that both operators
$N^\ell_{\overline {\alpha V}}$ and $J^\ell_{\overline {\alpha V}}$ 
are normalized to have integer eigenvalues.
The dot product is understood to have the following signature:
\begin{equation}\label{dotproduct}
 \beta V \cdot {\cal N}_{\overline {\alpha V}} \equiv
  \beta{\vec U} \cdot {\vec P}_{\overline {\alpha V}} -
 \sum_{r} ({\beta V})^r N^r_{\overline {\alpha V}}
 + \sum_{\ell:{\mbox{right}}}
 ({\beta T})^\ell J^\ell_{\overline {\alpha V}}
 -\sum_{\ell:{\mbox{left}}}
 ({\beta T})^\ell J^\ell_{\overline {\alpha V}}~.
\end{equation}
The dot product of the vectors $\beta {\vec U}$ and ${\vec P}_
{\overline {\alpha V}}$ has the Lorentzian metric
$\mbox{diag}((-)^6 , (+)^{22})$.

{}Now, the above characters are designed to have relatively simple modular 
transformation properties. Thus, various 
$\xi^{\overline {\alpha V}}_{\overline {\beta V}}$
are defined as follows:
\begin{eqnarray}
 \xi^{\overline {\alpha V}}_{\overline {\beta V}} =&&
   \xi^{\overline {\beta V}}_{\overline {-\alpha V}}
   \exp(2\pi i f({\overline {\alpha V}}, {\overline {\beta V}}))\nonumber\\
 &&(1
   +(M({\overline {\alpha V}}, {\overline {\beta V}}))^{-1/2} \delta_{\alpha^*,
   0}(1-\delta_{\beta^*,0}) +(M({\overline {\alpha V}}, {\overline {\beta V}}))
   ^{1/2} \delta_{\beta^*,0} (1-\delta_{\alpha^*,0}))\times\nonumber\\
 &&\prod_{\ell} (1+
   [2\sin(\pi {\overline {\alpha T}}^{\,\ell})]\delta_{{\overline {\beta T}}
   ^{\,\ell},0}
   +[2\sin(\pi {\overline {\beta T}}^{\,\ell})]^{-1} \delta_{
   {\overline {\alpha T}}^{\,\ell}, 0} ) ~,\\
  \xi^{\overline {\alpha V}}_{\overline {\beta V}} =&&
  \xi^{\overline {\alpha V}}_{\overline {\beta V -\alpha V +V_0}} ~,
\end{eqnarray} 
where the product over $\ell$ does {\em not} include terms with 
${\overline {\alpha T}}^{\,\ell}= {\overline {\beta T}}^{\,\ell}=0$.
$M({\overline {\alpha V}}, {\overline {\beta V}})$ is the
determinant of the metric of ${\cal I}({\overline {\alpha V}}, {\overline 
{\beta V}})$ which is the sublattice of ${I}({\overline  
{\alpha V}})$ invariant under the twist part of ${\overline {\beta V}}$;
note that in general ${\tilde {\cal I}}({\overline {\alpha V}},
{\overline {\beta V}}) \not=
{\cal I}^* ({\overline {\alpha V}}, {\overline {\beta V}})$.
 
Since $\xi^{\overline {\alpha V}}_{\overline {\beta V}}$ are non-zero 
integers, the phases $f({\overline {\alpha V}}, {\overline {\beta V}})$ are
zero or $1/2$ (mod 1). The above relations, and the requirement that
$\xi({\overline {\alpha V}}, {\bf n})$ 
be non-negative integers then unambiguously
fix the phases $f({\overline {\alpha V}}, {\overline {\beta V}})$ and
degeneracies $\xi({\overline {\alpha V}}, {\bf n})$. 

{}The above relations between different 
$\xi^{\overline {\alpha V}}_{\overline {\beta V}}$ are such that 
under the $S$- and $T$-modular transformations the characters
${\cal Z}^{\overline {\alpha V}}_{\overline {\beta V}}$ transform as follows:
\begin{eqnarray}\label{calZ}
 &&{\cal Z}^{\overline {\alpha V}}_{\overline {\beta V}}
 \stackrel{S}{\rightarrow} \exp(2\pi i 
 \varphi ({\overline {\alpha V}} ,{\overline {\beta V}}))
 {\cal Z}^{\overline {\beta V}}_{\overline {-\alpha V}} ~, \nonumber \\
 &&{\cal Z}^{\overline {\alpha V}}_{\overline {\beta V}}
 \stackrel{T}{\rightarrow} \exp(2\pi i
 ({1\over 2}\varphi ({\overline {\alpha V}} ,{\overline {\alpha V}})) 
  +{1\over 2})
 {\cal Z}^{\overline {\alpha V}}_{\overline {\beta V-\alpha V +V_0 }}~.
\end{eqnarray}
Here $V_0$ is the vector with $-1/2$ ({\em i.e.}, Ramond) entry for each 
world-sheet fermion and zero otherwise:
\begin{equation}
 V_0 =(-{1\over 2}(-{1\over 2}~0)^3 \vert\vert 0^{11})~.
\end{equation}
According to the above modular transformation properties of 
${\cal Z}^{\overline {\alpha V}}_{\overline {\beta V}}$, $V_0$ is always among
the generating vectors $\{V_i\}$.

{}The phases $\varphi ({\overline {\alpha V}}, {\overline {\beta V}})$
are defined
as follows:
\begin{equation}
 \varphi ({\overline {\alpha V}}, {\overline {\beta V}})=
 ({\overline {\alpha V}}-W
 ({\overline {\alpha V}}))\cdot
 ({\overline {\beta V}}-W({\overline {\beta V}}))+
 f({\overline {\alpha V}}, {\overline {\beta V}}) +
 \chi ({\overline {\alpha V}}, {\overline {\beta V}}) ~,
\end{equation}
where $\chi ({\overline {\alpha V}}, {\overline {\beta V}})$ phases are 
defined in Appendix A. The dot product of two vectors 
${\overline {\alpha V}}$ and ${\overline {\beta V}}$ is defined as in 
(\ref{dotproduct}). For example,
\begin{equation}
 V_i \cdot V_j =
  {\vec U}_i \cdot {\vec U}_j -
  \sum_{r} V^r_i V^r_j
  +\sum_{\ell:{\mbox{right}}} T^\ell_i T^\ell_j
  -\sum_{\ell:{\mbox{left}}} T^\ell_i T^\ell_j ~.
\end{equation} 
and there is a factor of ${1\over 2}$ for real fermions or twisted
real bosons.

\widetext
\section{Orbifold Rules}
\bigskip

{}In this section we derive the rules for constructing consistent string models
in the framework discussed in section II (This derivation for a special case
of asymmetric orbifolds was given in \cite{ZKHT}). The contribution to 
the orbifold one-loop 
partition function is a linear combination of the characters 
${\cal Z}^{\overline {\alpha V}}_{\overline {\beta V}}$
given above. Now we are ready to consider the coefficients 
$C^{\overline {\alpha V}}_{\overline {\beta V}}$ in ${\cal Z}$ 
(\ref{orbifold_partition_function}).

{}The coefficients $C^{\overline {\alpha V}}_{\overline {\beta V}}$
must be such that ${\cal Z}$ 
(\ref{orbifold_partition_function}) is modular invariant.
Taking into account the modular transformation properties (\ref{calZ}), we have
the following constraints on the coefficients 
$C^{\overline {\alpha V}}_{\overline {\beta V}}$ coming from the requirement 
of modular invariance of (\ref{orbifold_partition_function}):
\begin{eqnarray}\label{CmodinvS}
 S:~~~&&C^{\overline {\alpha V}}_{\overline {\beta V}}
 \exp(2\pi i \varphi({\overline {\alpha V}}\cdot
 {\overline {\beta V}}) ) =
 C^{\overline {\beta V}}_{\overline {-\alpha V}} ~,\\
\label{CmodinvT}
 T:~~~&&C^{\overline {\alpha V}}_{\overline {\beta V}}
 \exp(2\pi i({1\over 2}
 \varphi({\overline {\alpha V}},{\overline {\alpha V}}) +{1\over 2})) =
 C^{\overline {\alpha V}}_{\overline {\beta V-\alpha V +V_0 }}~.
\end{eqnarray}
In addition to (\ref{CmodinvS}) and (\ref{CmodinvT}) we require that, 
for any physical sector labeled by ${\overline {\alpha V}}$, the sum 
over $\beta$'s in (\ref{orbifold_partition_function}) form a proper 
projection with multiplicity
$\xi ({\overline {\alpha V}}, {\bf n})$. Specifically, this means that

\begin{equation}\label{fourier}
 {1\over {\prod_i m_i}}
 \sum_{\beta} C^{\overline {\alpha V}}_{\overline {\beta V}} 
 g^{-1} ({\overline {\beta V}}) = e^{2\pi i \alpha s} 
 \xi ({\overline {\alpha V}}, {\bf n}) \eta(
 {\overline {\alpha V}}, {\cal N}_{\overline {\alpha V}}, {\bf n}) ~,
\end{equation}
where $s_i$ is given by Eq.(\ref{supercurrent}). Now,
$\eta({\overline {\alpha V}}, {\cal N}_{\overline {\alpha V}}, 
{\bf n})$ takes values $0$ (${\em i.e.}$, projected out) or $1$ 
(${\em i.e.}$, kept), depending on the values 
of $\alpha_i$, ${\cal N}_{\overline {\alpha V}}$, ${\bf n}$.
This is precisely the physically sensible projection; space-time bosons 
contribute into the partition function with the weight plus one, whereas 
space-time fermions contribute with the weight minus one, each with degeneracy
$\xi ({\overline {\alpha V}}, {\bf n})$ 
due to the number of fixed points in the twisted sectors.
As a consequence, ${\cal Z}$ becomes
\begin{equation}
 {\cal Z}= \sum_{\alpha} \mbox{Tr} ( q^{H^L_{\overline {\alpha V}}} 
 ~{\overline q}^{H^R_{\overline {\alpha V}}}  e^{2\pi i \alpha s}
 \xi ({\overline {\alpha V}}, {\bf n})
 \eta({\overline {\alpha V}}, {\cal N}_{\overline {\alpha V}}, {\bf n}) )~.
\end{equation}

{}The formal solution to (\ref{fourier}) is given by
\begin{equation}\label{C_coeffs}
 C^{\overline {\alpha V}}_{\overline {\beta V}} =
 \exp(2\pi i [\beta\phi
 ({\overline {\alpha V}})+{\alpha s} ])~,
\end{equation}
where the phases $\phi_i ({\overline {\alpha V}})$ are such that
\begin{equation}\label{matching}
 m_i\phi_i ({\overline {\alpha V}}) =0~(\mbox{mod}~1)~.
\end{equation}

{}The phases $\phi_i ({\overline {\alpha V}})$ are constrained due to
(\ref{CmodinvS}) and (\ref{CmodinvT}):
\begin{eqnarray}\label{S}
S:~~~&&\beta \phi({\overline {\alpha V}}) +\alpha \phi({\overline {\beta V}})
 +\alpha s +\beta s +\varphi({\overline {\alpha V}},
 {\overline {\beta V}}) =0~(\mbox{mod}~1)~,\\
\label{T}
 T:~~~&&\alpha\phi({\overline {\alpha V}}) +\phi_0 ({\overline {\alpha V}})
 +\varphi({\overline {\alpha V}},
 {\overline {\alpha V}}) +{1\over 2} =0~(\mbox{mod}~1)~.
\end{eqnarray}

{}Let us introduce 
\begin{equation}
 \phi_j (V_i) \equiv -k_{ij} -s_j ~.
\end{equation}
Then the general solution to (\ref{S}) is given by 
\begin{equation}\label{phases}
 \phi_i ({\overline {\alpha V}}) =\sum_j
 k_{ij} \alpha_j + s_i -\varphi (V_i ,{\overline {\alpha V}}) 
 ~(\mbox{mod}~1)~,
\end{equation}
and the following constraints:
\begin{eqnarray}\label{k_ij}
 &&k_{ij}+k_{ji} =\varphi (V_i,V_j)~({\mbox{mod}~ 1})~,\\
 \label{phi_ij}
 &&\varphi({\overline {\alpha V}},{\overline {\beta V}})=
 \sum_i (\alpha_i \varphi(V_i, {\overline {\beta V}}) +
        \beta_i  \varphi(V_i, {\overline {\alpha V}})) -
 \sum_{i,j} \alpha_i \beta_j  \varphi (V_i,V_j)~({\mbox{mod}~ 1})~.      
\end{eqnarray}
For Eq.(\ref{phases}) to be compatible with Eq.(\ref{T}), the following 
constraints must also be satisfied:
\begin{eqnarray}\label{k_ii}
 &&k_{ii}+k_{i0} +s_i-{1\over 2} \varphi (V_i,V_i)=0~({\mbox{mod}~ 1})~,\\
 \label{phi_ii}
 &&{1\over 2} \varphi({\overline {\alpha V}},{\overline {\alpha V}})=
 \sum_i \alpha_i \varphi(V_i, {\overline {\alpha V}}) -
 {1\over 2}\sum_{i,j} \alpha_i \alpha_j  \varphi (V_i,V_j)
 +V_0 \cdot \Delta({\overline {\alpha V}})
 ~({\mbox{mod}~ 1})~.
\end{eqnarray}

{}Finally, let us rewrite the constraint (\ref{matching}) in terms of
$k_{ij}$ and $\varphi({\overline {\alpha V}},{\overline {\beta V}})$:
\begin{eqnarray}\label{mk}
 &&k_{ij} m_j=0~({\mbox{mod}~ 1})~,\\
 \label{mphi}
 &&m_i (\varphi(V_i,{\overline {\alpha  V}})-
 \sum_j \varphi(V_i, V_j) \alpha_j ) =0~({\mbox{mod}~ 1})~.
\end{eqnarray}
Note that not all of the above constraints are independent. Thus, 
(\ref{k_ii}) and (\ref{mk}) imply (\ref{k_ij}) for $i=j$; (\ref{phi_ii})
implies (\ref{phi_ij}) for $\alpha_i \equiv \beta_i$.

{}Let us summarize the above constraints.
The structure constants $k_{ij}$ must satisfy the following constraints:
\begin{eqnarray}
 \label{k1}
 &&k_{ij} +k_{ji} =\varphi(V_i ,V_j) ~({\mbox{mod}~ 1})~,~~~i\not= j~,\\
 \label{k3}
 &&k_{ii} +k_{i0} +s_i -{1\over 2} \varphi(V_i ,V_i) =0 ~({\mbox{mod}~ 1})~,\\
 \label{k2}
 &&k_{ij}m_j =0 ~({\mbox{mod}~ 1})~
\end{eqnarray}
Note that there is {\em no} summation over repeated indices here.
The phases $\varphi({\overline {\alpha V}},{\overline {\beta V}})$ must be such
that (${\overline {\alpha V}}\not={\overline {\beta V}}$ in (\ref{phi1})) 
\begin{eqnarray}\label{phi1}
 &&\varphi({\overline {\alpha V}},{\overline {\beta V}})=
 \sum_i (\alpha_i \varphi(V_i, {\overline {\beta V}}) +
        \beta_i  \varphi(V_i, {\overline {\alpha V}})) -
 \sum_{i,j} \alpha_i \beta_j  \varphi (V_i,V_j)~({\mbox{mod}~ 1})~,\\
 \label{phi2}
 &&{1\over 2} \varphi({\overline {\alpha V}},{\overline {\alpha V}})=
 \sum_i \alpha_i \varphi(V_i, {\overline {\alpha V}}) -
 {1\over 2}\sum_{i,j} \alpha_i \alpha_j  \varphi (V_i,V_j)
 +V_0 \cdot \Delta({\overline {\alpha V}})
 ~({\mbox{mod}~ 1})~,\\
 \label{phi3}
 &&m_i (\varphi(V_i,{\overline {\alpha  V}})-
 \sum_j \varphi(V_i, V_j) \alpha_j ) =0~({\mbox{mod}~ 1})~.      
\end{eqnarray}
{}Next we turn to identifying the spectrum of states. 
All the states are projected out of the sum in 
(\ref{orbifold_partition_function}) except those satisfying 
\begin{equation}\label{sgf}
 V_i \cdot {\cal N}_{\overline {\alpha V}} +f_i ({\overline {\alpha V}}, 
 {\bf n})
 +{{\alpha^* t^*}\over{2t_i}}{\vec Q}^2({\vec P}_{\overline {\alpha V}}) 
 =\sum_j k_{ij} \alpha_j + s_i -\varphi (V_i ,{\overline {\alpha V}}) 
 ~(\mbox{mod}~1)~.
\end{equation}
This is the spectrum generating formula. The definition of 
${\vec Q}({\vec P}_{\overline {\alpha V}})$ will be given in a moment.

{}The above rules can be simplified if
\begin{equation}
\label{simpli}
 W({\overline {\alpha V}}) \cdot W({\overline {\beta V}}) +
 W({\overline {\alpha V}}) \cdot \Delta({\overline {\beta V}}) +
 \Delta({\overline {\alpha V}}) \cdot W({\overline {\beta V}}) +
 f({\overline {\alpha V}}, {\overline {\beta V}}) +
 \chi({\overline {\alpha V}},{\overline {\beta V}}) = 0 ~.
\end{equation}
In this case $\varphi({\overline {\alpha V}},{\overline {\beta V}})$ becomes
\begin{equation}
\varphi({\overline {\alpha V}},{\overline {\beta V}})=
{\overline {\alpha V}}\cdot {\overline {\beta V}}+{\alpha V}
\cdot W({\overline {\beta V}}) + {\beta V} \cdot W({\overline {\alpha V}})
\end{equation}
This simplification allows us to rewrite the rules as follows.
First, the structure constants must satisfy the
following constraints (after a redefinition of the structure
constants $k_{ij} \rightarrow k_{ij}+W_i \cdot V_j$):
\begin{eqnarray}
 \label{k1z}
 &&k_{ij} +k_{ji} =V_i \cdot V_j ~({\mbox{mod}~ 1})~,~~~i\not= j~,\\
 \label{k2z}
 &&k_{ii} +k_{i0} +s_i -{1\over 2} V_i \cdot V_i =0 ~({\mbox{mod}~ 1})~,\\
 \label{k3z}
 &&(k_{ij}-W_i \cdot V_j)m_j =0 ~({\mbox{mod}~ 1})~.
\end{eqnarray}
The constraints (\ref{phi1}), (\ref{phi2}) and (\ref{phi3}) are
also simplified to:
\begin{equation}\label{mphiz}
 m_i V_i \cdot (W({\overline {\alpha V}}) -\sum_j \alpha_j W_ j)=
 0 ~({\mbox{mod}~ 1})~.
\end{equation}
and the spectrum generating formula now reads:
\begin{equation}\label{sgfz}
 V_i \cdot {\cal N}_{\overline {\alpha V}} +f_i ({\overline {\alpha V}},
 {\bf n})
 +{{\alpha^* t^*}\over{2t_i}}{\vec Q}^2({\vec P}_{\overline {\alpha V}})
 =\sum_j k_{ij} \alpha_j + s_i -V_i \cdot ({\overline {\alpha V}} -
 W({\overline {\alpha V}}))
 ~(\mbox{mod}~1)~.
\end{equation}
Here ${\vec Q}({\vec P}_{\overline {\alpha V}}) 
\in {\tilde I}({\overline {\alpha V}})$ is an arbitrary vector such that
${\vec P}_{\overline {\alpha V}}-\alpha^*
{\vec Q}({\vec P}_{\overline {\alpha V}}) 
\in I({\overline {\alpha V}})$. Also recall that
\begin{equation}\label{dotproduct1}
 V_i \cdot {\cal N}_{\overline {\alpha V}} \equiv
  {\vec U}_i \cdot {\vec P}_{\overline {\alpha V}} -
 \sum_{r} V^r_i N^r_{\overline {\alpha V}}
 + \sum_{\ell:{\mbox{right}}} T^\ell_i J^\ell_{\overline {\alpha V}}
 - \sum_{\ell:{\mbox{left}}} T^\ell_i J^\ell_{\overline {\alpha V}} ~.
\end{equation}
Note that in the twisted sectors ($t({\overline {\alpha V}}) \not= 1$), 
the states with quantum numbers ${\bf n}$ 
appear with the multiplicity $\xi ({\overline {\alpha V}}, {\bf n})$. 

{}The states that satisfy the spectrum generating formula include 
both on- and off-shell states.
The on-shell states must satisfy the additional constraint that
the left- and right-moving energies are equal. In the 
${\overline {\alpha V}}$ sector they are given by:
\begin{eqnarray}\label{LRenergy}
 E^L_{\overline {\alpha V}} =&&-1 +\sum_{\ell:~\mbox{left}}
 \{ {1\over 2} {\overline {\alpha T}}^{\,\ell} (1-{\overline 
 {\alpha T}}^{\,\ell})+
 \sum_{q=1}^{\infty} [(q+{\overline {\alpha T}}^{\,\ell} -1)n^\ell_q +
(q-{\overline {\alpha T}}^{\,\ell} ){\overline n}^{\,\ell}_q ]\} +\nonumber\\
 &&\sum_{q=1}^{\infty} q (n^0_q +{\overline n}^{\,0}_q )+{1\over 2}
 ({\vec P}^L_{\overline {\alpha V}}+\alpha{\vec U}^L)^2 ~,\\
 \label{LRenergy1}
 E^R_{\overline {\alpha V}} =&&-{1\over 2}+\sum_{\ell:~\mbox{right}}
 \{ {1\over 2} {\overline {\alpha T}}^{\,\ell} (1-{\overline {\alpha T}}
 ^{\,\ell})+
 \sum_{q=1}^{\infty} [(q+{\overline {\alpha T}}^{\,\ell} -1)m^\ell_q +
 (q-{\overline {\alpha T}}^{\,\ell} ){\overline m}^{\,\ell}_q ]\} +\nonumber\\
 &&\sum_{q=1}^{\infty} q (m^0_q +{\overline m}^{\,0}_q )+{1\over 2}
 ({\vec P}^R_{\overline {\alpha V}}+\alpha{\vec U}^R)^2 +\nonumber\\
 &&\sum_{r}
 \{ {1\over 2} ({\overline {\alpha V}}^{\, r})^2+
 \sum_{q=1}^{\infty} [(q+{\overline {\alpha V}}^{\, r} -{1\over 2} )k^r_q +
 (q-{\overline {\alpha V}}^{\, r}-{1\over 2 }){\overline k}^{\, r}_q ]\} ~.
\end{eqnarray}
Here $n^\ell_q$ and ${\overline n}^{\,\ell}_q$ are occupation numbers for the 
left-moving bosons $\phi^\ell_L$, 
whereas $m^\ell_q$ and ${\overline m}^\ell_q$ are those
for the right-moving bosons 
$\phi^\ell_R$. These take non-negative integer values.
$k^r_q$ and ${\overline k}^{\, r}_q$ are the occupation numbers for the 
right-moving fermions, and they take only two values: $0$ and $1$.
The occupation numbers are directly related to the boson and fermion number 
operators. For example, $N^r_{\overline {\alpha V}} =\sum_{q=1}^{\infty}
(k^r_q -{\overline k}^{\, r}_q)$; also, (for the left-moving degrees of
freedom, for example) 
$J^{\ell}_{\overline {\alpha V}}=\delta_{{\overline {\alpha T}}^{\,\ell} ,0}
{\cal J}^{\ell}-\sum_{q=1}^{\infty} 
(n^{\ell}_q -{\overline n}^{\, \ell}_q)$, where ${\cal J}^{\ell}$ is the 
angular momentum operator acting on the momentum states. 

{}Let us summarize the rules. To construct a consistent orbifold model, 
start with a $N=4$ space-time supersymmetric four-dimensional
heterotic string model \cite{Narain} with the internal momenta spanning an
even self-dual lattice
$\Gamma^{6,22}$ that possesses a $G=\otimes_i {\bf Z}_{t_i}$ symmetry
($t_i$ are co-primes). 
The sublattice $I_i \in \Gamma^{6,22}$ invariant under the ${\bf Z}_{t_i}$
twist must be such that $N_i =1$ or 
$t_i$. Now one can introduce a set
of vectors ${V_i}$ (which includes $V_0$) that correspond to a particular 
embedding of the orbifold group $G$. Each $\{V_i\}$ must satisfy 
(\ref{supercurrent}).
If we can find a set $\{V_i, k_{ij}\}$ that
satisfy the constraints (\ref{k1z}), (\ref{k2z}), (\ref{k3z}) and
(\ref{mphiz}), the constraint (\ref{simpli})
follows automatically and we have a consistent orbifold model. 
The complete spectrum (on- and off-shell states) 
of the model is given by  the spectrum generating formula (\ref{sgfz}), 
which together with the left/right energy formulae (\ref{LRenergy}) and
(\ref{LRenergy1}) determine
the on-shell physical spectrum. 
In the remaining sections, we shall use this simplified set of rules 
to construct various grand unified string models. 
Only when the set $\{V_i, k_{ij}\}$ does not
satisfy the simplified set of constraints do we
need to check if it satisfies the more general set of constraints
(\ref{k1}), (\ref{k3}), (\ref{k2}), (\ref{phi1}) and (\ref{phi2}).
We consider such an example in Appendix B.

\widetext
\section{A Three-Family $SO(10)_3$ Model}
\bigskip

{}In this section we shall use the simplified version of the 
rules derived in section III to construct the model described in 
Ref\cite{three}, ${\em i.e.}$, the $SO(10)_3$ model with 
three generations of chiral matter fields. We start from
an $N=4$ space-time supersymmetric
Narain model, which we will refer to as $N0$, 
with the lattice $\Gamma^{6,22}=\Gamma^{2,2} \otimes
\Gamma^{4,4} \otimes \Gamma^{16}$. Here $\Gamma^{2,2} =\{(p_R 
\vert\vert p_L ) \}$ is an even self-dual Lorentzian lattice with
$p_R ,p_L \in {\tilde 
\Gamma}^2$ ($SU(3)$ weight lattice), and $p_L - p_R \in \Gamma^2$
($SU(3)$ root lattice). Similarly, $\Gamma^{4,4} =\{(P_R 
\vert\vert P_L ) \}$ is an even self-dual Lorentzian lattice with
$P_R ,P_L \in {\tilde 
\Gamma}^4$ ($SO(8)$ weight lattice), $P_L - P_R \in \Gamma^4$
($SO(8)$ root lattice). $\Gamma^{16}$ is the ${\mbox{Spin}}(32)/{\bf 
Z}_2$ lattice. This model has $SU(3) \otimes SO(8) \otimes SO(32)$ gauge group.
It is useful to recall that there are four irreducible representations 
(irreps) in the $SO(2n)$ 
weight lattice: the identity ${\bf 0}$, which stands for the null vector,
the vector ${\bf v}$ , the spinor ${\bf s}$, and 
the anti-spinor ${\overline {\bf s}}$ (also known as the conjugate
${\bf c}$), with conformal dimensions $0$, $1/2$, $n/8$ and $n/8$
respectively. For $SO(2)$, we shall use $0, V={\pm 1}, S=1/2$ and $C=-1/2$ 
to designate the weights. The gauge bosons come from the ${\bf 0}$ weight.

{}Next, consider the model generated by the following set of vectors acting on 
the $N0$ model:
\begin{eqnarray}
 \label{n0tn1}
 &&V_0 =(-{1\over 2} (-{1\over 2}~ 0)^3 \vert\vert 0_r^3 \vert 0_r^{16} ) ~,\\
 &&V_1 =( 0 (0~{e_1\over 2})(0~a_1)(0~b_1) \vert\vert 0^3 \vert ({1\over
 2})_r^5 ~0_r^5 ~0_r^5 ~(-{1\over 2})_r )~,\\
 &&V_2 =( 0 (0~{e_2\over 2})(0~a_2)(0~b_2) \vert\vert 0^3 \vert 0_r^5 ~
({1\over 2})_r^5 ~0_r^5 ~(-{1\over 2})_r )~.
\end{eqnarray} 
where the subscript $r$ indicates real bosons. 
Here $e_1$ and $e_2$ are the simple roots of $SU(3)$ ($e_1 \cdot e_1 =
e_2 \cdot e_2 = -2 e_1 \cdot e_2 = 2$), whereas the four-dimensional real 
vectors ${\bf s'}=(a_1 ,b_1)$ and ${\bf c'}=(a_2 ,b_2)$ are fixed elements 
in the spinor and conjugate weights of $SO(8)$, respectively. 
The generating vectors 
$V_1$ and $V_2$ are order two shifts ($m_1 =m_2 =2$). The auxiliary vectors
$W({\overline {\alpha V}})$ are therefore null as there is no twisting of the 
lattice. So the constraint (\ref{mphiz}) is trivial. 

{}The matrix of the dot products $V_i \cdot V_j$ reads
\begin{equation}
 V_i \cdot V_j=\left( \begin{array}{ccc}
               -1 & 0 & 0\\
               0  & 0 & 1\\
               0  & 1 & 0
               \end{array}
        \right)~.
\end{equation}
Now, consider the constraints (\ref{k1z}), (\ref{k2z}) and (\ref{k3z}).
The structure constants $k_{11}=k_{01}=k_{10}$ and 
$k_{22}=k_{02}=k_{20}$ must be chosen to equal zero for the supersymmetry not
to be broken from $N=4$ to $N=0$. The structure constants 
$k_{00}$ and $k_{12}=k_{21}$ can be chosen to equal zero or half. For 
definiteness we can choose
\begin{equation}
 k_{ij}=0 ~.
\end{equation}
So we have a consistent model, which we will refer to as $N1$. To see 
its spectrum, we note that the model has $8$ sectors: $\alpha_i=0,1$, 
for $i=0,1,2$. Here, $\alpha_0=0$ corresponds to spacetime bosons and
$\alpha_0=1$ corresponds to spacetime fermions. 
The $SO(32)$ shifts are given in the $SO(10)^3 \otimes SO(2)$ basis.
For $SO(10)$, $({1\over 2})_r^5$ is a component in ${\bf s}$, while
${\bf v}=({\pm 1}, 0^4)$ (and its permutations).
It is convenient to 
consider $SO(32)$ weights in terms of $SO(10)^3 \otimes SO(2)$ weights.
Under $SO(32)\supset SO(10)^3 \otimes SO(2)$, we have
\begin{eqnarray}
 {\bf 0}&=&({\bf 0},{\bf 0},{\bf 0},0)+({\bf v},{\bf v},{\bf v},V)+
 ({\bf 0},{\bf 0},{\bf v},V)+({\bf v},{\bf v},{\bf 0},0)+ \nonumber \\
 &&({\bf 0},{\bf v},{\bf 0},V)+({\bf v},{\bf 0},{\bf v},0)+ 
 ({\bf v},{\bf 0},{\bf 0},V)+({\bf 0},{\bf v},{\bf v},0), \\
 {\bf s}&=&({\bf s},{\bf s},{\bf s},S)+({\bf c},{\bf c},{\bf c},C)+
 ({\bf s},{\bf c},{\bf c},S)+({\bf c},{\bf s},{\bf s},C)+ \nonumber \\
 &&({\bf c},{\bf s},{\bf c},S)+({\bf s},{\bf c},{\bf s},C)+ 
 ({\bf c},{\bf c},{\bf s},S)+({\bf s},{\bf s},{\bf c},C).
\end{eqnarray}
Let us first consider the ${\bf 0}$ sector. 
The spectrum generating formula (\ref{sgfz}) becomes,
\begin{eqnarray}
\label{sg0}
&V_0 \cdot {\cal N}={1\over 2}(N^0+N^1+N^2+N^3)={1\over 2}~({\mbox{mod}~1})\\
\label{sg1}
&V_1 \cdot {\cal N}=-{1\over 2}e_1\cdot p_R -{\bf s'}\cdot P_{R}
	+{\bf s}\cdot P_1 -{1 \over 2}Q =0~({\mbox{mod}~1})\\
\label{sg2}
&V_2 \cdot {\cal N}=-{1\over 2}e_1\cdot p_R -{\bf c'}\cdot P_{R}
	+{\bf s}\cdot P_2 -{1 \over 2}Q =0~({\mbox{mod}~1}).
\end{eqnarray}
where
$P_1$ and $P_2$ are the momenta of the first and the second $SO(10)$
respectively. Here $Q$ is the $SO(2)$ weight, ({\em i.e.}, the $U(1)$ charge) 
and ${\bf s}$ stands for the
fixed component $({1\over 2},{1\over 2},{1\over 2},{1\over 2},{1\over 2})$.
Let us concentrate on the massless spectrum. 
The right-moving vacuum energy in the $\overline {\alpha V}={\bf 0}$ 
sector is $-1/2$,
but Eq.(\ref{sg0}) indicates that at least one NS fermion must be excited.
The lowest energy states then correspond to one NS fermionic oscillator
excitations that contribute $1/2$ to the energy, reaching $E^R=0$.
$N^0={\pm 1}$ corresponds to a helicity state.
Together with the lowest left-moving spacetime oscillator excitation ({\em 
i.e.}, $n^0_1=1$ or ${\overline n}^{\,0}_1=1$ to reach $E^L=0$), we have
the graviton, the dilaton and the axion. 
To obtain the gauge bosons (excluding the ones in the 
supergravity multiplet), we need to consider only the 
left-moving part of the lattices, since all $p_R$ and $P_{SO(8)R}$
must be zero.
The spectrum generating formula removes all but the
$({\bf 0},{\bf 0},{\bf 0},0)$ in the ${\bf 0}$ of $SO(32)$, 
which gives the gauge bosons in
$SO(10)^3 \otimes SO(2)$. It is easy to see that all the
$SU(3) \otimes SO(8)$ states with $E^L=0$ in the $N0$ model remain intact.
If we excite one of the other fermions,
{\em i.e.}, 
$N^i={\pm 1}$ for $i=1,2$ or $3$, we obtain the scalar superpartners.

{}In the $\alpha_0=1$ ({\em i.e.}, $V_0$) sector, we have the same spectrum 
generating formula as above; here, $E^R=0$ already,
and the $N^i=0,1$ stands for 
the Ramond degeneracy. So we have $2^4/2=8$ states. Together with 
$n^0_1=1$ or ${\overline n}^{\,0}_1=1$, this yields $4$ gravitinos.
It is equally straightforward to get the rest of the massless states in
this sector, and to see that the other sectors, {\em i.e.}, the shifted 
sectors, give rise to massive states only.
So the resulting $N1$ model
is a Narain model with $N=4$ space-time supersymmetry and the gauge 
group $SU(3)\otimes SO(8) \otimes SO(10)^3 \otimes SO(2)$.
Thus, introducing the shift vectors $V_1$ and $V_2$ is equivalent to starting 
from the $N0$ Narain model and turning on appropriate Wilson lines.

{}Next, we want to perform a ${\bf Z}_6$ orbifold on the $N1$ model.
There are at least two equivalent ways to
reach the same final model. We shall present one approach here and another
approach in section V.
Let us first give a preview of what we shall do.
Since ${\bf Z}_6={\bf Z}_2 \otimes {\bf Z}_3$, we may consider
the ${\bf Z}_2$ and ${\bf Z}_3$ orbifolds separately. Let us call the 
${\bf Z}_3$ orbifold the $A1$ model and the ${\bf Z}_2$ orbifold the 
$A2$ model. Let us first describe the ${\bf Z}_3$ symmetry. 

{}The even self-dual Lorentzian lattice $\Gamma^{\prime 6,22}$ 
of the $N1$ model has a number of ${\bf Z}_3$ symmetries:\\
({\em i}) a ${\bf Z}_3$ symmetry under the rotation of the right-moving 
momenta $p_R$ (corresponding to the $\Gamma^{2,2}$ sublattice of the 
original lattice $\Gamma^{6,22}$) by $2\pi/3$;\\
({\em ii}) a ${\bf Z}_3$ symmetry under the permutation of the world-sheet
bosons corresponding to the three $SO(10)$s; and \\
({\em iii}) a ${\bf Z}_3 \otimes {\bf Z}_3$ symmetry of the 
$\Gamma^{4,4}$ sublattice of $\Gamma^{6,22}$.

{}To see this last symmetry explicitly, it is useful to
consider $SO(8)$ in its $SU(3) \otimes U(1) \otimes U(1)$ basis:
\begin{eqnarray}
\label{eight}
 && {\bf 8}_v = {\bf 1}(0, 1) + {\bf 1}(0, -1) + {\bf 3}({1 \over \sqrt{3}}, 0)
 + {\overline {\bf 3}} (-{1 \over \sqrt {3}} , 0) ~,\nonumber \\
 && {\bf 8}_s = {\bf 1}({\sqrt{3} \over 2}, {1\over 2}) +
 {\bf 1}(-{\sqrt{3} \over 2}, -{1 \over 2}) + {\bf 3}(-{1 \over {2 \sqrt {3}}}
 , {1 \over 2})
 + {\overline {\bf 3}} ({1 \over{2 \sqrt {3}}}, -{1 \over 2}) ~,\\
 && {\bf 8}_c = {\bf 1}({\sqrt{3}\over 2}, -{1 \over 2} ) +
 {\bf 1}(-{\sqrt{3}\over 2}, {1\over 2}) + {\bf 3}(-{1 \over{2 \sqrt{3}}}, 
 -{1 \over 2})
 + {\overline {\bf 3}} ({1 \over{2 \sqrt {3}}}, {1 \over 2}) ~.\nonumber
\end{eqnarray}
Notice that the charges on the $U(1) \otimes U(1)$ plane have an explicit
${\bf Z}_3$ symmetry (as they lie in a rescaled $SU(3)$ lattice). 
Under this ${\bf Z}_3$ rotation, it is easy to see
that ${\bf 8}_v \rightarrow {\bf 8}_s \rightarrow {\bf 8}_c$, which is
simply the well known triality of
the $SO(8)$ Dynkin diagram. This ${\bf Z}_3$ orbifold of $SO(8)$ yields
the group $G_2$. Since the central charge of the level-1 $G_2$ is
$c=14/5$, the rest of the central charge is in the corresponding coset.
It is easy to see that the ${\bf Z}_3$ orbifold of the $SO(8)_1$
current algebra can be written as the product of a $(G_2)_1$ current
algebra, an Ising model ($c=1/2$) and a tri-critical Ising model ($c=7/10$).
The other ${\bf Z}_3$ symmetry in $SO(8)$ that we are interested in
is simply that of the $SU(3)$ lattice specified above.
Under this ${\bf Z}_3$ rotation,
the three states of ${\bf 3}$ permute among themselves (similarly for
the ${\overline {\bf 3}}$). This ${\bf Z}_3$ orbifold reduces
$SO(8)$ to $U(1) \otimes U(1) \otimes SU(3)_1$. Here, the
$SU(3)_1$ in Eq.(\ref{eight}) is reduced to $ U(1) \otimes U(1)$, while the
$ U(1) \otimes U(1)$ there form the Cartan basis of the new $SU(3)_1$.
Under the diagonal ${\bf Z}_3$ ({\em i.e.}, simultaneous
${\bf Z}_3 \otimes {\bf Z}_3$) orbifold, $SO(8)_1$ goes to $SU(3)_3$.
Now we mod the $N1$ model by the above diagonal ${\bf Z}_3$ symmetry 
($({\em i}), ({\em ii}), ({\em iii})$) to 
construct a model with $SO(10)$ gauge subgroup realized via the corresponding
current algebra at level three. 
So, before the ${\bf Z}_2$ twist, the gauge group of the $A1$ model is
$SU(3)_1 \otimes SU(3)_3 \otimes SO(10)_3 \otimes U(1)$.
This model has nine
generations of chiral fermions of $SO(10)$. To cut the number of generations 
to three we will mod the resulting model by its ${\bf Z}_2$ symmetry
that corresponds to a symmetric ${\bf Z}_2$ twist of the $\Gamma^{4,4}$
sublattice. The ${\bf Z}_3$ orbifold is asymmetric, whereas the 
${\bf Z}_2$ orbifold is symmetric. 
Now we are ready for the explicit construction.

{}Let us start from the $N1$ model and consider the asymmetric 
orbifold model generated by the following vectors: 
\begin{eqnarray}
 \label{firsten}
 &&V_0 =(-{1\over 2} (-{1\over 2}~ 0)^3 \vert\vert 
   0^3 \vert 0^{5}~0_r^5 ~0_r) ~,\nonumber \\
 &&V_1 =( 0 (-{1\over 3}~{1\over 3})^3 \vert\vert 0~({1\over 3})^2 \vert 
 ({1\over 3})^5 ~0_r^5 ~({2\over 3})_r )~, \nonumber \\
 &&V_2 =( 0 (0~0)(-{1\over 2}~{1\over 2})^2 \vert\vert ({e_1\over 2})
 ({1\over 2})^2 \vert 0^5 ~0_r^5 ~0_r )~,\\
 &&W_1 =( 0 (0~{1\over 2})^3 \vert\vert 0~({1\over 2})^2 \vert 
 ({1\over 2})^5 ~0_r^5 ~0_r )~, \nonumber \\
 &&W_2 =( 0 (0~0)(0~{1\over 2})^2 \vert\vert 0
 ({1\over 2})^2 \vert 0^5 ~0_r^5 ~0_r )~.\nonumber
\end{eqnarray}  
So, $t_1=m_1=3$, $t_2=m_2=2$, and $t^*=6$.
Here the right- and left-moving bosons corresponding to the $SU(3)\otimes 
SO(8)$ subgroup are complexified, whereas the single boson corresponding to 
the $SO(2)$ subgroup is real. 
Let the real bosons $\phi^I_p$, $I=1,...,5$, correspond to the
$p^{\mbox{th}}$ $SO(10)$ subgroup, $p=1,2,3$. We may recombine these
fifteen real bosons into five real and five complex bosons, 
\begin{eqnarray} \label{basis10}
 \varphi^I = {1\over \sqrt{3}}(\phi^I_1 +\phi^I_2 +\phi^I_3)~, \\
  \Phi^I  = {1\over \sqrt{3}}(\phi^I_1 +\omega \phi^I_2 +\omega^2 \phi^I_3)~,
\end{eqnarray}
where $\omega =\exp (2\pi i /3)$. 
This is the eigen-basis of the ${\bf Z}_3$ twist, and this is the basis that
the above vectors act on. Thus, along with the 
${\bf Z}_3$ rotations of the complex bosons corresponding to the $SU(3)\otimes
SO(8)$ subgroup, the orbifold group element described by the $V_1$ vector also 
permutes the real bosons corresponding to the three $SO(10)$s, which in turn
is equivalent to modding out by their outer automorphism.

{}The matrix of the dot products $V_i \cdot V_j$ reads
\begin{equation}
 V_i \cdot V_j=\left( \begin{array}{ccc}
               -1 & -1/2 & -1/2\\
               -1/2 & -1/3 & -1/3\\
               -1/2 & -1/3 & 0
               \end{array}
        \right)~.
\end{equation}
Following from the constraints (\ref{k1z}), (\ref{k2z}) and (\ref{k3z}),
the structure constants $k_{ij}$ can be given by
\begin{equation}
 k_{ij}=\left( \begin{array}{ccc}
               k_{00} & 0  & k_{20}+1/2\\
               1/2 & 1/3 & 0\\
               k_{20} & 2/3 & k_{20}
               \end{array}
        \right)~.
\end{equation}
For definiteness we will take $k_{00}=0$ (the other choice
$k_{00}=1/2$ leads to the model with chirality of fermions opposite 
to that of the one under consideration). 
Here we choose $k_{20}=1/2$; as we shall see, with this choice, the 
resulting model has $N=1$ space-time supersymmetry. 
(The other choice $k_{20}=0$ leads to a model with no supersymmetry.) 
$\{ V_i, k_{ij}\}$ satisfies the simplified rules of section III.
The condition (\ref{simpli}) follows automatically.
(As a check, we note that 
$\chi({\overline {\alpha V}},{\overline {\beta V}}) = 0$ and 
$f({\overline {\alpha V}}, {\overline {\beta V}})=0$.)

{}We will describe the above model in two steps. 
First we consider the model (the $A1$ model),
generated by the vectors $\{V_0 ,V_1 \}$ on the $N1$ model. 
As we have mentioned already, it has
the gauge group $SU(3)_1 \otimes SU(3)_3 \otimes SO(10)_3 \otimes U(1)$ with
$N=1$ supersymmetry. As we shall see, it also has three generations of 
scalars in the adjoint of $SO(10)_3$ and 
nine generations of chiral ${\bf 16}$s of $SO(10)_3$ (There are other fields
in the massless spectrum which we will give in a moment). To cut the number of 
generations to three, we introduce the $V_2$ vector, {\em i.e.}, the 
$\{V_0 ,V_1,V_2 \}$ model. 
Sometimes, it is also illuminating to consider the $\{V_0 ,V_2 \}$ model
(the $A2$ model, {\em i.e.}, $\{V_0 ,V_2 \}$ acting on the $N1$ model). 

{}The space-time bosons and fermions come from the sectors 
$\overline {\alpha V}$ with $\alpha_0 =0$ and $\alpha_0 =1$, respectively.  
In the untwisted sectors
$\overline {\alpha V}$, $\alpha_0 =0,1$, $\alpha_1 =0$, the spectrum generating
formula (\ref{sgfz}) reads 
\begin{eqnarray}\label{UR}
 &&V_0 \cdot {\cal N}_{\overline {\alpha V}}= 
 {1\over 2} \sum_{r=0}^3
  N^r_{\overline {\alpha V}} ={1\over 2} ~
 (\mbox{mod~1})~,\\
 \label{UNS}
 &&V_1 \cdot {\cal N}_{\overline {\alpha V}}= 
 {1\over 3} (\sum_{r=1}^3 N^r_{\overline {\alpha V}}+
 \sum_{\ell=1}^3 J^\ell_{\overline {\alpha V}} - 
 \sum_{\ell=5}^{11} J^\ell_{\overline {\alpha V}} ) + {2\over 3}~Q
 =0~(\mbox{mod~1})~,\\
 &&V_2 \cdot {\cal N}_{\overline {\alpha V}}=
 {1\over 2} (\sum_{r=2}^3 N^r_{\overline {\alpha V}}+
 \sum_{\ell=2}^3
 J^\ell_{\overline {\alpha V}} + {1\over 2}~e_1\cdot p_L -
 \sum_{\ell=5}^{6} J^\ell_{\overline {\alpha V}} ) =0~(\mbox{mod~1})~,
\end{eqnarray} 
where $Q$ is the $U(1)$ charge ({\em i.e.}, the $SO(2)$ weight) and 
$(0\vert p_L) \in \Gamma^{2,2}$.
The right-moving vacuum energy in the ${\bf 0}$ sector is
$-1/2$, but according to (\ref{UR}) at least one NS fermion must be excited.
The lowest energy states then correspond to one NS fermionic oscillator
excitations that contribute $1/2$ to the energy. So, for the 
states to be massless, we must have $J^{\ell}_{\overline {\alpha V}}
=0$, $\ell=1,2,3$; otherwise, either the right-moving momenta are non-zero, 
or the right-moving oscillators are excited, and the resulting states 
are massive in both cases. The right-moving vacuum energy in the $V_0$
sector is already $0$. 

{}It is convenient to define $J_{SO(8)L}= J^5 +J^6$.
Then, in the appropriate basis, $J_{SO(8)L}$ measures the eigen-phase of
the twist operation on $P_{SO(8)L}$ of $\Gamma^{4,4}$. 
Similarly, let $J_{SO(10)}=\sum_{\ell=7}^{11} J^\ell$. 
So the spectrum 
generating formulae for the massless states reduce to
\begin{eqnarray}
 \label{aa0}
 {1\over 2}(N^0+N^1+N^2+N^3)={1\over 2}~({\mbox{mod}~1})~,\\
 \label{aa1}
 {1\over 3}(N^1+N^2+N^3-J_{SO(8)L}-J_{SO(10)})+{2\over 3}~Q=0~
 (\mbox{mod~1})~,\\
 \label{aa2}
 {1\over 2}(N^2+N^3 +e_1\cdot p_L -J_{SO(8)L})=0~(\mbox{mod~1})~,
\end{eqnarray}
and the left-moving energy becomes
\begin{equation}
 E^L=-1 +{1\over 2}~(p_L^2+P_{SO(8)L}^2+P_{SO(10)}^2+Q^2)+
 {\mbox{oscillators}}~.
\end{equation}
Let us first consider the $\{V_0 ,V_1 \}$ model, {\em i.e.}, the $A1$ model. 
Under the ${\bf Z}_3$ twist, $J_{SO(8)L}=0, ~1, ~2~ (\mbox{mod~3})$ and 
$J_{SO(10)}=0, ~1, ~2 ~(\mbox{mod~3})$. To find the gauge bosons in this 
model, let $N^0={\pm 1}$ in the ${\bf 0}$ sector. 
States in the original root lattice of $SO(10)^3$ give $E^L=0$ already, so,
$Q=0$ and $P_{SO(8)L}=0$. Of the three $J_{SO(10)}=0, 1, 2$ eigenstates, 
only the $J_{SO(10)}=0$
({\em i.e.}, ${\bf Z}_3$ invariant) states are kept by Eq.(\ref{aa1}). This 
gives the gauge bosons of $SO(10)_3$. 
Similarly, we can consider the original root lattice of $SO(8)$. 
In the $SU(3) \otimes U(1) \otimes U(1)$ basis, the ${\bf 28}$ of $SO(8)$
becomes
\begin{eqnarray}
 \label{ir28}  
  {\bf 28} &=& {\bf 1}(0,0) + {\bf 1}(0,0) + {\bf 3}({1 \over \sqrt{3}}, 1)
  + {\overline {\bf 3}} (-{1 \over \sqrt {3}} , -1)+ \nonumber \\
  && {\bf 3}({1 \over \sqrt{3}}, -1)
  + {\overline {\bf 3}} (-{1 \over \sqrt {3}} , 1)
  + {\bf 3}(-{2 \over \sqrt {3}}, 0)
  + {\overline {\bf 3}} ({2 \over \sqrt {3}}, 0)+ {\bf 8}(0,0)~.
\end{eqnarray}
The eigenstates of the ${\bf Z}_3$ twist can be organized 
in the basis $SO(8)\supset SU(3)_3$,
\begin{equation}
 \label{irr28}
  {\bf 28}={\bf 8} +{\bf 10}+ {\overline {\bf 10}} ~,
\end{equation}
which have $J_{SO(8)L}=0, 1, 2$ respectively. Again, only the $J_{SO(8)L}=0$ 
states are kept. The remaining $SU(3) \otimes U(1)$ gauge bosons in the 
original $N1$ model are left untouched. 
If, instead of the $N^0$ excitation, we have $N^i={\pm 1}$ for $i=1, 2$ or $3$,
then we have scalar fields. Then Eq.(\ref{aa1}) implies either 
$J_{SO(8)L}=1$, giving ${\bf 10}$ of $SU(3)_3$, or $J_{SO(10)}=1$,
giving $SO(10)$ adjoint Higgs fields.

{}This model has $N=1$ space-time supersymmetry. This is easiest to see
if we count the number of gravitinos in the $V_0$ sector. 
The only quantum numbers of the Ramond fermions allowed by
Eqs.(\ref{aa0},~\ref{aa1}) are
$(N^0,N^1,N^2,N^3)=(0,1,1,1)$ or $(1,0,0,0)$, yielding only one gravitino.
Next, let us consider the chiral fields in this sector.
Let us choose $N^0=0$. For $(N^1,N^2,N^3)=(1,0,0),(0,1,0)$ and $(0,0,1)$, 
Eq.(\ref{aa1}) requires  $J_{SO(8)L}=1$. This gives
the particle states in the ${\bf 10}$ of $SU(3)_3$.
The conjugate states in the ${\bf 10}$ come from
the choice $N^0=1$, together with 
$(N^1,N^2,N^3)=(1,1,0),(0,1,1)$ and $(1,0,1)$.
Together they form $3$ copies of ${\bf 10}_L$ of $SU(3)_3$, where the 
left-handed chirality label is by convention. 
This choice of convention fixes the chiralities of the rest of the spectrum.

{}It is straightforward to obtain the remaining 
massless states in the untwisted sector of the $A1$
model. To summarize, we have: \\
$\bullet$ The $N=1$ supergravity multiplet together with
the dilaton-axion supermultiplet;\\
$\bullet$ The $N=1$ Yang-Mills supermultiplet in the adjoint of $SU(3)_1
\otimes SU(3)_3 \otimes SO(10)_3 \otimes U(1)$;\\
$\bullet$ Three $N=1$ chiral supermultiplets in $({\bf 1}, {\bf 10}, {\bf 1})
(0)_L$ (The bold face in parentheses indicates the irreps of $SU(3)_1
\otimes SU(3)_3 \otimes SO(10)_3$, and the $U(1)$ charge is given in 
parentheses in regular font; the subscript indicates the space-time
chirality: $L$ and $R$ for left- and right-handed fermions respectively);\\
$\bullet$ Three $N=1$ Higgs supermultiplets in $({\bf 1}, {\bf 1}, {\bf 45})
(0)$.

{}To obtain the massless states in the untwisted sector of the final model, 
we now impose the $V_2$ constraint.
The $e_1/2$ shift in the $V_2$ vector breaks
$SU(3)_1$ to $SU(2)_1\otimes U(1)$, since the gauge bosons corresponding
to the four roots containing an $e_2$ factor are removed by the 
constraint (\ref{aa2}). 
Now, under the ${\bf Z}_6 ={\bf Z}_3 \otimes {\bf Z}_2$ twist, 
$J_{SO(8)L}=0,~1,~2,~3,~4,~5$. To see that
the ${\bf Z}_2$ twist on the $SO(8)$ lattice breaks $SU(3)_3$ further to 
$SU(2)_3 \otimes U(1)$, let us write the $SU(3)_3$ irreps. in the
$SU(2)_3 \otimes U(1)$ basis,
\begin{eqnarray}
 \label{the8}
  {\bf 3} &=& {\bf 2}(+1)+{\bf 1}(-2)~, \nonumber \\
  {\bf 8} &=& {\bf 1}(0) +{\bf 3}(0) + {\bf 2}(+3) +{\bf 2}(-3)~, 
\end{eqnarray} 
where the $U(1)$ charge is again given in the parentheses. 
Recall that the ${\bf 8}$ of $SU(3)_3$ in 
Eq.(\ref{irr28}) has $J_{SO(8)L}=0~ (\mbox{mod~3})$. Under the
${\bf Z}_6$ twist, the ${\bf 1}$ and the ${\bf 3}$ have $J_{SO(8)L}=0$, 
while the ${\bf 2}$s have $J_{SO(8)L}=3$. This property may be seen by going
back to Eq.(\ref{ir28}).
So the final gauge symmetry becomes
$SU(2)_1\otimes U(1)\otimes SU(2)_3\otimes U(1)\otimes SO(10)_3\otimes U(1)$.
Recall that the ${\bf 10}$ of $SU(3)_3$ 
in Eq.(\ref{irr28}) has 
$J_{SO(8)L}=1~(\mbox{mod~3})$. Under $SU(3)_3 \supset SU(2)_3 \otimes U(1)$:
\begin{equation}\label{Ten}
 {\bf 10} ={\bf 1}(-6)+{\bf 2}(-3)+{\bf 3}(0)+{\bf 4}(+3)~.  
\end{equation}
where the ${\bf 1}$ and the ${\bf 3}$ have $J_{SO(8)L}=4$, 
while the ${\bf 2}$ and the ${\bf 4}$ have $J_{SO(8)L}=1$.
So the constraint (\ref{aa2}) cuts the three copies of
 $({\bf 1}, {\bf 10}, {\bf 1})(0)_L$ to
one copy of the ${\bf 1}$ and ${\bf 3}$ (with $N^1={\pm 1}$) 
and two copies of ${\bf 2}$ and the ${\bf 4}$ (with $N^2 + N^3={\pm 1}$).
It also cuts the $({\bf 1}, {\bf 1}, {\bf 45})(0)$ 
from three copies to one copy.
The resulting massless spectrum in the untwisted sector of the final model
is summarized in the first column of Table I.

{}Now we turn to the twisted sectors; again, we consider the $A1$ model first.
In the twisted sectors ${\overline {\alpha V}}$, where
$\alpha_1=1,2$, we note that the states in the $\alpha_1=2$ sectors are 
conjugates of the states in the corresponding $\alpha_1=1$ sectors. This 
allows us to concentrate on the $\alpha_1=1$ sectors.
Also, states in the $\alpha_0 =1$ sectors are the fermionic 
superpartners of the states in the $\alpha_0 =0$ sectors. So we need 
to consider only the $\alpha_0 = \alpha_1 =1$ sector, where
\begin{equation}
{\overline {V_0+V_1}}=( -{1\over 2} ({1\over 6}~{1\over 3})^3 \vert\vert 
  0~({1\over 3})^2 \vert
 ({1\over 3})^5 ~0_r^5 ~({2\over 3})_r ) \\
\end{equation}
{}In the $A1$ model, $\alpha^* =1$ and $t^*=t_1=m_1=3$. 
The invariant sublattice $I({\overline {V_0+V_1}})$ is given by $\Gamma^2
\otimes \Gamma^6$, where $\Gamma^2$ is the $SU(3)$ lattice and
$\Gamma^6= \{(\sqrt{3}{\bf q} \vert Q) \}$, where
$({\bf q} \vert Q)=({\bf 0} \vert 0), ({\bf v} \vert V),
({\bf s} \vert S), ({\overline {\bf s}} \vert {\overline S})$.
The dual lattice is 
${\tilde I}({\overline {V_0+V_1}}) ={\tilde \Gamma}^2 \otimes 
{\tilde \Gamma}^6$, where ${\tilde \Gamma}^6 =\{({\bf q}/\sqrt{3} \vert Q) \}$.
The momenta in the twisted sector belong to the shifted
dual lattices ${\tilde \Gamma}^2 \otimes {\tilde \Gamma}^6_s$
where $ {\tilde \Gamma}^6_s = {\tilde \Gamma}^6 +({\bf 0}\vert 2/3)$.

{}The determinant of the metric of the invariant sublattice is $M
({\overline {\alpha V}})=3^6$, and the number of fixed points is given by 
\begin{equation}
 \xi ({\overline {\alpha V}})=(2\sin(\pi /3) )^{10} (M({\overline 
 {\alpha V}}))^{-1/2} =9~.
\end{equation}
We see that the only non-trivial contribution to the number of 
fixed points in the twisted sector comes from the symmetric
${\bf Z}_3$ twist in $\Gamma^{4,4}$.
This twist contributes $9=3_R \times 3_L$ fixed points, {\em i.e.},
they come from both left- and right-moving world-sheet 
conformal fields corresponding to irreps ${\bf 3}_L \otimes {\bf 3}_R$
of the underlying current subalgebra $SU(3)_L \otimes 
SU(3)_R$. In this ${\overline {V_0+V_1}}$
twisted sector, the spectrum generating formula reads:
\begin{eqnarray}
 \label{tt0}
  &&V_0 \cdot {\cal N} =
  {1\over 2}(N^0+N^1+N^2+N^3) = {1\over 2} ~(\mbox{mod~1})~,\\
 \label{tt1}
  &&V_1 \cdot {\cal N} +f_i ({\overline {\alpha V}}, n_1)
  +{1 \over 2}{\vec Q}^2({\vec P}_{\overline {\alpha V}})
  =0  ~(\mbox{mod~1})~.
\end{eqnarray}
where $n_1=0,~1,~2$.
Knowing $\xi ({\overline {\alpha V}})=9$, it follows 
from Eqs.(\ref{xigo}), (\ref{fixedpointphases}) and (\ref{xito}) that
$\xi ({\overline {\alpha V}}, n_1)=9$ for $n_1=0$, and
zero otherwise. So we need consider only the $n_1=0$ term in Eq.(\ref{tt1}).
Recall that ${\vec P}_{\overline {\alpha V}}
\in {\tilde I}({\overline {\alpha V}})$; also,
${\vec Q}({\vec P}_{\overline {\alpha V}})
\in {\tilde I}({\overline {\alpha V}})$ is an arbitrary vector such that
${\vec P}_{\overline {\alpha V}}-\alpha^*
{\vec Q}({\vec P}_{\overline {\alpha V}}) \in I({\overline {\alpha V}})$.
For $\alpha_1 =1$ we can choose 
${\vec Q}({\vec P}_{\overline {\alpha V}})={\vec P}_{\overline {\alpha V}}$.
Again, let us consider only the massless states.
Since $E^R=0$ already, we can ignore the right-moving $(N^1, N^2, N^3)$
and $J^\ell$ in Eqs.(\ref{tt0}, \ref{tt1}). So Eq.(\ref{tt0}) implies
$N^0=1$, {\em i.e.}, anti-particle states.
Since all the $SU(3)$ non-trivial states in $\Gamma^{2,2}$
are massive, we may also ignore them. 
The resulting momenta ${\bf p}$ in this twisted sector 
are in ${\tilde \Gamma}^6_s$,
{\em i.e.},  ${\bf p}=({\bf q}/\sqrt{3} \vert {2 \over 3}+Q)$.
Now the left-moving energy is given by
\begin{equation}
 E^L= -{2 \over 9}+{1 \over 2}({2 \over 3}+Q)^2+{1 \over 6}{\bf q}^2~.
\end{equation}
and the constraint (\ref{tt1}) reduces to
\begin{equation}
 {2 \over 3}Q + {1\over 2}{\vec P}^2_L =0 ~(\mbox{mod~1})~.
\end{equation}
where ${\vec P}_L=({\bf q}/\sqrt{3} \vert Q)$. Here $J_{SO(8)L}=J_{SO(10)}=0$, 
since their excitations are too heavy. Note that 
${\bf p}={\vec P}_L+({\bf 0} \vert {2 \over 3})$.
The only possible massless states come from ${\vec P}_L=({\bf 0} \vert 0),
({\bf v}/\sqrt{3} \vert -1)$ and 
$( {\bf c}/\sqrt{3} \vert C)$. 
It is easy to check that these choices of ${\vec P}_L$ (or ${\bf p}$) 
also satisfy the above spectrum generating formulae, {\em i.e.}, the 
level-matching condition\cite{vafa} is sufficient. 
Now we can put them together with the ${\alpha}_1=2$ sector, resulting in, 
among other fields, $9$ right-handed anti-spinors of $SO(10)$.
We shall rewrite them as left-handed $SO(10)$ spinors.

{}The hidden sector gauge group is $SU(3)_1$, and no massless fields with
its quantum numbers appear. The right-moving
fixed points correspond to the number of generations of chiral fermions, 
whereas the $3$ left-moving fixed points appear in some irrep
of the gauge group $SU(3)_3$. That they form the ${\bf 3}$ of $SU(3)_3$ 
is easily seen in the alternative construction given in the next section.
Here an examination of their scattering may be used to fix their quantum 
numbers. However, as we shall see, it is easiest to use the $SU(3)_3$ 
anomaly-free condition to uniquely fix their quantum numbers.
In summary, we have three $N=1$ chiral supermultiplets in 
$({\bf 1}, {\overline {\bf 3}},
{\bf 16})(-1)_L$, $({\bf 1}, {\overline {\bf 3}}, {\bf 10})(+2)_L$ and 
$({\bf 1}, {\overline {\bf 3}}, {\bf 1})(-4)_L$, where the radius of the 
$U(1)$ boson has been shifted from $1/2$ to $1/6$. 
This completes the massless spectrum of the $A1$ model.
Note that the $SU(3)_3$ chiral anomaly in the twisted
sectors is cancelled by that in the untwisted sector as a
${\bf 10}_L$ of $SU(3)_3$ has $27$ times the anomaly contribution
of a ${\bf 3}_L$. The model is also $U(1)$ anomaly-free
due to the underlying $E_6$ structure of the $SO(10)_3 \otimes U(1)$
matter fields as can be seen from the branching 
${\bf 27} ={\bf 16}(-1) +{\bf 10} (+2) +{\bf 1}(-4)$ 
under $E_6 \supset SO(10) \otimes U(1)$.
Here, $SU(3)_3 \otimes U(1)$ may be interpreted as
the horizontal symmetry.

{}Now we return to the final model. The ${\bf Z}_2$ invariant states in 
the above twisted sector form the ${\bf Z}_3$ twisted sector of the 
final model, namely the $T3$ sector (${\em i.e.}$, $\alpha^*=2,~4$, with 
$t^*=6$). Under the $Z_6$ twist, we have additional twisted sectors, 
namely $T6$ (${\em i.e.}$, $\alpha^*=1,~5$) and $T2$
(${\em i.e.}$, $\alpha^*=3$). The number 
of fixed points $\xi({\overline {\alpha V}}, {\bf n})$ for each twisted
sector and for each ${\bf n}$ is given in the table below. 

\bigskip
\begin{equation}
\begin{array}{ccccccccc}
 ~~~~ & {\bf n}=(n_1,n_2) & ~~(0,0)~~ & ~~(0,1)~~ & ~~(1,0)~~ 
  & ~~(1,1)~~  & ~~(2,0)~~ & ~~(2,1)~~ &~~ \\ 
 & & & & & & & &  \\
 & T6 & 1 & 0 & 0 & 0 & 0 & 0 & \\
  & & & & & & & & \\
 & T3 & 5 & 4 & 0 & 0 & 0 & 0 & \\
  & & & & & & & & \\
 & T2 & 2(6) & 0 & 1(5) & 0 & 1(5) & 0 & \\
 & & & & & & & &
\end{array}
\end{equation}
\medskip

{}Here the number of fixed points in the $T2$ sector before the 
inclusion of the invariant lattice volume factor are given in parentheses.
It is easy to see that the $V_1$ constraint changes nothing in the spectrum,
as expected. So, all we need is to impose the $V_2$ constraint
on the ${\overline {V_0+V_1}}$ sector:
\begin{equation}
 V_2 \cdot {\cal N} +f_2 ({\overline {V_0+V_1}}, {\bf n})
 +3{\vec Q}^2({\vec P}_{\overline {V_0+V_1}})
 =0  ~(\mbox{mod~1})~,
\end{equation} 
which reduces to the constraint $f_2({\overline {V_0+V_1}}, {\bf n})=0$, 
which is satisfied only for ${\bf n}=(0,0)$. From 
the above table, we see that this gives 
$\xi({\overline {V_0+V_1}}, {\bf n}=(0,0))=5$ fixed points. 
Actually, this result is easily seen in the standard orbifold construction.
Of the original $9$ fixed points, the one at the origin is invariant under the
${\bf Z}_2$ twist. The remaining $8$ fixed points form $4$ pairs, and
the ${\bf Z}_2$ twist permutes the $2$ fixed points in each pair.
Forming $4$ symmetric and $4$ antisymmetric
combinations, we have $9=5(1)+4(-1)$ (where the
${\bf Z}_2$ phases are given in parentheses); that is, $5$ of the original
$9$ are invariant under the  ${\bf Z}_2$ twist.
So these $5$ copies of the $SO(10)_3$ chiral matter fields survive, 
while the other $4$ are projected out.
They transform in the irreps of $SU(2)_3 \otimes U(1)$.
We can see this result from yet another viewpoint.
We may treat the $9$ fixed points as
$9=3_R \times 3_L$. The ${\bf Z}_2$ phases of
these fixed points can be understood in terms of the branching $3_{R,L} =
1_{R,L} (1)+2_{R,L} (-1)$ (The ${\bf Z}_2$ phases are given in parentheses).
Therefore, we have $9=5(1)+4(-1)$ (Here $5=2_R \times 2_L +1_R \times 1_L$
and $4=2_R \times 1_L + 1_R \times 2_L$).
So we end up with $2$ copies of 
$({\bf 1}, {\bf 2}, {\bf 16})
(0,-1,-1)_L$ and one copy of $({\bf 1}, {\bf 1}, {\bf 16})(0,+2,-1)_L$, plus
the corresponding  vector and singlet irreps of
$SO(10)_3$ (the $U(1)$ charges are normalized to
$(1/\sqrt{6} , 1/3\sqrt{6}, 1/6)$).

{}In the final model, we have other twisted sectors besides the $T3$ sector.
Let the $V_2$ and $V_0+V_2$ sectors form the $T2$ sector, 
while the $V_1+V_2$, $V_0+V_1+V_2$,
$2V_1+V_2$ and $V_0+2V_1+V_2$ sectors form the $T6$ sector.
Let us first consider the $V_0+V_1+V_2$ sector :
\begin{equation}
 {\overline {V_0+V_1+V_2}}=( -{1\over 2} ({1\over 6}~{1\over 3})
  (-{1\over 3} {5 \over 6})^2 \vert\vert {e_1\over 2}
  ~({5\over 6})^2 \vert
 ({1\over 3})^5 ~0_r^5 ~({2\over 3})_r ) \\
\end{equation}
The sublattice invariant
under the ${\bf Z}_6$ twist is the same as that for the ${\bf Z}_3$ twist,
 $I({\overline {V_0+V_1+V_2}}) = \Gamma^2 \otimes \Gamma^6$.
So the number of fixed points in the $T6$ sector is one. From the table 
above, $\xi({\overline {V_0+V_1+V_2}}, {\bf n})=1$ for 
${\bf n}=(0,0)$ only, and zero, otherwise.
Again, since the ground state has $E^R=0$ already, no right-moving 
excitation is allowed for massless states; so we may ignore all 
right-moving quantum numbers except $N^0$. The left-moving energy is
\begin{equation}
 E^L= -{11\over 36} +{1\over 2}~(e_1+{\tilde p}_L)^2
  +{1\over 2}~{\tilde P}_{SO(10)}^2+{1\over 2}({e_1\over 2} +Q)^2 + ~...~,
\end{equation}
where contributions that yield only massive states are ignored. 
Here $({\tilde p}_L\vert {\tilde P}_{SO(10)} \vert Q)$ is in the
dual invariant lattice
${\tilde I}({\overline {V_0+V_1+V_2}}) ={\tilde \Gamma}^2 \otimes
{\tilde \Gamma}^6$, where ${\tilde \Gamma}^6 =\{({\bf q}/\sqrt{3} \vert Q) \}$.
The only possible massless states in this twisted sector come from 
the shifted momenta
$({1\over 2}~e_1+{\tilde p}_L\vert {\bf q}/\sqrt{3} \vert {2 \over 3}+Q)$.
where ${\tilde p}_L=-{\tilde e}^1$ and $(-e_1+{\tilde e}^1)$, while
$({\bf q}/\sqrt{3} \vert Q)= ({\bf 0} \vert 0),
({\bf v}/\sqrt{3} \vert -1)$ and
$( {\bf c}/\sqrt{3} \vert -1/2)$.
It is straightforward to check that they all satisfy the 
corresponding spectrum generating formulas for this sector, with
$N^0=0$. Here ${1\over 2}~e_1+{\tilde p}_L$ correspond to $U(1)$ charges,
so they are singlets under $SU(2)_1 \otimes SU(2)_3$.
Together with the ${\overline {V_0+2V_1+V_2}}$ sector,
the massless chiral fields are
$({\bf 1}, {\bf 1}, {\bf 16})(\pm 1, -1, -1)_R$, plus  the
corresponding vector and singlet irreps of $SO(10)_3$, where the radius
of the first $U(1)$ is ${1 /\sqrt {6}}$.
Note that these states are right-handed, and come in pairs ($\pm 1$ of the
first $U(1)$ charge). So, effectively, we have a total of $3=5-2$
left-handed chiral families of ${\bf 16}$'s.

{}Finally, let us consider the space-time fermions in the $T2$ sector,
\begin{equation}
 {\overline {V_0+V_2}}=( -{1\over 2} (-{1\over 2}~0)
  (0~ {1 \over 2})^2 \vert\vert {e_1\over 2}
  ~({1\over 2})^2 \vert 0^5 ~0_r^5 ~0_r ) \\
\end{equation}
and $W({\overline {V_0+V_2}})=W_2$. It is convenient to first impose the
$V_0$ and $V_2$ constraints in the spectrum generating formula, and then
project onto ${\bf Z}_3$ invariant states ({\em i.e.}, impose the $V_3$
constraint). This is equivalent to
considering the twisted sector of the $A2$ model, 
or the ${\bf Z}_2$ orbifold of the $N1$ model. The sublattice
invariant under the ${\bf Z}_2$ twist is given by the sublattice of
$\Gamma^{2,2}\otimes \Gamma^{16}$ invariant under the Wilson lines $U_1$ and
$U_2$. The metric of this sublattice has determinant $16$. Therefore, the
number of fixed points is $4_R \times 4_L /\sqrt{16} =2_R \times 2_L$.
The ${\bf Z}_2$ orbifold breaks $SU(3) \otimes SO(8)$ to
$SU(2)_1 \otimes U(1) \otimes SU(2)^4$. The $2_L$ fixed points form a
doublet of one of the $SU(2)$ factor in $SU(2)^4$, say the last $SU(2)$.
Again the ground state $E^R=0$, while the left-moving energy is
\begin{equation}
 \label{a2tt}
 E^L= -{3\over 4} +{1\over 2}({e_1 \over 2}+ p_L)^2
  +{1\over 2}(n^5_1 +{\overline n}^5_1+n^6_1 + {\overline n}^6_1)~+~...
\end{equation}
where only terms that can contribute to the massless spectrum are included.
(Note that all the states in non-trivial irreps of $SO(10)_3 \otimes U(1)$ 
are massive.)
At $E^L= -{1\over 2}$, we have $2$ copies of (off-shell)
$({\bf 2}, {\bf 1}, {\bf 1}, {\bf 1},{\bf 2})(0)$. They come from
$p_L=0$ and $p_L=-e_1$, with shifted momenta ${e_1 \over 2}+ p_L=
\pm {e_1/2}$, forming a doublet of $SU(2)_1$. To obtain the physical 
massless states, we need $E^L=0$. There are two types
of such states: ({\em i}) we can excite one of the four oscillator modes 
indicated in (\ref{a2tt}), yielding $8$ massless states; and ({\em ii}) 
we can change the momentum to $p_L=e_2$ and $p_L=-e_1-e_2$; since the 
shifted momenta $\pm (e_2+ e_1/2)$ are now orthogonal to the shift
${e_1/2}$, they are $SU(2)_1$ singlets; so they form a pair of $U(1)$
charged states. 
For these massless states, we see that they all satisfy the $V_0$ and 
$V_2$ constraints,
\begin{eqnarray}
  {1\over 2}(N^0+N^1) = {1\over 2} ~(\mbox{mod~1})~, \nonumber \\
  {e_1 \over 2}p_L-{1\over 2}J_{SO(8)L}={1\over 2}  ~(\mbox{mod~1})~.
\end{eqnarray}
Collecting these states, we have two massless sets of
$({\bf 2}, {\bf 2}, {\bf 2}, {\bf 2},{\bf 1})(0)$. 
and $({\bf 1}, {\bf 1}, {\bf 1}, {\bf 1},{\bf 2})(\pm 3)$. Note that 
their quantum number assignments are uniquely fixed by the underlying 
${\bf Z}_3$ symmetry (${\em i.e.}$, the cyclic permutation of the 
first three $SU(2)$s in $SU(2)^4$) and the counting of states.
To obtain the states in the final model,
consider the action of the ${\bf Z}_3$ twist, {\em i.e.}, the $V_1$
constraint,
\begin{equation}
  f_1 +{1\over 3}(N^1 - J_{SO(8)L}) = {2\over 3} ~(\mbox{mod~1}).
\end{equation}
To reach the gauge symmetry of the final model, the ${\bf Z}_3$ twist 
breaks the last $SU(2)$ to $U(1)$ and converts the middle 
three $SU(2)$s to $SU(2)_3$. It also leaves the first $SU(2)$ alone.
Under the ${\bf Z}_3$ rotation, the fixed points have phases
$2_{L,R}=1_{L,R}(\omega) + 1_{L,R}({\omega}^2)$. 
The $({\bf 2}, {\bf 2}, {\bf 2})$
of $SU(2)^3$ become ${\bf 2}, ~{\bf 2}$, and ${\bf 4}$ of $SU(2)_3$.
So the resulting ${\bf Z}_3$-invariant
massless states are $({\bf 2}, {\bf 2})_L$ and $({\bf 2}, {\bf 4})_L$
(in $SU(2)_1\otimes SU(2)_3$) plus a pair of singlets. To clarify the
notation, here we point out that by the subscript $L$ in the case
of fermions in a real irrep of a gauge
group (say, $({\bf 2},{\bf 2}))_L$ of $SU(2)_1 \otimes SU(2)_3$) we mean 
a two component field.

{}This concludes the construction of the $3$-family $SO(10)$ grand 
unified model. The massless spectrum is summarized 
in the first column in Table I. 
There, the states are grouped into sectors, the untwisted sector
$U$ ($\alpha^*=0$), the twisted sectors $T3$ ($\alpha^*=2,~4$),
$T6$ ($\alpha^*=1,~5$) and $T2$ ($\alpha^*=3$).
This grouping implements the selection rules for string couplings,
since the sum of $\alpha^*$s of all the incoming particles at each vertex 
must be zero (mod $6$). For example, a particle in the $T2$ sector coupling 
to a particle in the $T3$ sector can only go to a particle in the $T6$ sector.

\widetext
\section{An Alternative Construction}
\bigskip

{}In this section we will describe the model considered in the previous 
section in a different basis in which the quantum numbers of the states 
are more transparent. This is not surprising, since it
is known that some twists can be re-expressed as shifts.
Again, let us start with
the $N1$ Narain model (that was obtained in the previous section by
starting with the $N0$ model and turning on Wilson lines). This model has 
$N=4$ space-time supersymmetry and the gauge group $SU(3) \otimes SO(8) \otimes
SO(10)^3 \otimes SO(2)$. Recall that the internal momenta span an even 
self-dual Lorentzian lattice $\Gamma^{\prime 6,22}$ described in the previous
section. Let us work in the basis where the right-moving momenta corresponding
to the $\Gamma^{4,4}$ sublattice of the original lattice $\Gamma^{6,22}=
\Gamma^{2,2} \otimes \Gamma^{4,4} \otimes \Gamma^{16}$ are described in terms
of the branching $S0(8)\supset SU(3)\otimes U(1)^2$, used in the 
previous section, whereas the left-moving $SO(8)$ momenta
are written in terms of the branching $SO(8) \supset SU(2)^4$. The 
$SO(8)$ weights ${\bf 0}, {\bf v}, {\bf s}$ and ${\bf c}$ read:
\begin{eqnarray}
 \label{twotfour}
 {\bf 1} =({\bf 1} ,{\bf 1}, {\bf 1}, {\bf 1}) +
          ({\bf 2} ,{\bf 2}, {\bf 2}, {\bf 2})~, \nonumber \\
 {\bf 8}_v =({\bf 2} ,{\bf 2}, {\bf 1}, {\bf 1}) +
          ({\bf 1} ,{\bf 1}, {\bf 2}, {\bf 2})~, \nonumber \\
 {\bf 8}_s =({\bf 1} ,{\bf 2}, {\bf 2}, {\bf 1}) +
          ({\bf 2} ,{\bf 1}, {\bf 1}, {\bf 2})~, \\
 {\bf 8}_c =({\bf 2} ,{\bf 1}, {\bf 2}, {\bf 1}) +
          ({\bf 1} ,{\bf 2}, {\bf 1}, {\bf 2})~.\nonumber
\end{eqnarray}
Note that under cyclic permutations of the first three $SU(2)$s the irreps
${\bf 8}_v$, ${\bf 8}_s$ and ${\bf 8}_c$ are permuted. This is the same as one
of the ${\bf Z}_3$ symmetries
that we considered in the previous section, namely,
the triality symmetry of the $SO(8)$ Dynkin diagram. The second ${\bf Z}_3$ 
twist that we considered in the previous section rotates the bosons 
corresponding to the $SU(3)$ subgroup by $2\pi/3$. In the $SU(2)^4$ basis this
twist is represented by a shift, namely, the ${\bf Z}_3$ shift $-\sqrt{2}/3$
(note that $\sqrt{2}$ is the simple root of $SU(2)$). 
It is convenient to introduce new bosons
corresponding to the first three $SU(2)$s.
Let $\eta_1$, $\eta_2$, $\eta_3$ and 
$\eta_4$ be the real bosons of $SU(2)^4$. 
Next, let $\Sigma =(\eta_1 + \omega^2 \eta_2 + \omega \eta_3 )/ \sqrt{3}$, 
so its complex conjugate $\Sigma^\dagger =(\eta_1 + \omega
\eta_2 + \omega^2 \eta_3 )/ \sqrt{3}$ (here $\omega =\exp(2\pi i /3)$),
and $\rho =(\eta_1 + \eta_2 + \eta_3 )/ \sqrt{3}$. In this
basis, we have one complex boson
$\Sigma$, and two real bosons $\rho$ and $\eta_4$. The outer
automorphism of the first three $SU(2)$s that permutes them is now a ${\bf 
Z}_3$ twist of $\Sigma$($\Sigma^\dagger$) that are eigenvectors with
eigenvalues $\omega$ ($\omega^2$). Meantime, $\rho$ and $\eta_4$ are invariant
under this twist, and therefore unaffected. The other ${\bf Z}_3$ twist 
mentioned earlier corresponds to shifting the momentum lattice of $\eta_4$ by
$-\sqrt{2}/3$. This shift has no effect on $\rho$ and $\Sigma$ 
($\Sigma^\dagger$). 

{}Next let us consider the ${\bf Z}_2$ twist in this new basis. It breaks 
$SO(8)$ to $SU(2)^4$. In the $SU(2)^4$ language this can be achieved by a 
${\bf Z}_2$ shift $\sqrt{2} /2$ in one of the $SU(2)$s. For this shift to be 
compatible with the above ${\bf Z}_3$ twist, we must shift the last $SU(2)$
boson $\eta_4$. As we shall see, this shift, combined with
the above outer automorphism of the first three $SU(2)$s and accompanied by a 
${\bf Z}_3$ shift in the last $SU(2)$ as described above, exactly corresponds
to the ${\bf Z}_6$ orbifold of $\Gamma^{4,4}$ considered in the 
previous section (provided that
the right-movers are twisted as before).

{}Thus, in this new basis, the set of vectors acting on the $N1$ model 
considered in the previous section are given by
\begin{eqnarray}
\label{alt10m}
 &&V_0 =(-{1\over 2} (-{1\over 2}~ 0)^3 \vert\vert 0 \vert 0
 0_r 0_r \vert 0^{5}~0_r^5 ~0_r) ~,\nonumber \\
 &&V_1 =( 0 (-{1\over 3}~{1\over 3})^3 \vert\vert 0 \vert ({2\over 3})
  0_r (-{\sqrt{2}\over 3})_r  \vert 
 ({1\over 3})^5 ~0_r^5 ~({2\over 3})_r )~, \nonumber \\
 &&V_2 =( 0 (0~0)(-{1\over 2}~{1\over 2})^2 \vert\vert {e_1\over 2}
 \vert 0 0_r ({\sqrt{2}\over 2})_r \vert 0^5 ~0_r^5 ~0_r )~,\\
 &&W_1 =( 0 (0~{1\over 2})^3 \vert\vert 0 \vert ({1\over 2}) 0_r 
 0_r \vert ({1\over 2})^5 ~0_r^5 ~0_r )~,\nonumber \\
 &&W_2 =( 0 (0~0)(0~{1\over 2})^2 \vert\vert 0 \vert 0
  0_r 0_r \vert0^5 ~0_r^5 ~0_r )~. \nonumber
\end{eqnarray}
Here we have separated the left-moving $SU(3)$ 
complex boson, the first left-moving entry, from the 
left-moving $SU(2)^4$ bosons by a single vertical line. The second left-moving
entry corresponds to the $\Sigma$ complex boson. The third and fourth entries
are for $\rho$ and $\eta_4$, respectively. The rest of them are the same as
in the previous section, and correspond to the $SO(10)^3 \otimes SO(2)$ 
bosons. Note that we have deliberately chosen the monodromy of the ${\bf Z}_3$
twist that acts on the left-moving $SO(8)$ bosons to be $2/3$; the other 
choice, $1/3$, gives an equivalent model, but only the former choice
can be analyzed using the simplified rules of section III;
in this case, one can check that the condition (\ref{simpli}) is satisfied.

{}First we consider the model generated by the vectors $\{V_0 ,V_1 \}$. Let us 
first consider the untwisted sector of this model. The gauge bosons come in 
the adjoint of $SU(3)_1 \otimes SU(3)_3 \otimes SO(10)_3 \otimes U(1)$.
The appearance of the $SU(3)_1 \otimes SO(10)_3 \otimes U(1)$ gauge bosons 
should be clear from the previous section. Let us concentrate on the 
$SU(3)_3$ gauge bosons. Note that the $SU(2)^4$ subgroup of $SO(8)$ breaks 
to $SU(2)_3 \otimes U(1)$ as a result of the outer automorphism twist 
performed on the first three $SU(2)$s and the shift in the momentum
lattice of the last $SU(2)$ which breaks it to $U(1)$. However, there are
additional gauge bosons that come from the original $SO(8)$ gauge bosons which
in the $SU(2)^4$ basis read $({\bf 2}, {\bf 2}, {\bf 2}, {\bf 2})$. Under the
${\bf Z}_3$ twist that permutes the first three $SU(2)$s we have,
in the branching $SU(2)^3 \supset SU(2)_3$, 
\begin{equation}
 ({\bf 2}, {\bf 2}, {\bf 2}) = {\bf 4}(1) + {\bf 2}(\omega) +{\bf 2}
 (\omega^2) ~,
\end{equation}
where the ${\bf Z}_3$ phases are given in parentheses.
Introducing $J_{SU(2)}$ for the first three $SU(2)$s,
this means $J_{SU(2)}=0, 1, 2$ for the above three irreps respectively.
Under the ${\bf Z}_3$ shift that breaks the fourth 
$SU(2)$ to $U(1)$, we have ${\bf 2} = (+3) +(-3)$, and
${\bf 3} = (0)+(+6)+(-6)$,    
where the normalization of this $U(1)$ charge is $1/3\sqrt{2}$.
Now, Eq.(\ref{irr28}) can be rewritten in the $SU(2)_3\otimes U(1)$ basis,
\begin{eqnarray}
 \label{irr28a}
  {\bf 28} &=& {\bf 8} +{\bf 10}+ {\overline {\bf 10}} \nonumber \\
  &=& [{\bf 1}(0)(0)+{\bf 3}(0)(0)+{\bf 2}(3)(2)+{\bf 2}(-3)(1)] \nonumber \\
  && +[{\bf 1}(-6)(0)+ {\bf 3}(0)(1)+{\bf 2}(-3)(2)+{\bf 4}(3)(0)] \\
  && +[{\bf 1}(6)(0)+ {\bf 3}(0)(2)+{\bf 2}(3)(1)+{\bf 4}(-3)(0)] \nonumber
\end{eqnarray}
where the first set of parentheses gives the $U(1)$ charge and the second set gives the
$J_{SU(2)}$ value. The spectrum generating formula reads, for both the 
${\bf 0}$ and the $V_0$ sectors (with $k_{00}=0$ for definiteness):
\begin{eqnarray}
 &&V_0 \cdot {\cal N}={1\over 2} \sum_{r=0}^3 N^r ={1\over 2} ~
 (\mbox{mod~1})~,\\
 &&V_1 \cdot {\cal N}=
 {1\over 3} (\sum_{r=1}^3 N^r+
 \sum_{\ell=1}^3
 J^\ell - J_{SU(2)} -J_{SO(10)}+2Q -
 \sqrt{2} p(\eta_4) )=0~(\mbox{mod~1})~.
\end{eqnarray}
Here $p(\eta_4)$ is the momentum of the $\eta_4$ real boson, and
the above $U(1)$ charge is equal to $3 {\sqrt{2}} p(\eta_4)$.
For gauge bosons in the ${\bf 0}$ sector, $N^0={\pm 1}$.
Since $\sum_{\ell=1}^3 J^\ell = J_{SO(10)} =0$ for 
massless states, the above formula reduces to
\begin{equation}
 -{1\over 3}(\sqrt{2} p(\eta_4)+ J_{SU(2)}) =0~(\mbox{mod~1})~.
\end{equation}
So, among the ${\bf 28}$ of $SO(8)$, only the ${\bf 2}(3)(2) +{\bf 2}(-3)(1)$
plus the $SU(2)_3 \otimes U(1)$ gauge bosons satisfy this constraint. From 
Eq.(\ref{the8}), we see that the resulting gauge group is $SU(3)_3$.
It is also easy to see that we have three copies
of the Higgs fields in the adjoint of $SO(10)_3$.

{}Although there are no massless states 
in the irreps of $SU(3)_1$, there are, however, chiral fermions in the irrep
of $SU(3)_3$. Let us consider these states in the $V_0$ sector.
Consider the 
left-handed states ({\em i.e.}, the states with the left-handed space-time
chirality) which have $N^0 =0$.
Then there are three massless state solutions to $\sum_{r=1}^3 N^r =
1~(\mbox{mod~2})$, namely, $(N^1, N^2, N^3)=
(1,0,0), (0,1,0), (0,0,1)$;
thus, there are three copies of the states 
we are discussing. The above constraint becomes
\begin{equation}
 {1\over 3}(1-\sqrt{2} p(\eta_4) -J_{SU(2)}) =0~(\mbox{mod~1})~.
\end{equation} 
Of the states in Eq.(\ref{irr28a}), we see that only
the decuplet ${\bf 10}$ of $SU(3)_3$ survives. 
Thus, in the untwisted sector we have three families of left-handed 
${\bf 10}$s of $SU(3)_3$.

{}Next we turn to discussing the twisted sectors. Let us concentrate on the 
${\overline {V_0+V_1}}$ twisted sector (The ${\overline {V_0+2V_1}}$ sector
gives the conjugates of the states appearing in 
${\overline {V_0+V_1}}$). With our choice of $k_{00}=0$, the states in this 
sector are right-handed. The sublattice $I(V_1)\subset\Gamma^{\prime 6,22}$ 
invariant under the twist part of $V_1$ is given by
\begin{equation}
\label{amil}
 I(V_1) =\Gamma^2 \otimes \Delta \otimes \Gamma^6 ~. 
\end{equation}    
Here $\Gamma^2$ and $\Gamma^6$ are the same as in the previous section.
$\Delta = \{(\sqrt{{3\over 2}} n, \sqrt{{1\over 2}} n),~n\in 
{\bf Z} \}$. The determinant of the metric of this lattice is $M(V_1)=
3\times 3 \times 3^5 =3^7$, where the three factors come from the $\Gamma^2$,
$\Delta^2$ and $\Gamma^6$ sublattices, respectively. Thus, the number of
fixed points in the twisted sector ${\overline {V_0+V_1}}$ is given by
\begin{equation}
 \xi({\overline {V_0+V_1}}) =[2\sin(\pi/3)]^9/{\sqrt{M(V_1)}} = 3~.
\end{equation}
The momenta in the ${\overline {V_0+V_1}}$ sector belong to the shifted
lattice ${\tilde I}(V_1) +({\bf 0})(0\vert \sqrt{2}/3)({\bf 0} \vert 2/3)$.
Here ${\tilde I}(V_1) ={\tilde \Gamma}^2 \otimes {\tilde \Delta} \otimes
{\tilde \Gamma}^6$ is the lattice dual to $I(V_1)$. ${\tilde \Gamma}^2$ and
${\tilde \Gamma}^6$ are the same as in the previous section, and 
${\tilde \Delta} =\{(n/\sqrt{6} , n/\sqrt{2}),~n\in 
{\bf Z} \}$. As in the previous section, we have states in the irreps of 
$SO(10)_3 \otimes U(1)$, namely, ${\overline {\bf 16}}(+1)$, ${\bf 10}
(-2)$ and ${\bf 1}(+4)$ (with the $U(1)$ charge radius 
$1/6$). These states come in three copies (due to the three right-moving 
fixed points), and they also have $SU(3)_3$
quantum numbers. For the states to be 
massless, the $SU(2)_3 \otimes U(1) (\subset SU(3)_3)$ quantum numbers must
be ${\bf 1}(-2)$ (This state has $p(\eta_4)=p(\rho)=0$, where
$p(\rho)$ is the momentum of the $\rho$ boson), and ${\bf 2}(+1)$
(These states have $p(\eta_4)=1/\sqrt{2}$ and $p(\rho) =\pm 1/\sqrt{6}$).
These states combine into ${\bf 3}$ of $SU(3)_3$. Thus, in the twisted sectors
we have three families of left-handed $({\overline {\bf 3}}, 
{\bf 16})(-1)$, $({\overline {\bf 3}},{\bf 10})(+2)$, $({\overline {\bf 3}},
{\bf 1})(-4)$ (irreps of $SU(3)_3 \otimes SO(10)_3 \otimes U(1)$).
This concludes the discussion of the $\{ V_0 ,V_1 \}$ model, which is 
the same as the $A1$ model constructed in the previous section.

{}Next we turn to the final model generated by the vectors 
$\{ V_0 ,V_1 ,V_2 \}$ (\ref{alt10m}).
As in the previous section, to preserve $N=1$ space-time supersymmetry we 
must choose $k_{20} =1/2$. Then in the untwisted sectors, we simply 
keep the  ${\bf Z}_2$ invariant states in the untwisted sector in the
$A1$ model. Since the $V_2$ vector breaks the 
original $SU(3)\otimes SO(8)$ gauge group to 
$SU(2) \otimes U(1) \otimes SU(2)^4$, the gauge group in the final model
is $SU(2) \otimes U(1) \otimes SU(2)_3 \otimes U(1) \otimes SO(10)_3 \otimes
U(1)$. Out of the three copies of matter fields in the untwisted sectors
of the $\{ V_0 ,V_1 \}$ model, two copies have the ${\bf Z}_2$ phase $-1$,
while the other copy has the ${\bf Z}_2$ phase $+1$. The singlet and the 
triplet of $SU(2)_3$ in the branching (\ref{Ten}) have the ${\bf Z}_2$ phase
$+1$, whereas the doublet and the quartet have the ${\bf Z}_2$ phase $-1$.
Therefore, we have one copy of the singlet and the triplet, and two copies
of the doublet and the quartet (with the corresponding $U(1)$ charges). Also,
only one copy of the $SO(10)_3$ adjoint Higgs field survives.
   
{}Now consider the ${\overline {V_0+V_1}}$ twisted sector.
There are two sources of ${\bf Z}_2$ phases.
One is the fixed points, and the other one is the ${\bf 3}$ of $SU(3)_3$.
In the branching ${\bf 3} = {\bf 1} (-2) +{\bf 2}(+1)$ (under the breaking
$SU(3)_3 \supset SU(2)_3 \otimes U(1)$) the singlet has the ${\bf Z}_2$ phase
$+1$ (since $p(\eta_4)/\sqrt{2}=0$ for this state), whereas the doublet has
the ${\bf Z}_2$ phase $-1$ (since $p(\eta_4)/\sqrt{2}=1/2$ for these states).
Similarly, one of the fixed points has the ${\bf Z}_2$ phase $+1$, while the
other two have the ${\bf Z}_2$ phase $-1$. This means that, say, 
out of the three copies of the left-handed states 
$({\overline {\bf 3}}, {\bf 16})$ in the $T3$ sector of the $\{V_0 ,
V_1 \}$ model, two copies of $({\bf 2}, {\bf 16})_L$ and one copy of 
$({\bf 1}, {\bf 16})_L$ survive the ${\bf Z}_2$ projection. That is, the 
$T3$ sector gives rise to five left-handed spinors of $SO(10)_3$ (as
well as their vector and singlet counterparts with the corresponding $U(1)$
charges). The above discussion can be seen directly from the
rules of section III. The $V_2$ constraint in the $V_1$ sector becomes:
\begin{equation}
 V_2 \cdot {\cal N}_{V_1} +f_2 (V_1 ,{\bf n}) =1/2~(\mbox{mod~1})~. 
\end{equation}
Now, $\xi(V_1 ,(n_1, n_2) =(0,0))=2$, $\xi(V_1 ,(n_1, n_2) =(0,1))=1$, 
and all the others are zero. Their fixed point phases are
$f_2 (V_1 ,(n_1 ,n_2) =(0,0))=0$ and $f_2 (V_1 ,(n_1, n_2) =(0,1))=1/2$,
respectively. In the above
discussion, the ${\bf Z}_2$ phases of the fixed points were defined as those
that add up with the $V_2 \cdot {\cal N}_{V_1}$ phases (which are the phases
coming from the ${\bf 3}$ of $SU(3)_3$) to give zero. That is, the fixed point
phases in our notation are given by $f_2 (V_1 ,{\bf n}) -1/2$, so that $2$ 
fixed points have the ${\bf Z}_2$ phase $-1$, and $1$ fixed point has the 
${\bf Z}_2$ phase $+1$, as inferred earlier. So the above spectrum generating
formula gives us five generations of the
$SO(10)_3$ chiral matter fields.

{}Next consider the $T6$ sector. The sublattice invariant under the
twist part of $V_1+V_2$ (${\bf Z}_6$ twist) is the same as $I(V_1)$.   
In these sectors the number of fixed points, therefore, can be seen to be $1$.
Let us consider the $V_0 +V_1 +V_2$ twisted sector for definiteness. The 
states in this sector are left-handed, as can be seen from the spectrum 
generating formula. The number of massless chiral ${\overline {\bf 16}}$s 
coming from this sector is two. These have opposite $U(1)$ charges $\pm 1$
(with all
the other quantum numbers the same), where this $U(1)$ is the one that comes
from the $SU(3)_1 \supset SU(2)_1 \otimes U(1)$ breaking (and is normalized to
$1/\sqrt{6}$). This means that the $T6$ sector gives rise to $2$ 
right-handed families of of $SO(10)_3$ (as
well as their vector and singlet counterparts with the corresponding $U(1)$
charges). The net number of generations from the $T3$ and $T6$ sectors is,
therefore, $5-2=3$.

{}Finally, let us discuss the $T2$ sector. The invariant sublattice
$I(V_2) =\Gamma^{2,18} \otimes \Gamma^4$. Here $\Gamma^{2,18}$ is an even 
self-dual Lorentzian lattice which can be obtained from 
$\Gamma^{2,2} \otimes \Gamma^{16}$ by turning on the Wilson lines similar to
those that we use to construct $\Gamma^{\prime 6,22}$ from 
$\Gamma^{6,22}$ but without the entries corresponding to the $\Gamma^{4,4}$
sublattice of $\Gamma^{6,22}$. Recall that $\Gamma^4$ is the root lattice
of $SO(8)$, and the determinant of its metric is $4$. Therefore, the
number of fixed points in the $T2$ sector is $\xi(V_2)=
[2\sin(\pi/2]^2 / \sqrt{4}=2$. One of these fixed points has the ${\bf Z}_3$ 
phase $\omega$, and the other one has the ${\bf Z}_3$ phase $\omega^2$. The 
momenta in the $T2$ sector belong to the shifted lattice
${\tilde I}(V_2)+({\bf 0} \vert\vert e_1 /2 \vert {\bf 0},{\bf 0},{\bf 0},0)
(0,0,0,1/\sqrt{2})$ (Here the first factor is a shift in the $\Gamma^{2,18}$
lattice, while the second factor, which is written in the $SU(2)^4$ basis,
is the shift in the ${\tilde \Gamma}^4$ lattice). Note that all the states with
the $SO(10)_3 \otimes SO(2)$ quantum numbers are massive. The only massless 
states come from the irreps of  $SU(2)\otimes U(1) \otimes SU(2)^4$ \ 
(with the $U(1)$ radius $1/\sqrt{6}$):
$({\bf 1} \vert {\bf 1}, {\bf 1} ,{\bf 1} ,{\bf 2})(\pm 3)$ and 
$({\bf 2} \vert {\bf 2}, {\bf 2} ,{\bf 2} ,{\bf 1})(0)$ 
After the ${\bf Z}_3$ projection we have (in the $SU(2)_1 \otimes SU(2)_3$
irreps) one $({\bf 2} ,{\bf 2})$, one 
$({\bf 2} ,{\bf 4})$, and a singlet (see the first column in Table I). 
This completes our analysis of the alternative construction of the model.

\widetext
\section{The Moduli Space}
\bigskip

{}In section IV we could start from the most general
even self-dual Lorentzian 
lattice $\Gamma^{4,4}$ that admits a ${\bf Z}_6 \otimes {\bf Z}_6$
symmetry so that both complex bosons can be simultaneously twisted
(Then the particular choice of the lattice with enhanced $SO(8)$ gauge
symmetry corresponds to a special point in the moduli space of such lattices). 
Since the ${\bf Z}_2 \otimes {\bf Z}_2$ symmetry is automatic, we need 
only to consider the ${\bf Z}_3 \otimes {\bf Z}_3$ symmetry. 
The most general
lattice that possesses such a symmetry can be written in the
following ${E_I, I=1,2,3,4}$ basis which proves to be useful for our purposes. 
Let us also define ${\tilde E}^I$, so that $E_I \cdot
{\tilde E}^J ={\delta_I}^J$.
Now, $\Gamma^{4,4} = \{(P_R \vert\vert P_L ) \}$ with
\begin{equation}
\label{modumom}
 P_{L,R} ={\tilde E}^I (M_I -2B_{IJ}N^J) \pm E_I N^I /2 ~,
\end{equation}
where $M_I ,N^I \in {\bf Z}$, $I=1,2,3,4$; $B_{IJ}$ and
$G_{IJ} =E_I \cdot E_J$ are the anti-symmetric and the metric
constant background fields, respectively.
Recall that ${\tilde G}^{IJ}={\tilde E}^J \cdot {\tilde E}^J$ is the 
inverse of $G_{IJ}$.
Next we define the action of the ${\bf Z}_3$ twist
$\Theta$ on these momenta as $\Theta E_I = {\Theta_I}^J E_J$ and
$\Theta {\tilde E}^I = {\Theta^I}_J {\tilde E}^J$, where 

\begin{equation}\label{Theta}
 {\Theta_I}^J =\left( \begin{array}{cccc}
               0 & 1 & 0 & 0\\
               -1 & -1 & 0 & 0\\
               0 & 0 & 0 & 1\\
               0 & 0 & -1 & -1
               \end{array}
        \right)~,~~~
 {\Theta^I}_J =\left( \begin{array}{cccc}
               -1 & 1 & 0 & 0\\
               -1 & 0 & 0 & 0\\
               0 & 0 & -1 & 1\\
               0 & 0 & -1 & 0
               \end{array}
        \right)~.
\end{equation}
Note that they are block diagonal and ${\Theta^I}_J ={(\Theta^{-1})_J}^I$.

{}An orbifold twist must leave the conformal dimensions invariant; 
that is, under the $\Theta$ twist,
\begin{equation}
   P_{L,R}~~{\rightarrow}~~ P_{L,R}'
\end{equation}
we must have $P_L^2/2= (P_L')^2/2$ and $P_R^2/2= (P_R')^2/2$.
This requirement imposes strong constraints on
$B_{IJ}$ and $G_{IJ}$. It is straightforward to solve these constraints, 
and we have
\begin{equation}
	2B_{IJ}=\left( \begin{array}{cccc}
               0 & d & a & c\\
               -d & 0 & -a-c & a\\
               -a & a+c & 0 & b\\
               -c & -a & -b & 0
               \end{array}
        \right)~,~~~
 G_{IJ} =\left( \begin{array}{cccc}
               2u & -u & f & h\\
               -u & 2u & -f-h & f\\
               f & -f-h & 2v & -v\\
               h & f & -v & 2v
               \end{array}
        \right)~.
\end{equation}
where $2B_{IJ}$ are defined modulo one.
Thus, the moduli space of the $\Gamma^{4,4}$ lattice with the ${\bf Z}_3$
(and the ${\bf Z}_6$) symmetry is eight-dimensional
(four parameters from $B_{IJ}$, and the other four from $G_{IJ}$).
In the $N0$ model, the gauge symmetry is $SU(3) \otimes R \otimes SO(32)$,
where the gauge symmetry group $R$ coming from $\Gamma^{4,4}$ 
depends on the values of the moduli. 
A generic point in the above moduli space has $R=U(1)^4$. 
There are also special points in this moduli
space with enhanced gauge symmetry. These are of particular interest 
since the $R=U(1)^4$ leads to an empty messenger group $M$ in the 
corresponding $N=1$ model, which is phenomenologically unappealing. 
For this reason, we will confine our attention to a one-dimensional
subspace of the above moduli space that includes the points of enhanced
gauge symmetry. To describe this subspace it suffices to put $G_{11}=
G_{33}=2$, $G_{13}=0$;
$2B_{IJ}=-{1\over 2} G_{IJ}$ if $I>J$, and
$2B_{IJ}={1\over 2} G_{IJ}$ if $I<J$. So, there is only one free
parameter left, $h \equiv G_{14}$. Let us consider the region 
$0\leq h \leq 1$. We may express the vectors $E_I$ (and ${\tilde E}^I$)
in terms of the $SU(3)$ root and weight vectors $e_i$ and ${\tilde e}^i$:
\begin{eqnarray}
 \begin{array}{rlrlrlrl}
  E_1 &= (e_1 , 0) &  E_2 &= (e_2, 0) &
  E_3 &= (-h {\tilde e}^2 ,g e_1 )~~ &  E_4 &= (h {\tilde e}^1 ,g e_2 ) \\
  {\tilde E}^1 &= ({\tilde e}^1 ,-{h e_2}/ 3g )~~ &
  {\tilde E}^2 &= ({\tilde e}^2 ,{h e_1}/ 3g )~~ &
  {\tilde E}^3 &= (0,{{\tilde e}^1}/g ) &
  {\tilde E}^4 &= (0,{{\tilde e}^2}/ g )
 \end{array}
\end{eqnarray}
{}Here $g\equiv \sqrt{1-{h^2 /3}}$. Note that the
states with $P_R =0$ and $P^2_L =2$ are massless. For $h \not=0,1$
the only solutions to these conditions are those with $P_L =E_I N^I$, 
$N^3,N^4=0$, which give six gauge bosons of $SU(3)$. 
The other two gauge bosons 
are the Cartan generators, coming from the oscillator excitations. 
There are two more gauge bosons coming from the oscillator excitations 
of the other two world-sheet bosons.
So, for $0<h<1$, we have $R=SU(3)\otimes U(1)^2$. 
For $h=0$ the gauge symmetry is enhanced to $SU(3)^2$,
as $N^3$ and $N^4$ in Eq.(\ref{modumom}) need no longer be
zero for the states to be massless.
Another special point of enhanced gauge
symmetry is $h=1$ where $R=SO(8)$.
This can be readily verified by writing the irreps of
$SO(8)$ in the $ SU(3) \otimes U(1)^2$ basis (\ref{eight}).

{}So the $N0$ model has gauge symmetry $SU(3) \otimes R \otimes SO(32)$,
where $R=SU(3)^2, SU(3)\otimes U(1)^2$ and $SO(8)$ for $h=0$, $0<h<1$ and
$h=1$ respectively. Away from this one-dimensional moduli space,
$R=U(1)^4$ generically. Next we introduce the 
Wilson lines that break the gauge symmetry to 
$SU(3) \otimes R \otimes SO(10)_3\otimes U(1)$, where $R$ is untouched:
\begin{eqnarray}
 &&V_1 =(0~(0~e_1/2)(0~a_1)(0~b_1) \vert\vert 0~c_1 ~d_1 \vert
 {\bf s}~{\bf 0}~{\bf 0}
 \vert {C})~,\\
 &&V_2 =(0~(0~e_2/2)(0~a_2)(0~b_2) \vert\vert 0~c_2 ~d_2 \vert
 {\bf 0}~{\bf s}~{\bf 0}
 \vert {C})~.
\end{eqnarray}
Here $(a_1 ,b_1) =(-h{\tilde E}^4+E_1+E_3)/2$, and 
$(c_1 ,d_1) =-h{\tilde E}^4/2$ for $0\leq h \leq 1/2$;
$(a_1 ,b_1) =((h-1){\tilde E}^4-E_1+E_3)/2$
and $(c_1 ,d_1) =(h-1){\tilde E}^4/2$ for $1/2 < h \leq 1$.
Similarly, $(a_2 ,b_2) =(h{\tilde E}^3+E_2+E_4)/2$, and 
$(c_2 ,d_2) =h{\tilde E}^3/2$ for $0\leq h \leq 1/2$;
$(a_2 ,b_2) =(-(h-1){\tilde E}^3-E_2+E_4)/2$, and 
$(c_2 ,d_2) =-(h-1){\tilde E}^3/2$ for $1/2 h \leq 1$.
Mathematically, this definition of the Wilson lines $V_1$ and $V_2$ is 
discontinuous. Physically, however, there is no discontinuity as the latter 
belongs to $\Gamma^{6,22}$ (Recall that a lattice shift is defined up to
a lattice spacing as the latter does not affect the corresponding orbifold
model).
Using exactly the same ${\bf Z}_3\otimes {\bf Z}_2$ orbifold
set $\{V_i, k_{ij}\}$ given in section IV, we obtain
the corresponding $3$-family $SO(10)$ grand unified model with
gauge symmetry $SU(2)_1\otimes U(1)\otimes M\otimes SO(10)_3\otimes U(1)$.
Here $M=U(1)^2, U(1)$ and $SU(2)_3\otimes U(1)$ for 
$R=SU(3)^2, SU(3)\otimes U(1)^2$ and $SO(8)$ respectively.
For $R=U(1)^4$, $M$ is empty. 
The $h=1$ point with $SO(8)$ has been analyzed in great detail already. 
The massless spectra of the $M=U(1)^2$ model, the $M=U(1)$ model and the
empty $M$ model are given in Table II.  

{}Note that we can also obtain the $M=U(1)$ model by
giving a non-zero vacuum expectation value to the appropriate Higgs field 
in either the $M=SU(2)_3\otimes U(1)$ model 
(the $({\bf 1},{\bf 2},{\bf 1})(0,-3,0)$) or 
the $M=U(1)^2$ model ($({\bf 1},{\bf 1})(0,0,-6,0)$).
Similarly, the empty $M$ model can be obtained from the $M=U(1)$ model via
spontaneous symmetry breaking in the corresponding effective field theory.
We may also obtain $SU(5)$ grand unified models by giving the $SO(10)$
adjoint (and possibly other) Higgs fields non-zero vacuum expectation values.
A typical grand unification scale is a few orders of magnitude below the
Planck scale.
Presumably, the effective field theory approach is adequate if the 
vacuum expectation values are at the grand unification scale or below.
When the vacuum expectation values approach the Planck or string scale,
the string construction is clearly preferable.
A more complete exploration of the full moduli space is clearly desirable.

{}The hidden sector $SU(2)_1$ of the above set of models 
is asymptotically free while $M$ is not. At the
string scale, the hidden $SU(2)_1$ coupling $\alpha_2$ is three times that
of the $SO(10)_3$. So, for typical values of the $SO(10)$ grand unification
coupling, $\alpha_2$ becomes large at a rather low scale, within a few orders
of magnitude above the electroweak scale. If $M$ is empty, the hidden
sector physics will have negligible impact on the physics in the
visible sector. So this is phenomenologically undesirable. So hopefully,
$M$ is non-trivial at the string scale, and the spontaneous symmetry breakings
of the $M$ sector occur at relatively low scales.

\widetext
\section{Another $SO(10)$ Model}
\bigskip

{}Next we turn to a variation of the above 
$SU(2)_1 \otimes SU(2)_3 \otimes SO(10)_3 \otimes U(1)^3$ model. 
Consider the model generated by the following vectors acting on the $N1$ 
model, in the same basis as in the last section:
\begin{eqnarray} \label{alts}
 &&V_0 =(-{1\over 2} (-{1\over 2}~ 0)^3 \vert\vert 0^2
  0_r 0_r \vert 0^{5}~0^5_r ~0_r) ~,\nonumber \\
 &&V_1 =( 0 (-{1\over 3}~{1\over 3})^3 \vert\vert {e_1\over 3}
 ({2\over 3}) 0_r 0_r\vert
 ({1\over 3})^5 ~0^5_r ~({2\over 3})_r )~, \nonumber \\
 &&V_2 =( 0 (0~0)(-{1\over 2}~{1\over 2})^2 \vert\vert {e_1\over 2}
  0 0_r ({\sqrt{2}\over 2})_r \vert 0^5 ~0^5_r ~0_r )~,\\
 &&W_1 =( 0 (0~{1\over 2})^3 \vert\vert 0~{1\over 2}~0_r 0_r \vert
 ({1\over 2})^5 ~0^5_r ~0_r )~, \nonumber \\
 &&W_2 =( 0 (0~0)(0~{1\over 2})^2 \vert\vert 0^2
  0_r 0_r \vert 0^5 ~0^5_r ~0_r )~.\nonumber
\end{eqnarray}
This model has the same gauge symmetry
$SU(2)_1 \otimes SU(2)_3 \otimes SO(10)_3 \otimes U(1)^3$.
It differs from the above model by shifting the momentum lattice of the
$SU(3)$ boson instead of the last of the $SO(8)$ bosons. As a consequence,
this model has a spectrum very similar to that of the one we discussed 
earlier, but the assignments of the $M$ and $U(1)$ quantum numbers are 
different. Without going
into details, we will simply make a few comments on this model. Note that 
the $V_2$ vector is the same as that in Eq.(\ref{alt10m}) and, therefore, 
its effect on the $N1$ model is the same as before. The $V_1$ vector is 
different from that in  Eq.(\ref{alt10m}),
however. It breaks the original gauge group $SU(3) \otimes SO(8) \otimes
SO(10)^3 \otimes SO(2)$ to $U(1)^2 \otimes G_2 \otimes SO(10)_3 \otimes U(1)$.
The $G_2$ gauge bosons can be seen as follows. The $SU(2)^4$ gauge group is
reduced to $SU(2)_3 \otimes SU(2)_1$. But there are additional states coming
from the original $({\bf 2}, {\bf 2}, {\bf 2}, {\bf 2})$ of the $SO(8)$ 
gauge bosons (see Eq.(\ref{twotfour})).
The only states that survive the ${\bf Z}_3$ projection are those with the
$({\bf 4}, {\bf 2})$ quantum number (in $SU(2)_3 \otimes SU(2)_1$). The above
gauge bosons form the adjoint of $G_2 (\supset SU(2)_3 \otimes SU(2)_1$):
${\bf 14} =({\bf 3},{\bf 1}) +({\bf 1},{\bf 3}) +({\bf 4}, {\bf 2})$. The
further ${\bf Z}_2$ projection removes the $({\bf 4}, {\bf 2})$ states,
and we are left with the $SU(2)_3 \otimes SU(2)_1$ subgroup of $G_2$.
The number of chiral generations in the $T3$ and $T6$ sectors is still $5_L$
and $2_R$. The number of the $SU(2)_1$ doublets is still six, so that this
subgroup (which is still the hidden sector) is asymptotically free. The
complete massless spectrum of this model is given in the second column in
Table I.

{}Note that both the hidden $SU(2)_1$ sector as well as the $SU(2)_3$ in the 
M sector in this model comes from the $SO(8)$ lattice in the original 
$N1$ model. In the moduli space discussed in the previous section, we see 
that $SO(8)$ is a special point. When we move away from that point in the 
moduli space, the gauge symmetry $SU(2)_1 \otimes SU(2)_3$ in the final 
model will be reduced. In that case, the biggest non-abelian gauge group 
one can get is $SU(2)_4$, not very interesting phenomenologically
since it turns out not to be asymptotically free.
One may gain further insight into the relation between this and the first 
$SO(10)$ models by considering a more general moduli space.

{}Let us make another comment concerning the construction of this model. 
The above construction, 
{\em i.e.}, the set $\{V_i,k_{ij}\}$, satisfies the simplified rules.
Then the condition (\ref{simpli}) follows. This same model may be obtained 
by a different set of $\{V_i,k_{ij}\}$. 
This alternative set turns out not to satisfy the simplified rules. So in
that case, we must use the full set of rules. This issue is discussed in
Appendix B.

\widetext
\section{The Third $SO(10)$ Model}
\bigskip

{}Again we start with with a Narain model, called the $N2$ model, with 
$\Gamma^{6,22}=\Gamma^{6,6} \otimes \Gamma^{16}$, where
$\Gamma^{16}$ is the ${\mbox{Spin}}(32)/{\bf Z}_2$ lattice, and
$\Gamma^{6,6}$ is the $E_6$ lattice, {\em i.e.}, $p_R ,p_L \in {\tilde
\Gamma}^6$ ($E_6$ weight lattice), and $p_L - p_R \in \Gamma^6$
($E_6$ root lattice). We choose this particular lattice for later 
convenience. Recall that, under $E_6\supset SU(3)^3$,
\begin{eqnarray}
 {\bf 27} &=& ({\bf 3},{\bf 3},{\bf 1})+ ({\overline {\bf 3}},{\bf 1},
  {\bf 3}) + ({\bf 1}, {\overline {\bf 3}}, {\overline {\bf 3}}) ~,\\
 {\bf 78} &=& ({\bf 8},{\bf 1},{\bf 1}) +({\bf 1},{\bf 8},{\bf 1}) +
 ({\bf 1},{\bf 1},{\bf 8})+ ({\bf 3},{\overline {\bf 3}},{\bf 3})+
 ({\overline {\bf 3}},{\bf 3}, {\overline {\bf 3}}) ~, 
\end{eqnarray}
so we can write ${\bf p} \in \Gamma^6$ as
\begin{equation}
 \label{psu3}
 {\bf p}=({\bf q}_1+ a {\bf w}_1,{\bf q}_2+ a {\overline {\bf w}}_2,
  {\bf q}_3+ a {\bf w}_3) ~,
\end{equation}
where ${\bf q}_i \in \Gamma^2$ ($SU(3)$ root lattice), ${\bf w}_i$
(${\overline {\bf w}}_i$) is in the ${\bf 3}$ (${\overline {\bf 3}}$)
weight of $SU(3)$, and $ a=0,~1,~2$.
To convert this model to a $N=4$ supersymmetric model with gauge symmetry
$SU(3)^2 \otimes U(1)^2 \otimes SO(10)^3 \otimes SO(2)$, we introduce,
besides $V_0$, two Wilson lines,
\begin{eqnarray} \label{N3mod}
 &&V_1 =(0(0~{e_1\over 2})^2 (0~0) \vert\vert 0^2 ~{e_1\over 2}\vert
{\bf s}~{\bf 0}~{\bf 0} \vert C) ~,\\
 &&V_2 =(0(0~{e_2\over 2})^2 (0~0) \vert\vert 0^2 ~{e_2\over 2}\vert
{\bf 0}~{\bf s}~{\bf 0} \vert C) ~,
\end{eqnarray}
where $e_i$ are $SU(3)$ roots, ${\bf s}$ is a component in the spinor 
weight of $SO(10)$, and $C=-1/2$ in $SO(2)$. Under the action of these 
Wilson lines, we see that states with $a=1, 2$
in (\ref{psu3}) are removed, and the $E_6$ gauge symmetry
is broken down to $SU(3)^2 \otimes U(1)^2$. Since no new gauge bosons
are introduced by the other sectors, we end up with the gauge symmetry
$SU(3)^2 \otimes U(1)^2 \otimes SO(10)^3 \otimes SO(2)$.
The model remains $N=4$ supersymmetric, since
the Wilson lines do not break supersymmetry.
Let us call this the $N3$ model.

{} To obtain the $3$-family $SO(10)$ model, we introduce the following 
set of vectors to act on the $N3$ model,
\begin{eqnarray} 
  \label{third10}
 &&V_0 =(-{1\over 2} (-{1\over 2}~ 0)^3 \vert\vert 0^3
  \vert 0^{5}~0^5_r ~0_r) ~,\nonumber \\
 &&V_1 =( 0 (-{1\over 3}~{1\over 3})^3 \vert\vert 0~{e_1\over 3}~{2\over 3}
  \vert ({1\over 3})^5 ~0^5_r ~({2\over 3})_r )~, \nonumber \\
 &&V_2 =( 0 (0~0)(-{1\over 2}~{1\over 2})^2 \vert\vert ({e_1\over 2})^2
   0 \vert 0^5 ~0^5_r ~0_r )~,\\
 &&W_1 =( 0 (0~{1\over 2})^3 \vert\vert 0~0~{1\over 2} \vert
 ({1\over 2})^5 ~0^5_r ~0_r )~, \nonumber \\
 &&W_2 =( 0 (0~0)(0~{1\over 2})^2 \vert\vert 0^3
   \vert 0^5 ~0^5_r ~0_r )~.\nonumber
\end{eqnarray}
Here, it is important to specify the basis the above vectors are acting on.
As before, the $SO(10)$ bosons are in the basis (\ref{basis10}). The new
feature is the basis of the right-moving bosons in $\Gamma^{6,6}$, namely,
$X_1,~X_2$ and $X_3$. The above vectors act on the basis $X_{+}$, $X_{-}$ 
and $X_3$, where
\begin{equation} \label{rbasis}
 X_{\pm} ={1\over \sqrt{2}}(X_1 {\pm} X_2)~.
\end{equation}
This means that $V_2$ acts as a ${\bf Z}_2$
outer automorphism on the first two right-moving complex bosons.
The final gauge symmetry of this model 
is $SU(2)_1\otimes SO(10)_3 \otimes U(1)^4$.
It is now straightforward to work out its massless spectrum of this,
which is given in the third column in Table I.

\widetext
\section{$E_6$, $SU(5)$ and Others}
\bigskip

{}$E_6$ grand unified string models were recently constructed 
by Erler\cite{erler}. His models involve $(E_6)_2$ models with an
even number of families. To obtain an odd number of families, it is natural 
to go to level-$3$. Below is a brief description of a 
$3$-family $(E_6)_3$ model, which is closely related to the third $SO(10)$
model given in section VIII. We shall first construct
a $N=4$ supersymmetric model with gauge symmetry
$SU(3)^2 \otimes E_6^3$, and then orbifold it. We shall also show how
$3$-family $SU(5)$ models can be obtained from the above $SO(10)$ models.
On the way, we shall also present a $3$-family $SU(6)$ model.

{}Let us start with the above $N2$ model, 
with $\Gamma^{6,22}=\Gamma^{6,6} \otimes \Gamma^{16}$, where 
$\Gamma^{16}$ is the ${\mbox{Spin}}(32)/{\bf Z}_2$ lattice, and
$\Gamma^{6,6}$ is the $E_6$ lattice.
To convert this model to a model with gauge symmetry 
$SU(3)^2 \otimes E_6^3$, we introduce, besides $V_0$, two Wilson lines,
\begin{eqnarray}
 \label{wilsone6}
 &&V_1 =(0(0~0)^3 \vert\vert 0^2 ~{e_1\over 2}\vert {\bf s}~{\bf 0}~{\bf 0} 
  \vert C) ~,\\
 \label{wilsone61}
 &&V_2 =(0(0~0)^3 \vert\vert 0^2 ~{e_2\over 2}\vert {\bf 0}~{\bf s}~{\bf 0}
  \vert C) ~,
\end{eqnarray}
The left-moving parts of these two Wilson lines are identical to the 
Wilson lines (\ref{N3mod}) introduced above. Again, the gauge bosons 
corresponding to the 
states with $ a =1, 2$ in Eq.(\ref{psu3}) are projected out. So
the gauge bosons in the $\alpha V={\bf 0}$ sector form the gauge symmetry
$SU(3)^2 \otimes U(1)^2 \otimes SO(10)^3 \otimes SO(2)$, just like the $N3$
model above. The difference comes from the introduction of additional gauge 
bosons by the new sectors. Recall that under $E_6\supset SO(10)\otimes U(1)$,
\begin{equation}
 {\bf 78} = {\bf 1}(0)+{\bf 45}(0)+{\bf 16}(3)+ {\overline {\bf 16}}(-3)~.
\end{equation}
It is easy to see that the $V_1$, $V_2$ and $V_1+V_2$ sectors provide
the necessary ${\bf 16}(3)$ and ${\overline {\bf 16}}(-3)$ gauge bosons 
to the three $SO(10)s$ respectively (A little care is needed to 
disentangle the three $U(1)s$). So the resulting model, to be referred 
to as the $N4$ model, has a gauge symmetry $SU(3)^2 \otimes E_6^3$.

{} Next, we perform a ${\bf Z}_6$ orbifold on this $E_6^3$ model, 
{\em i.e.}, we let the set of vectors (\ref{third10}) act on this $N4$
model. Now $V_1$, {\em i.e.}, the ${\bf Z}_3$
twist, on $(E_6)^3$ converts it to $(E_6)_3$, yielding
$9$ chiral families, {\em i.e.}, ${\bf 27}_L$ of $(E_6)_3$. 
Again, $V_2$ cuts the $9$ left-handed families to five, while 
introducing $2$ right-handed families. The massless spectrum of this 
$E_6$ model is shown in the first column in Table III. The analysis is
very similar to earlier discussions, so we shall not repeat them. Instead,
we shall make a number of related remarks.

{}We note that the third $SO(10)_3$ model constructed above can also 
be obtained 
from this $E_6$ model by spontaneous symmetry breaking. Classically, a
typical vacuum expectation value of the adjoint Higgs of the $E_6$ model 
lies in a flat direction. Choosing an appropriate vacuum expectation value 
breaks the $E_6$ gauge symmetry to $SO(10)\otimes U(1)$, yielding the 
third $SO(10)_3$ model given in the third column in Table I.

{}Note that we may reach the $N4$ model via
other paths. For example, we may start with
$\Gamma^{6,22}=(\Gamma^{2,2})^3 \otimes (\Gamma^{8})^2$,
where $\Gamma^{2,2}$ is the same $SU(3)$ lattice
in the $N0$ model and $\Gamma^{8}$ is the $E_8$ lattice,
and then introduce, besides $V_0$, the following vector:
\begin{equation}
V_1 =( 0 (0~0)^3 \vert\vert 0^2~{\tilde e}^1 \vert
 ({1\over 3})_r~({1\over 3})_r~({2\over 3})_r~0^5_r \vert
 (-{1\over 3})_r~(-{1\over 3})_r~(-{2\over 3})_r~0^5_r) \nonumber
\end{equation}
This $V_1$ may be rewritten in a more suggestive way in the $E_8 \supset
E_6 \otimes SU(3)$ basis:
\begin{equation}
V_1 =( 0 (0~0)^3 \vert\vert 0^2~{\tilde e}^1 \vert
 (-{\tilde e}^1)~{\bf 0} \vert {\tilde e}^1~{\bf 0}) \nonumber
\end{equation}
where the ${\bf 0}$ indicates the identity weight of $E_6$.
The ${\bf 0}$ sector gives rise to the gauge bosons of 
$SU(3)^2 \otimes SU(3)^3 
\otimes E_6 \otimes E_6$. The $V_1$ and $2V_1$ sectors supply the 
remaining gauge bosons $({\bf 3}, {\overline {\bf 3}},{\bf 3})$ and
$({\overline {\bf 3}},{\bf 3}, {\overline {\bf 3}})$ in $SU(3)^3$, so
the final gauge symmetry is $SU(3)^2 \otimes E_6^3$.

{}In each of the above $SO(10)$ models, we may also introduce a Wilson line
to break it to $SU(5)$. This is the string equivalence of spontaneous 
symmetry breaking. To be specific, we can add to any of the above sets of 
Wilson lines the following vector,
\begin{equation}
 \label{tent5}
 V_3 =( 0 (0~0)^3 \vert\vert {\tilde e}^2 ~0^2~ \vert
 ({1\over 3}~{1\over 3}~{1\over 3}~{1\over 3}~{2\over 3})^3_r \vert 0_r)~.
\end{equation}
Acting on any of the $N=4$ supersymmetric models that contain the gauge 
symmetry $SO(10)^3$, {\em i.e.}, the $N1$ or the $N3$ model, this Wilson 
line breaks $SO(10)$ to $SU(5)\otimes U(1)$. To illustrate this construction,
let us consider the case where this Wilson line is added to the 
set (\ref{N3mod}) 
acting on the $E_6\otimes SO(32)$ model. This results in the $N5$ model, 
{\em i.e.}, an $N=4$ supersymmetric model with gauge symmetry 
$SU(3)^2 \otimes SU(5)^3 \otimes U(1)^6$. 
Now, we let the set (\ref{third10}) of twists act on this $N5$ model. 
This results in a $3$-family $SU(5)$ model, with gauge symmetry
$SU(2)_1 \otimes SU(5)_3 \otimes U(1)^5$.
Its massless spectrum is given in the third column in Table III. 
It is just as easy to add the Wilson
line $V_3$ to the other cases to break $SO(10)_3$ to 
$SU(5)_3 \otimes U(1)$.

{}If we add the $V_3$ Wilson line (\ref{tent5}) to the set of Wilson lines 
(\ref{wilsone6}) and (\ref{wilsone61})
that yields the $E_6$ model, we get an $SU(6)$ model instead, 
since this Wilson line happens to break $E_6$ to $SU(6)\otimes U(1)$.
So, after the ${\bf Z}_6$ twist (\ref{third10}) acting on this model, the 
final orbifold model is a $3$-family
$SU(6)$ model, with gauge symmetry $SU(2)_1 \otimes SU(2)_3 \otimes 
SU(6)_3 \otimes U(1)^3$.
The massless spectrum of this model is given in the second column in 
Table III. Note that the (semi)hidden sector is enhanced in this model
compared with the $E_6$ model that we started with. Also note that the spectrum
of this $SU(6)$ model is similar to that of the $SO(10)$ model given in the
second column of Table I. In particular, the massless spectrum of the
$SU(6)$ model 
can be obtained from that of the 
$SO(10)$ model via replacing $SO(10)_3 \otimes U(1)$
(the last $U(1)$) by $SU(6)\otimes U(1)$ (the last $U(1)$). Similarly,
if in the Wilson line (\ref{tent5}) we shifted (by ${\tilde e}^2$)
not the first entry but the second one, we would get an $SU(6)$ model whose
spectrum can be obtained from that of the $SO(10)$ model given in the first
column of Table I via replacing $SO(10)_3 \otimes U(1)$
by $SU(6)\otimes U(1)$.

{}Adding certain Wilson lines can sometimes enhance the 
hidden/messenger sector
gauge group (while changing the observable sector gauge group).
For example, if instead of the above $V_3$ 
Wilson line we add
\begin{equation}
 \label{tent6}
 V_3 =( 0 (0~0)^3 \vert\vert {\tilde e}^2~(-{\tilde e}^2) ~0~ \vert
 ({1\over 3}~{1\over 3}~{1\over 3}~{1\over 3}~{2\over 3})^3_r \vert 0_r)
\end{equation}
to the $N1$ model, the resulting Narain model will have the gauge group
$SU(6) \otimes SU(5)^3 \otimes U(1)^5$, and the corresponding ${\bf Z}_6$
orbifold model (whose observable sector would contain the gauge group
$SU(5)_3$) will have an enhanced hidden/messenger sector. This may be 
important for phenomenology, in particular, supersymmetry breaking in the
hidden sector. The analysis of such models is underway and will be reported 
elsewhere.

{}If one prefers, one may modify the above Wilson lines (or add new 
Wilson lines) to break the the grand unified gauge group further to 
$SU(3) \otimes SU(2)\otimes U(1)$. One may also choose to break 
$SO(10)$ to $SU(2)^2\otimes SU(4)$. This can also be done by giving the 
adjoint Higgs appropriate expectation values in the effective field theory.
By now, the procedure should be clear.

\widetext
\section{Summary and Remarks}
\bigskip

{}We have seen that the rules for asymmetric orbifold model-building can be 
considered as a generalization of 
those for the free fermionic string models. These rules
are summarized at the end of
section III. They are useful to keep track of the various phases
in the partition function, and become very handy for working out
the spectrum when the orbifolds get complicated.
Using these rules, we construct the three-family grand unified string 
models. We consider three types of ${\bf Z}_6$ orbifolds, namely the
three sets of vectors (\ref{firsten}), (\ref{alts}) and (\ref{third10}),
each acting on an appropriate $N=4$ supersymmetric model. If the $N=4$
model has gauge group $SO(10)^3$ ($SU(5)^3$), the resulting model had a
gauge group $SO(10)_3$ ($SU(5)_3$). The $SO(10)_3$ models with an 
asymptotically-free hidden sector (namely $SU(2)_1$) are presented 
in Tables I and II.
Using the third set of vectors (\ref{third10}), we 
also obtain a $3$-family $(E_6)_3$ model and a $3$-family
$SU(6)_3$ model. They are presented in Table III. Other models with
smaller gauge groups can be obtained from the above models via either the 
string approach ({\em i.e.}, changing moduli or Wilson lines) or 
the field theory approach ({\em i.e.}, spontaneous symmetry breaking).
We give such an $SU(5)$ example in Table III.

{}Although the only way to obtain level-$3$ $SO(10)$ current algebra is via 
an outer automorphism used in our construction, it has been
pointed out that $SO(10)$ at higher levels (say, $K=4$) can be 
constructed via different (more ``economical'') embeddings \cite{dien}. 
However, it seems unlikely that we can get a three-family $SO(10)_4$ model,
especially if we impose
any of the other rather weak phenomenological requirements. So
a distinct feature of grand unification in string theory is the very limited
number of possibilities. At the same time, the models are quite complicated.
This complicated structure hopefully bodes well for phenomenology.

{}There are still a number of important questions concerning grand
unification in string theory. In particular, the complete moduli space
that has a grand unified gauge symmetry with $3$-families, 
an asymtotically-free hidden sector, 
and an intermediate/messenger sector should be classified. 
The three types of asymmetric orbifolds considered in this
work may not exhaustive.
Also, the couplings in each case should be worked out, following the
method already developed\cite{fuse}. Both problems are relatively 
straightforward. A more challenging problem is the string dynamics.
With recent advances in understanding non-perturbative
dynamics, presumably this is also within reach.

\acknowledgements

This work was supported in part by the National Science Foundation.


\newpage
\appendix \section{World-Sheet Fermions and Bosons}
\bigskip

\subsection{Free Fermions}
\medskip

{}Consider a single free left-moving complex fermion with the monodromy
\begin{equation}\label{monofer}
 \psi_v (z e^{2\pi i} )=e^{-2\pi i v}\psi_v (z)~,~~~-{1\over{2}}\leq v<
 {1\over{2}}~.
\end{equation}
The field $\psi_v (z)$ has the following mode expansion
\begin{equation}
 \psi_v (z) =\sum_{n=1}^{\infty} \{ b_{n+v-1/2} z^{-(n+v)} +
 d^{\dagger}_{n-v-1/2} z^{n-v-1} \}~.
\end{equation}
Here $b^{\dagger}_r$ and $d^{\dagger}_s$ are creation operators, and $b_r$ and
$d_s$ are annihilation operators. The quantization conditions read
\begin{equation}
 \{ b^{\dagger}_r ,b_{r^\prime} \}=\delta_{r r^\prime} ,~~~
 \{ d^{\dagger}_s ,d_{s^\prime} \}=\delta_{s s^\prime} ,~~~
 \mbox{others vanish}.
\end{equation}
The Hamiltonian $H_v$ and fermion number operator $N_v$ are 
given by 
\begin{eqnarray}
 &&H_v =\sum_{n=1}^{\infty} \{ (n+v-{1\over 2})
   b^{\dagger}_{n+v-1/2} b_{n+v-1/2}
   +(n-v-{1\over 2})d^{\dagger}_{n-v-1/2} d_{n-v-1/2} \} 
   +{v^2 \over{2}}-{1\over{24}}~,\\
 &&N_v=\sum_{n=1}^{\infty} \{ b^{\dagger}_{n+v-1/2} b_{n+v-1/2}
   -d^{\dagger}_{n-v-1/2} d_{n-v-1/2} \} ~.
\end{eqnarray}
Note that the vacuum energy is ${v^2 \over{2}}-{1\over{24}}$.
Also note that for a Ramond fermion ($v=-1/2$) the vacuum is 
degenerate: There are two ground states $\vert 0\rangle$ and $b^{\dagger}_0
\vert 0\rangle$.
 
{}The operator $-N_v$ is the generator of $U(1)$ rotations.
The corresponding characters read
\begin{eqnarray}\label{fermionZ}
 Z^v_u =&&\mbox{Tr}(q^{H_v} g^{-1} (u))= \mbox{Tr}(q^{H_v} \exp 
 (-2\pi i u N_v ))= \nonumber\\
 &&q^{{v^2 \over{2}}-{1\over{24}}} \prod_{n=1}^{\infty} (1+q^{n+v-1/2}
 e^{-2\pi i u} ) (1+q^{n-v-1/2} e^{2\pi i u} ) ~.
\end{eqnarray}

{}Under the generators of modular transformations ($q=\exp(2\pi i \tau$))
\begin{equation}
 S:\tau\rightarrow -1/\tau ~,~~~T:\tau\rightarrow \tau+1~,
\end{equation}
the characters (\ref{fermionZ}) transform as
\begin{eqnarray}
 &&Z^v_u \stackrel{S}{\rightarrow} e^{2\pi ivu} Z^u_{-v} ~,\\
 &&Z^v_u \stackrel{T}{\rightarrow} e^{2\pi i({v^2 \over{2}}-{1\over{24}})}
   Z^v_{u-v-1/2} ~.
\end{eqnarray}

{}In the cases where $v=-1/2$ (Ramond fermion) or  
$v=0$ (Neveu-Schwarz fermion) in (\ref{monofer}) the complex fermion $\psi_v
(z)$ can be represented in terms of a linear combination of two real 
fermions. The corresponding characters for real fermions then are square roots 
of the characters $Z^v_u$ for the complex fermions ($v$ and $u$ being
$-1/2$ or $0$). A more detailed discussion of the real fermion characters
is given in \cite{KLST}.

\subsection{Twisted Bosons}
\medskip

{}Consider a single free left-moving complex boson with the monodromy
\begin{equation}\label{monobos}
 \partial \phi_v (z e^{2\pi i} )=e^{-2\pi i v}\partial \phi_v (z)~,
 ~~~0\leq v<1~.
\end{equation}
The field $\partial \phi_v (z)$ has the following mode expansion
\begin{eqnarray}
 i\partial \phi_v (z) =&&\delta_{v,0} p z^{-1} + (1-\delta_{v,0} ) \sqrt{v}
 \, b_v z^{-v-1} + \nonumber\\
 &&\sum_{n=1}^{\infty} \{ {\sqrt{n+v}}\,
 b_{n+v} z^{-n-v-1} +{\sqrt{n-v}}\, d^{\dagger}_{n-v} z^{n-v-1} \}~.
\end{eqnarray}
Here $b^{\dagger}_r$ and $d^{\dagger}_s$ are creation operators, and $b_r$ and
$d_s$ are annihilation operators. The quantization conditions read
\begin{equation}
 [ b_r ,b^{\dagger}_{r^\prime} ]=\delta_{r r^\prime} ,~~~
 [ d_s ,d^{\dagger}_{s^\prime} ]=\delta_{s s^\prime} ,~~~
 [x^{\dagger} ,p]=[x,p^{\dagger} ]=i,~~~
 \mbox{others vanish}.
\end{equation}
The Hamiltonian $H_v$ and angular momentum operator $J_v$ are given by 
\begin{eqnarray}
 H_v& = &\delta_{v,0} pp^{\dagger} + (1-\delta_{v,0})vb^{\dagger}_{v} b_{v} + 
 \sum_{n=1}^{\infty} \{ (n+v)
   b^{\dagger}_{n+v} b_{n+v} 
   +(n-v)d^{\dagger}_{n-v} d_{n-v} \} +\nonumber\\
 &&{v(1-v) \over{2}}-{1\over{12}} ~,\\
 J_v &=&\delta_{v,0}i(xp^{\dagger}-x^{\dagger} p)-
 (1-\delta_{v,0})b^{\dagger}_{v} b_{v} -
   \sum_{n=1}^{\infty} \{ b^{\dagger}_{n+v} b_{n+v}
   -d^{\dagger}_{n-v} d_{n-v} \} ~.
\end{eqnarray}
Note that the vacuum energy is ${v(1-v) \over{2}}-{1\over{24}}$.

{}The operator $J_v$
is the generator of $U(1)$ rotations.
The corresponding characters read ($v+u\not=0$):
\begin{eqnarray}\label{bosonX}
 X^v_u =&&\mbox{Tr}(q^{H_v} g^{-1} (u))= \mbox{Tr}(q^{H_v} \exp 
 (2\pi i u J_v ))= \nonumber\\
 &&q^{{v(1-v) \over{2}}-{1\over{12}}} 
 (1-(1-\delta_{v,0} ) q^v e^{-2\pi i u} )^{-1} 
 \prod_{n=1}^{\infty} (1-q^{n+v}
 e^{-2\pi i u} )^{-1} (1-q^{n-v} e^{2\pi i u} )^{-1} ~.
\end{eqnarray}

{}Under the generators of modular transformations
the characters (\ref{bosonX}) transform as
\begin{eqnarray}
 X^v_u &\stackrel{S}{\rightarrow}& 
 ( 2\sin(\pi u) \delta_{v,0} +[2\sin(\pi v)]^{-1}
  \delta_{u,0} + (1-\delta_{vu,0})
 e^{-2\pi i(v-1/2)(u-1/2)} ) X^u_{-v} ~,\\
 X^v_u &\stackrel{T}{\rightarrow}&
 e^{2\pi i({{v(1-v)}\over 2} -{1\over{12}})} X^v_{u-v} ~. 
\end{eqnarray}

{}In the cases where $v=-1/2$ or $v=0$ in (\ref{monobos}),
the complex boson  $\phi_v
(z)$ can be represented in terms of a linear combination of two real 
bosons. The corresponding characters for real bosons then are square roots 
of the characters $X^v_u$ for the complex bosons ($v$ and $u$ being
$-1/2$ or $0$). The twisted boson characters with a different overall
normalization were discussed in Ref\cite{RELATE}.

\subsection{Chiral Lattices}
\medskip

{}Consider $d$ free left-moving real bosons with the monodromy
\begin{equation}
 \phi^I_v (z e^{2\pi i} )=\phi^I_v (z) + v^I ~, 
\end{equation}
where $I=1,2,...,d$, and $v^I$ is the $I^{\mbox{th}}$ component of the shift
vector $v$.
The field $\phi^I_v (z)$ has the following mode expansion:
\begin{equation}
 i\phi^I_v (z) =ix^I+(p^I +v^I)\ln(z) -\sum_{n\not=0} {1\over{\sqrt{n}}} 
 a^I_{n} z^{-n} ~.
\end{equation}
Here $a^I_n,~n>0$, are annihilation operators, and $a^I_n,~n<0$, are creation 
operators. In the following
the eigenvalues of the momentum operator $p^I$ are taken to
span an even lattice $\Gamma^d$ (with metric $g_{ij}$).
The quantization conditions read
\begin{equation}
 [ a^I_n ,a^J_{n^\prime} ]=\delta^{IJ} \delta_{n n^\prime} ,
 ~~~[x^I ,p^J]=i\delta^{IJ},~~~
 \mbox{others vanish}.
\end{equation}
The Hamiltonian operator is given by
\begin{equation}
 H_v ={(p+v)^2 \over 2} + \sum_{n=1}^{\infty} n 
 a^I_{-n} a^I_{n} -{d\over{24}}~.
\end{equation}

{}The momentum operator $p$ is the generator of translations. 
The corresponding characters read
\begin{eqnarray}\label{Y_characters}
 Y^v_u =&&\mbox{Tr} (q^{H_v} 
 g^{-1} (u))= \mbox{Tr} (q^{H_v} \exp(-2\pi i pu)) =\nonumber\\
 &&{1\over{\eta^d (q)}} \sum_{p\in\Gamma^d} q^{{1\over2}(p+v)^2}
 \exp(-2\pi i pu)~.
\end{eqnarray}

{}Let $w_a \in{\tilde\Gamma}^d,~a=1,...,M-1$, be a 
set of vectors such that $\Gamma^d_0 \oplus \Gamma^d_1 \oplus ...
\oplus \Gamma^d_{M-1} ={\tilde \Gamma}^d$, where $w_0$ is the null vector 
($w^I_0 \equiv 0$), and $\Gamma^d_a \equiv \{w_a +p \vert p\in\Gamma^d\}$,
$a=0,1,...,M-1$
(Thus, $w_a \notin 
\Gamma^d$ for $a\not=0$; also note  that $M=\det(g_{ij})$).
Consider the set of characters $Y^{v+w_a }_u$:
\begin{eqnarray}\label{chiral_characters}
 &&Y^{v+w_a}_u \stackrel{T}{\rightarrow} \exp(2\pi i({1\over2} (w_a +v)^2
 -{d\over{24}})) Y^{v+w_a}_{u-v} ~,\\
 &&Y^{v+w_a}_u \stackrel{S}{\rightarrow} \sum_{b=0}^{M-1} 
 S_{ab} (v,u)Y^{u+w_b}_{-v} ~,
\end{eqnarray}
where
\begin{equation}
 S_{ab} (v,u)=
 M^{-{1\over 2}} \exp(2\pi i (w_a +v )(w_b +u))~.
\end{equation}

{}Let $N$ the 
smallest positive integer such that $\forall a~Nw^2_a \in 2{\bf Z}$.
If $N=1$ (in which case $\Gamma^d$ is an even self-dual lattice
with $M=\det(g_{ij})=1$), 
we will use the characters ${Y}^v_u$ defined in (\ref{Y_characters})
whose modular transformations are particularly simple for $N=1$:
\begin{eqnarray}\label{self-dual}
 &&{Y}^v_u \stackrel{T}{\rightarrow} \exp(2\pi i({1\over2} v^2
 -{d\over{24}})) {Y}^{v}_{u-v} ~,\\
 &&{Y}^{v}_u \stackrel{S}{\rightarrow} \exp(2\pi ivu) {Y}^u_{-v} ~.
\end{eqnarray}

{}If $N>1$, the set of characters $Y^{v+w_a}_u$ is such that the 
$T$-transformation is diagonal (with respect to $a$), 
whereas the $S$-transformation is not. 
There exists a basis such that 
both $S$- and $T$-transformations act as permutations. In particular, 
consider the case where $N$ is a prime. 
In the discussion of asymmetric orbifolds we will use the set of characters 
\begin{eqnarray}\label{CHIRAL_CHARACTERS}
 &&Y^{0, v}_{\sigma , u} \equiv Y^v_u ~,\\ 
 &&Y^{\lambda , v}_{\sigma , u} \equiv \sum_{a=0}^{M-1} 
 \exp(-2\pi i \lambda (uw_a +{1\over2}\sigma w^2_a )) Y^{v+\lambda w_a}_u ~,
 ~~~\lambda\not=0~,
\end{eqnarray}
where $\lambda$ and $\sigma$ are integers taking values between $0$ and $N-1$, 
such that $\lambda+\sigma\not=0$.
The modular transformation properties of $Y^{\lambda , v}_{\sigma , u}$ read 
\begin{eqnarray}
 &&Y^{\lambda , v}_{\sigma , u}
 \stackrel{T}{\rightarrow} \exp(2\pi i({1\over 2}v^2 -{d\over{24}}))
 Y^{\lambda , v}_{\sigma-\lambda , u-v}~,\\
 &&Y^{\lambda , v}_{\sigma , u}
 \stackrel{S}{\rightarrow} \{M^{-{1\over 2}}
\delta_{\lambda ,0} +
 M^{1\over 2} \delta_{\sigma ,0} + (1-\delta_{\lambda\sigma,0})
 \exp(2\pi i \chi(\lambda, \sigma))\}
 \exp(2\pi i vu) Y^{\sigma , u}_{-\lambda , -v}~,
\end{eqnarray}
where $Y^{\lambda , v}_{\sigma , u} \equiv Y^{\lambda+N , v}_
{\sigma , u} \equiv Y^{\lambda , v}_{\sigma+N  , u}$, and
\begin{equation}
 \exp(2\pi i \chi(\lambda, \sigma) ) 
 \equiv 
 M^{-{1\over 2}}\sum_{a=0}^{M-1} 
 \exp(-2\pi i {1\over 2} \lambda \sigma w^2_a ) ~,~~~\lambda\sigma\not=0 ~.
\end{equation} 
Note that $\chi(\lambda, \sigma)$ are real numbers, and $\chi(\lambda, \lambda)
\equiv -d/8$.

{}To illustrate the above discussion we note that the root lattices 
of simply-laced Lie groups are even. The groups that have prime $N$
are the following: ({\em i})
$SU(n)$, $n$ is an odd prime, and $N=n$; 
({\em ii}) $E_6$, $N=3$; 
({\em iii}) SO(8n), $N=2$; ({\em iv}) $E_8$, $N=1$.  

{}Similar considerations apply to right-moving chiral lattices, and also
Lorentzian lattices. In the latter case all the scalar products of vectors
are understood with respect to the Lorentzian signature. 


\newpage
\section{Construction of the Alternative $SO(10)_3$}
\bigskip

{}Let us make a comment concerning the alternative $SO(10)_3$ model 
presented in section VII. There the construction,
{\em i.e.}, the set $\{V_i,k_{ij}\}$, satisfies the simplified rules,
and the condition (\ref{simpli}) follows. This same model may be 
obtained by a different set of $\{V_i,k_{ij}\}$
acting on the $N1$ model. The vectors are, besides $V_0$,

\begin{eqnarray}
 &&V_1 =( 0 (-{1\over 3}~{1\over 3})^3 \vert\vert {e_1\over 3}
  0 ({1\over 3}) \vert
 ({1\over 3})^5 ~0^5_r ~({2\over 3})_r )~, \nonumber \\
 &&V_2 =( 0 (0~0)(-{1\over 2}~{1\over 2})^2 \vert\vert {e_1\over 2}
  ({1\over 2})^2 \vert 0^5 ~0^5_r ~0_r )~,\nonumber \\
 &&W_1 =( 0 (0~{1\over 2})^3 \vert\vert 0^2 ~({1\over 2}) \vert
 ({1\over 2})^5 ~0^5_r ~0_r )~,\nonumber \\
 &&W_2 =( 0 (0~0)(0~{1\over 2})^2 \vert\vert 0
 ({1\over 2})^2 \vert 0^5 ~0^5_r ~0_r )~.\nonumber
\end{eqnarray}

{}With the appropriate choice of $k_{ij}$, this gives the same model as the
set $\{V_i,k_{ij}\}$ given in section VII, with its massless spectrum shown 
in the second column in Table I.

{}The invariant sublattice $I(V_1)$ is the same in both cases,
given by Eq.(\ref{amil}). However, the invariant sublattices for the 
other sectors are different in these two constructions.
For the set $\{V_i,k_{ij}\}$ given in section VII, $I(V_1+V_2)=I(V_1)$
and $I(V_2)=\Gamma^4 \otimes \Gamma^{2,18}$, where $\Gamma^{2,18}$ is 
even self-dual. For the set $\{V_i,k_{ij}\}$ given here,
$I(V_1+V_2)=\Gamma^2 \otimes \Gamma^6$, and 
$I(V_2)=\Gamma^{2,2}\otimes \Gamma^{16}$, 
where $\Gamma^{16}$ corresponds to the
$SO(10)^3\otimes SO(2)$ lattice, which is even but not self-dual. Now
this set $\{V_i,k_{ij}\}$ does not satisfy the simplified rules, but it
does satisfy the more general set of rules given in section III. 
To check these points, it is convenient to have
the values for $\chi({\overline {\alpha V}},{\overline {\beta V}})$,
$f({\overline {\alpha V}}, {\overline {\beta V}})$ and 
$\xi^{\overline {\alpha V}}_{\overline {\beta V}}$ in both cases.
They are given in Table IV below. There, the first column corresponds to the
$\{V_i,k_{ij}\}$ given in section VII and the second column corresponds
to the $\{V_i,k_{ij}\}$ given in this appendix.


\newpage
\widetext

\begin{table}[t]
\begin{tabular}{|c||l|l|l|} 
 & & & \\ 
M & $SU(2)^2 \otimes SO(10)\otimes U(1)^3$ &
  $SU(2)^2 \otimes SO(10) \otimes U(1)^3$ &
  $SU(2) \otimes SO(10) \otimes U(1)^4$ \\  \hline
 & & & \\
   & $({\bf 1},{\bf 1},{\bf 45})(0,0,0)$ & $({\bf 1},{\bf 1},{\bf 45})(0,0,0)$
   & $ ({\bf 1},{\bf 45})(0,0,0,0)$ \\
   & $({\bf 1},{\bf 3},{\bf 1})(0,0,0)$  &  $({\bf 1},{\bf 3},{\bf 1})(0,0,0)$
   & $ ({\bf 1},{\bf 1})(0,0,0,0)_L$ \\
 $U$ & $ ({\bf 1},{\bf 1},{\bf 1})(0,-6,0)_L$ & 
     $ ({\bf 1},{\bf 1},{\bf 1}) (0,+6,0)_L$ 
   & $ ({\bf 1},{\bf 1})(0,+6,0,0)_L$ \\
   & $2 ({\bf 1},{\bf 4},{\bf 1})(0,+3,0)_L$ & 
     $2 ({\bf 1},{\bf 1},{\bf 1})({\pm 3},-3,0)_L$ &
     $2 ({\bf 1},{\bf 1})(0,-3,{\pm 3},0)_L$ \\
   & $2 ({\bf 1},{\bf 2},{\bf 1})(0,-3,0)_L$ &
     $2 ({\bf 2},{\bf 2},{\bf 1})(0,0,0)_L$ & \\
 & & & \\ \hline
 & & & \\
   & $2 ({\bf 1},{\bf 2},{\bf 16})(0,-{1},-{1})_L$ &
     $2 ({\bf 1},{\bf 1},{\bf 16})({\pm 1},+1,-{1})_L$ &
     $2 ({\bf 1},{\bf 16})(0,+{1},{\pm 1},-1)_L$ \\
   & $2 ({\bf 1},{\bf 2},{\bf 10})(0,-{1},+2)_L$ &
     $2 ({\bf 1},{\bf 1},{\bf 10})({\pm 1},+1,+2)_L$ &
     $2 ({\bf 1},{\bf 10})(0,+{1},{\pm 1},+{2})_L$ \\
  $T3$ & $2 ({\bf 1},{\bf 2},{\bf 1})(0,-{1},-{4})_L$ &
     $2 ({\bf 1},{\bf 1},{\bf 1})({\pm 1},+1,-{4})_L$ & 
     $2 ({\bf 1},{\bf 1})(0,+{1},{\pm 1},-4)_L$ \\
   & $ ({\bf 1},{\bf 1},{\bf 16})(0,+{2},-{1})_L$ &
     $ ({\bf 1},{\bf 1},{\bf 16})(0,-{2},-{1})_L$ & 
     $ ({\bf 1},{\bf 16})(0,-{2},0,-{1})_L$ \\
   & $ ({\bf 1},{\bf 1},{\bf 10})(0,+{2},+{2})_L$ &
     $ ({\bf 1},{\bf 1},{\bf 10})(0,-{2},+{2})_L$ &
     $ ({\bf 1},{\bf 10})(0,-{2},0,+{2})_L$ \\
   & $({\bf 1},{\bf 1},{\bf 1})(0,+{2},-{4})_L$ &
      $({\bf 1},{\bf 1},{\bf 1})(0,-{2},-{4})_L$ &
      $ ({\bf 1},{\bf 1})(0,-{2},0,-4)_L$ \\
 & & & \\  \hline
 & & & \\
  & $ ({\bf 1},{\bf 1},{\overline {\bf 16}}) (\pm 1,+{1},+{1})_L$ &
      $ ({\bf 1},{\bf 2},{\overline {\bf 16}}) ( 0,-{1},+{1})_L$ &
      $ ({\bf 1},{\overline {\bf 16}}) (\pm 1,-{1},0,+{1})_L$ \\
 $T6$  &$ ({\bf 1},{\bf 1},{\bf 10})(\pm 1,+{1},-{2})_L$ &
      $ ({\bf 1},{\bf 2},{\bf 10})( 0,-{1},-{2})_L$ &
      $ ({\bf 1},{\bf 10})(\pm 1,-{1},0,-{2})_L$ \\
   &  $({\bf 1},{\bf 1},{\bf 1})(\pm 1,+{1},+{4})_L$ &
      $ ({\bf 1},{\bf 2},{\bf 1})( 0,-{1},+{4})_L$ &
      $ ({\bf 1},{\bf 1})(\pm 1,-{1},0,+{4})_L$ \\
 & & & \\  \hline 
 & & & \\
   &  $({\bf 2},{\bf 2},{\bf 1})(0,0,0)_L$ &
      $({\bf 2},{\bf 1},{\bf 1})(\pm 3,0,0)_L$ &
  $({\bf 2},{\bf 1})(0,0,\pm {3},0)_L$ \\
 $T2$ &  $({\bf 2},{\bf 4},{\bf 1})(0,0,0)_L$ &
       $({\bf 1},{\bf 4},{\bf 1})(0,+3,0)_L$ &
 $({\bf 1},{\bf 1})(\pm {3},+{3},0,0)_L$ \\
   & $({\bf 1},{\bf 1},{\bf 1})(\pm 3,-3,0)_L$ &
   $ 2 ({\bf 1},{\bf 2},{\bf 1})(0,-3,0)_L$ & \\
   &   & $({\bf 1},{\bf 2},{\bf 1})(0,+3,0)_L$ & \\
 & & & \\ \hline
& & & \\
 $U(1)$ & $(1/ \sqrt{6},~1/{3\sqrt{2}},~1/6)$ &
    $(1/ \sqrt{6},~1/{3\sqrt{2}},~1/6)$ &
  $(1/ \sqrt{6},~1/{3\sqrt{2}}, ~1/\sqrt{6},~1/6)$~ \\
\end{tabular}
\caption{The massless spectra of $SO(10)$ models from three different 
${\bf Z}_6$ orbifolds : 
(1) and (2) $SU(2)_1\otimes SU(2)_3 \otimes SO(10)_3\otimes U(1)^3$,
and (3) $SU(2)_1 \otimes SO(10)_3 \otimes U(1)^4$. The $U(1)$ normalization 
radii are given at the bottom of the table.
The graviton, dilaton and gauge supermultiplets are not shown.}

\end{table}

\newpage
\widetext

\begin{table}[t]
\begin{tabular}{|c|l|l|l|} 
 & & & \\
 M & $SU(2) \otimes SO(10) \otimes U(1)^4$ & 
      $SU(2) \otimes SO(10) \otimes U(1)^3$ &
  $SU(2) \otimes SO(10) \otimes U(1)^2$ \\ \hline
 & & & \\
 R & $SU(3)\otimes SU(3)$ & $SU(3)\otimes U(1)^2$ & $U(1)^4$ \\ \hline   
 & & & \\
 & & & \\
   & $ ({\bf 1},{\bf 45})(0,0,0,0)$ 
    & $ ({\bf 1},{\bf 45})(0,0,0)$
    & $ ({\bf 1},{\bf 45})(0,0)$ \\
   & $2 ({\bf 1},{\bf 1})(0,+12,0,0)_L$ 
    & $2 ({\bf 1},{\bf 1})(0,+12,0)_L$
    & $4 ({\bf 1},{\bf 1})(0,0)$ \\
 $U$ & $2 ({\bf 1},{\bf 1})(0,0,+12,0)_L$ 
    &  $2 ({\bf 1},{\bf 1})(0,0,0)$ & \\
   & $3 ({\bf 1},{\bf 1})(0,-6,0,0)_L$ 
    &  $3 ({\bf 1},{\bf 1})(0,-6,0)_L$ & \\
   & $3 ({\bf 1},{\bf 1})(0,0,-6,0)_L$ & & \\ 
 & & & \\
\hline
 & & & \\
   &  $2 ({\bf 1},{\bf 16})(0,+{2},+{2},-1)_L$ 
    & $3 ({\bf 1},{\bf 16})(0,+{2},-1)_L$ 
     & $5 ({\bf 1},{\bf 16})(0,-1)_L$  \\
   & $2 ({\bf 1},{\bf 10})(0,+2,+2,+{2})_L$ 
    & $3 ({\bf 1},{\bf 10})(0,+2,+{2})_L$ 
    & $5 ({\bf 1},{\bf 10})(0,+{2})_L$ \\
   & $2 ({\bf 1},{\bf 1})(0,+{2},+{2},-4)_L$ 
    & $3 ({\bf 1},{\bf 1})(0,+{2},-4)_L$
    & $5 ({\bf 1},{\bf 1})(0,-4)_L$ \\
   & $ ({\bf 1},{\bf 16})(0,-{4},-{4},-{1})_L$ 
    & $2 ({\bf 1},{\bf 16})(0,-{4},-{1})_L$ & \\
 $T3$ & $ ({\bf 1},{\bf 10})(0,-{4},-{4},+{2})_L$ 
    & $ 2 ({\bf 1},{\bf 10})(0,-{4},+{2})_L$ & \\
   & $ ({\bf 1},{\bf 1})(0,-{4},-{4},-4)_L$ 
   &  $ 2 ({\bf 1},{\bf 1})(0,-{4},-4)_L$ & \\
   & $ ({\bf 1},{\bf 16})(0,-{4},+{2},-{1})_L$ & &\\
   & $ ({\bf 1},{\bf 10})(0,-{4},+{2},+{2})_L$ & & \\
   & $ ({\bf 1},{\bf 1})(0,-{4},+2,-{4})_L$ & & \\
   & $ ({\bf 1},{\bf 16})(0,+{2},-{4},-{1})_L$ & & \\
   & $ ({\bf 1},{\bf 10})(0,+{2},-{4},+{2})_L$ & & \\
   & $ ({\bf 1},{\bf 1})(0,+{2},-{4},-{4})_L$  & & \\
 & & & \\
 \hline
  & & & \\
   &  $ ({\bf 1},{\overline {\bf 16}}) (\pm 1,+{1},+{1},+{1})_L$ 
    & $ ({\bf 1},{\overline {\bf 16}}) (\pm 1,+{1},+{1})_L$
    & $ ({\bf 1},{\overline {\bf 16}}) (\pm 1,+{1})_L$ \\
 $T6$ &  $ ({\bf 1},{\bf 10})(\pm 1,+{1},+{1},-{2})_L$ 
     &  $ ({\bf 1},{\bf 10})(\pm 1,+{1},-{2})_L$ 
      &  $ ({\bf 1},{\bf 10})(\pm 1,-{2})_L$ \\
   & $ ({\bf 1},{\bf 1})(\pm 1,+{1},+{1},+{4})_L$ 
    & $ ({\bf 1},{\bf 1})(\pm 1,+{1},+{4})_L$ 
    & $ ({\bf 1},{\bf 1})(\pm 1,+{4})_L$ \\
 & & & \\
 \hline
 & & & \\
   &   $2 ({\bf 2},{\bf 1})(0,-{3},-{3},0)_L$ 
    & $2 ({\bf 2},{\bf 1})(0,{\pm 3},0)_L$ 
    & $6 ({\bf 2},{\bf 1})(0,0)_L$ \\
 $T2$ &  $  ({\bf 2},{\bf 1})(0,\pm {9},+{3},0)_L$ 
    & $ ({\bf 2},{\bf 1})(0,\pm {9},0)_L$ & \\
   &  $  ({\bf 2},{\bf 1})(0,+{3},\pm {9},0)_L$ & & \\
   &  $ ({\bf 1},{\bf 1})(\pm 3,-{3},-{3},0)_L$ 
    &  $ ({\bf 1},{\bf 1})(\pm 3,-{3},0)_L$
    &  $ ({\bf 1},{\bf 1})(\pm 3,0)_L$ \\ 
 & & & \\
\hline
 & & & \\
 $U(1)$ & $(1/ \sqrt{6},~1/{6\sqrt{2}}, ~1/{6\sqrt{2}},~1/6)$
   & $(1/ \sqrt{6},~1/{6\sqrt{2}},~1/6)$
   & $(1/ \sqrt{6},~1/6)$ \\
\end{tabular}
\caption{The massless spectra of the three models connected by a continuous
parameter:
(1) $SU(2)_1 \otimes SO(10)_3\otimes U(1)^4$,
(2) $SU(2)_1 \otimes SO(10)_3 \otimes U(1)^3$, 
and (3) $SU(2)_1 \otimes SO(10)_3 \otimes U(1)^2$. For these models
the gauge group $R$ in the original moduli space is 
also shown. 
The $U(1)$ normalization
radii are given at the bottom of the Table. 
The gravity, dilaton and gauge supermultiplets are not shown.}

\end{table}


\newpage
\widetext

\begin{table}[t]
\begin{tabular}{|c|l|l|l|}
 & & & \\
 M & $SU(2) \otimes E_6 \otimes U(1)^3$ &
      $SU(2)^2 \otimes SU(6) \otimes U(1)^3$ &
  $SU(2) \otimes SU(5) \otimes U(1)^5$ \\ \hline
 & & & \\
   & $ ({\bf 1},{\bf 78})(0,0,0)$
    & $ ({\bf 1},{\bf 1},{\bf 35})(0,0,0)$
    & $ ({\bf 1},{\bf 24})(0,0,0,0,0)$ \\
  $U$ & $2 ({\bf 1},{\bf 1})(0,-{3},\pm 3)_L$
    & $2 ({\bf 1},{\bf 1},{\bf 1})(-{3},\pm 3,0)_L$
    & $2 ({\bf 1},{\bf 1})(0,-{3},\pm 3,0,0)_L$ \\
   & $ ({\bf 1},{\bf 1})(0,+6,0)_L$
    &  $ ({\bf 1},{\bf 1},{\bf 1})(+6,0,0)_L$
    & $ ({\bf 1},{\bf 1})(0,+6,0,0,0)_L$ \\
   &  &  $ 2({\bf 2},{\bf 2},{\bf 1})(0,0,0)_L$ 
   & $2 ({\bf 1},{\bf 1})(0,0,0,0,0)_L$ \\
 & &$({\bf 1},{\bf 3},{\bf 1})(0,0,0)$ & \\
\hline
 & & & \\
   &  $ ({\bf 1},{\bf 27})(0,-{2},0)_L$
    & $ ({\bf 1},{\bf 1},{\overline {\bf 6}})(-{2},0,\pm {1})_L$
     & $ ({\bf 1},{\overline {\bf 5}})(0,-{2},0 -{1},+3)_L$  \\
   & $2 ({\bf 1},{\bf 27})(0,+1,\pm {1})_L$
    & $ ({\bf 1},{\bf 1},{\bf 15})(-2,0,0)_L$
    & $ ({\bf 1},{\bf 10})(0,-2,0,-1,-1)_L$ \\
   &   & $2 ({\bf 1},{\bf 1},{\overline {\bf 6}})(+1,\pm 1,\pm 1)_L$
    & $ ({\bf 1},{\bf 1})(0,-2,0,-1,-5)_L$ \\
   &  & $2 ({\bf 1},{\bf 1},{\bf 15})(+1,\pm 1,0)_L$ 
   &  $ ({\bf 1},{\bf 5})(0,-2,0,+2,+2)_L$ \\
 $T3$ & &  & $  ({\bf 1},{\overline {\bf 5}})(0,-2,0,+{2},-2)_L$ \\
   & &  &  $ ({\bf 1},{\bf 1})(0,-2,0,-4,0)_L$ \\
   & & & $2 ({\bf 1},{\overline {\bf 5}})(0,+1,\pm 1,-1,+3)_L$  \\
   & & & $2 ({\bf 1},{\bf 10})(0,+1,\pm 1,-1,-1)_L$ \\
   & & & $2 ({\bf 1},{\bf 1})(0,+1,\pm 1,-1,-5)_L$ \\
   & & & $2 ({\bf 1},{\bf 5})(0,+1, \pm 1,+2,+2)_L$ \\
   & & & $2 ({\bf 1},{\overline {\bf 5}})(0,+1,\pm 1,+{2},-2)_L$ \\
   & & & $2 ({\bf 1},{\bf 1})(0,+1, \pm 1,-4,0)_L$ \\
 & & & \\
 \hline
  & & &\\
   &  $ ({\bf 1},{\overline {\bf 27}}) (\pm 1,-1,0)_L$
    & $ ({\bf 1}, {\bf 2},{\bf 6}) (-{1},0, \pm 1)_L$
    & $ ({\bf 1},{\bf 5}) (\pm 1,-{1},0,+1,-3)_L$ \\
   & &  $ ({\bf 1},{\bf 2},{\overline {\bf 15}})(-{1},0,0)_L$
      &  $ ({\bf 1},{\bf 10})(\pm 1,-1,0,+1,+1)_L$ \\
    $T6$ & & & $({\bf 1},{\bf 1})(\pm 1,-1,0,+1,+5)_L$ \\
   & & & $({\bf 1},{\bf 5})(\pm 1,-1,0,-2,+2)_L$ \\
   & & & $ ({\bf 1},{\overline {\bf 5}})(\pm 1,-1,0,-{2},-2)_L$ \\
   & & & $ ({\bf 1},{\bf 1})(\pm 1,-1,0,+4,0)_L$ \\
 & & & \\
 \hline
 & & & \\
   $T2$ & $({\bf 2},{\bf 1})(0,0,{\pm 3})_L$ 
    & $({\bf 2},{\bf 1},{\bf 1})(0,{\pm 3},0)_L$
    & $({\bf 2},{\bf 1})(0,0,{\pm 3},0,0)_L$ \\
  &  $ ({\bf 1},{\bf 1})(\pm {3},+{3},0)_L$
    & $({\bf 1},{\bf 4},{\bf 1})(+{3},0,0)_L$
    & $({\bf 1},{\bf 1})(\pm {3},+{3},0,0,0)_L$ \\
  & & $2({\bf 1},{\bf 2},{\bf 1})(-3,0,0)_L$ & \\
  & & $({\bf 1},{\bf 2},{\bf 1})(+3,0,0)_L$ & \\
\hline
 & & & \\
 $U(1)$ & $(1/ \sqrt{6}, ~1/{3\sqrt{2}}, ~1/\sqrt{6})$
   & $(1/{3\sqrt{2}}, 1/\sqrt{6}, 1/\sqrt{6})$
   & $(1/\sqrt{6}, 1/{3\sqrt{2}},1/\sqrt{6},1/{6},1/{2\sqrt{15}})$ \\
\end{tabular}
\caption{The massless spectra of the $E_6$, $SU(6)$ and $SU(5)$ models 
from the same ${\bf Z}_6$ orbifold:
(1) $SU(2)_1 \otimes (E_6)_3\otimes U(1)^3$,
(2) $SU(2)_1 \otimes SU(2)_3 \otimes SU(6)_3 \otimes U(1)^3$,
and (3) $SU(2)_1 \otimes SU(5)_3 \otimes U(1)^5$.
The $U(1)$ normalization
radii are given at the bottom of the Table.
The gravity, dilaton and gauge supermultiplets are not shown.}

\end{table}


\begin{table}
\begin{tabular}{|c|c|c|c|c|c|c|c|c|c|c|c|c|}

 ${\overline {\alpha V}} / {\overline {\beta V}}$ 
 & \multicolumn{2}{c|}{${\bf 0}$} &
 \multicolumn{2}{c|}{${\overline {2V_1 +V_2}}$} &
 \multicolumn{2}{c|}{$V_1$} &
 \multicolumn{2}{c|}{$V_2$} &
 \multicolumn{2}{c|}{${\overline {2V_1}}$} &
 \multicolumn{2}{c|}{${\overline {V_1 +V_2}}$} \\ \hline

 & 0 & 0 & 0 & 0 & 0 & 0  & 0 & 0 & 0 & 0 & 0 & 0 \\
 ${\bf 0}$ & 0 & 0 & 0 & 0 & 0 & 0  & 0 & 0 & 0 & 0 & 0 & 0 \\
 & 1 & 1 & 1 & 1 & 1 & 1 & 1 & 1 & 1 & 1 & 1 & 1 \\ \hline

 & 0 & 0 & $-{1 \over 4}$ & 0 & ${1 \over 4}$ & 0 & 0 & 0 & $-{1 \over 4}$
 &  0 & ${1 \over 4}$ & 0 \\
 ${\overline {2V_1 +V_2}}$ & 0 & 0 & 0 & 0 & 0 & 0 & ${1 \over 2}$ &
 ${1 \over 2}$  & 0 & 0 & 0 & 0 \\
 & 1 & 2 & 1 & 2 & 1 & 2 & 1 & 2 & 1 & 2 & 1 & 2 \\ \hline

 & 0 & 0 & ${1 \over 4}$ & 0 & $-{1 \over 4}$ &  $-{1 \over 4}$ &
 0 & 0 & ${1 \over 4}$ & ${1 \over 4}$ & $-{1 \over 4}$ & 0 \\
 $V_1$ & 0 & 0 & 0 & 0 & 0 & 0 & ${1 \over 2}$ &
 ${1 \over 2}$ & 0 & 0 & 0 & 0 \\
 & 3 & 3 & 1 & 1 & 3 & 3 & 1 & 1 & 3 & 3 & 1 & 1 \\ \hline

 & 0 & 0 & 0 & 0 & 0 & 0 & ${1 \over 2}$ &
    0 & 0 & 0 & 0 & 0 \\
 $V_2$ & 0 & 0 & ${1 \over 2}$ & ${1 \over 2}$ & ${1 \over 2}$ & ${1 \over 2}$
 & 0 & 0 & ${1 \over 2}$ & ${1\over 2}$ & ${1 \over 2}$ & ${1\over 2}$ \\
 & 2 & 4 & $-{1}$ & $-{2}$ & $-{1}$ & $-{2}$ & 2 & 4 & $-{1}$ & $-{2}$
 & $-{1}$  & $-{2}$ \\ \hline

 & 0 & 0 & $-{1 \over 4}$ & 0 & ${1 \over 4}$ & ${1 \over 4}$ & 0 & 0 &
 $-{1 \over 4}$ & $-{1 \over 4}$ & ${1 \over 4}$ & 0 \\
 ${\overline {2V_1}}$  & 0 & 0 & 0 & 0 & 0 & 0 & ${1 \over 2}$ &
 ${1 \over 2}$ & 0 & 0 & 0 & 0 \\
 & 3 & 3 & 1 & 1 & 3 & 3 & 1 & 1 & 3 & 3 & 1 & 1 \\ \hline

 & 0 & 0 & ${1 \over 4}$ & 0 & $-{1 \over 4}$ & 0 & 0 & 0 & ${1 \over 4}$
 &  0 & $-{1 \over 4}$ & 0 \\
 ${\overline {V_1 +V_2}}$ & 0 & 0 & 0 & 0 & 0 & 0 & ${1 \over 2}$ &
 ${1 \over 2}$ & 0 & 0 & 0 & 0 \\
 & 1 & 2 & 1 & 2 & 1 & 2 & 1 & 2 & 1 & 2 & 1 & 2

\end{tabular}
\caption{ The values $ \chi({\overline {\alpha V}},{\overline {\beta V}})$,
$f({\overline {\alpha V}}, {\overline {\beta V}})$ and
$\xi^{\overline {\alpha V}}_{\overline {\beta V}}$ for the original
construction (first column) and the alternative construction (second column).}

\end{table}


\end{document}